\documentclass[aps,amsmath,amssymb,showpacs,showkeys]{revtex4-2}
\usepackage[dvips]{graphicx}
\usepackage{times}
\usepackage{braket}
\usepackage{xcolor}
\usepackage{orcidlink}
\usepackage{hyperref}
\usepackage{color,soul}
\hypersetup{
  colorlinks=true,
  urlcolor=blue,
  linkcolor=red,
  citecolor=blue
}
\begin{document}
\title{Propagation and Fluxes of Ultra-High Energy Cosmic Rays in  
$\boldsymbol{f(R)}$ Gravity Theory}

\author{Swaraj Pratim Sarmah\orcidlink{0009-0003-2362-2080}}
\email[Email: ]{swarajpratimsarmah16@gmail.com}

\author{Umananda Dev Goswami\orcidlink{0000-0003-0012-7549}}
\email[Email: ]{umananda2@gmail.com}

\affiliation{Department of Physics, Dibrugarh University, Dibrugarh 786004, 
Assam, India}

%\date{}
\begin{abstract}
In this work, we study the effect of diffusion of ultra-high energy 
cosmic ray (UHECR) protons in the presence of turbulent 
magnetic fields (TMFs) in the light of the $f(R)$ theory of gravity. The 
$f(R)$ theory of gravity is a successful modified theory of gravity in 
explaining the various aspects of the observable Universe including its 
current state of expansion. 
For this work, we consider two most studied $f(R)$ gravity models, viz., the 
power-law model and the Starobinsky model. With these two models, we study the 
diffusive character of the propagation of UHECR protons in terms of their 
density enhancement. The density enhancement is a measure of how the 
density of CRs changes due to their diffusion in the intergalactic medium and 
interaction with the cosmic microwave background (CMB) radiation. 
Ankle, instep and Greisen-Zatsepin-Kuzmin (GZK) cutoff are all 
spectrum characteristics that extragalactic UHECRs acquire when they 
propagate through the CMB. We analyse all these characteristics through the 
diffusive flux as well as its modification factor. Model dependence of the 
modification factor is minimal compared to the diffusive flux. We compare the 
UHECR proton spectra calculated for the considered $f(R)$ gravity 
models with the available data of the Telescope Array (TA) and Pierre 
Auger Observatory (PAO). We see that both models of $f(R)$ gravity 
predict energy spectra of UHECRs with all experimentally observed features, 
which lay well within the range of combined data of both experiments 
throughout the energy range of concern. It is to be noted that our present 
work is only to investigate the possible effects of $f(R)$ gravity theory on 
the UHECRs propagation, using pure proton composition as a simplified case 
study since protons are least affected by magnetic fields. However, at this 
stage, our results cannot be used to favor or disfavor $f(R)$ cosmology over 
$\Lambda$CDM cosmology as more work is needed in this regard.
\end{abstract}

%\pacs{}

\maketitle                                                                      

\section{Introduction}
The discovery of cosmic rays (CRs) by V.\ F.\ Hess in 1912 \cite{Hess} is one 
of the most significant milestones in the history of modern physics. 
CRs are charged ionizing particles, mostly consisting of protons, helium, 
carbon and other heavy ions up to iron emanating from outer space. 
Although the discovery of CRs occurred more than 110 years ago, the 
origin, acceleration, and propagation mechanisms of CRs are still not 
clearly known \cite{harari,molerach,berezinskyGK}, especially in the higher 
energy range i.e.\ the energy range $E\geq 0.1$ EeV ($1$ EeV = $10^{18}$ eV). 
The sources of such usually referred ultra-high energy CRs (UHECRs) are not 
established yet \cite{berezinskyGG, nagano, Bhattacharjee, Olinto}. However, 
in the energy range $E \leq 0.1$ EeV, it is assumed that the sources are of 
galactic origin and they are accelerated by supernova explosions 
\cite{s.mollerach}, while those well above this range ($\sim 1$ EeV 
and above) are most probably extragalactic in origin and plausibly to 
accelerate in gamma-ray ($\gamma$-ray) bursts or in active galaxies 
\cite{harari}. 

The energy spectrum of CRs has an extraordinary range of energies. It 
extends over many orders of magnitude from GeV energies up to $100$ 
EeV and exhibits a power-law spectrum. There is a small spectral 
break known as the knee at about $4$ PeV ($1$ PeV = $10^{15}$ eV) and a 
flattening at the ankle at about $5$ EeV. In this spectrum, a strong cutoff 
near $50$ EeV, which is called the Greisen-Zatsepin-Kuzmin (GZK) 
cutoff \cite{Greisen,Zatsepin} appears due to the interaction with cosmic 
microwave background (CMB) photons.

The intergalactic medium (IGM) contains turbulent magnetic fields (TMFs), which
impact significantly the propagation of extragalactic UHECRs. In the 
presence of any random magnetic field, the propagation of a charged particle 
depends on how much distance is traveled by that particle compared with the 
scattering length $\lambda =3D/c$ in the medium, where $D$ denotes
the diffusion coefficient and $c$ is the speed of light in free space 
\cite{Supanitsky}. If the traveled distance of the charged particle is 
much smaller than the scattering length, then the propagation is 
ballistic in nature while that is diffusive if the distance is much larger 
than the scattering length. Consideration of an extragalactic TMF and also 
taking into account the finite density of sources in the study of the 
propagation of UHECRs may result in a low-energy magnetic horizon effect, 
which may allow the observations to be consistent with a higher spectral index 
\cite{s.mollerach, mollerach2013, wittkowski}, closer to the values 
anticipated from diffusive shock acceleration. Other hypotheses rely on the
assumption of acceleration of heavy nuclei by extragalactic sources, which 
then interact with the infrared radiation present in those environments to 
photodisintegration, producing a significant number of secondary nucleons that 
might explain the light composition seen below the ankle \cite{unger, globus}. 
In the presence of an intergalactic magnetic field, the propagation of 
UHECRs can be studied from the Boltzmann transport equation or by using 
some simulation methods. In Ref.\ \cite{Supanitsky}, the author presents a 
system of partial differential equations to describe the propagation of UHCRs 
in the presence of a random magnetic field. In that paper, the author 
considered the Boltzmann transport equation and obtained the partial 
differential equations for the number density as well as for the flux of 
particles. A diffusive character of the propagation of CRs is also obtained 
in that paper. In Ref.\ \cite{batista} (see also 
Ref.\ \cite{kalashev}), an astrophysical 
simulation framework is proposed for studying the propagating extraterrestrial 
UHE particles. In their work, authors presented a new and upper version of 
publicly available code CRPropa 3. It is a code for the efficient development 
of astrophysical predictions for UHE particles. Ref.\ \cite{berezinkyGre} 
presented an analytical solution of the diffusion equation for high-energy 
CRs in the expanding Universe. A fitting to the diffusion coefficient $D(E)$ 
obtained from numerical integration was presented in Ref.\ \cite{harari} for 
both Kolmogorov and Kraichnan turbulence. Authors of 
Ref.\ \cite{molerach} studied the effects of diffusion of CRs in the magnetic 
field of the local supercluster on the UHECRs from a nearby extragalactic 
source. In that study, the authors found that a strong enhancement 
at certain energy ranges of the flux can help to explain the features of the 
CR spectrum and the composition in detail. In Ref.\ \cite{berezinskyGG}, the 
authors demonstrated the energy spectra of UHECRs as observed by Fly’s Eye 
\cite{bird}, HiRes \cite{hires} and AKENO \cite{agasa} from the idea of the UHE 
proton's interaction with CMB photons. A detailed analytical study of the 
propagation of UHE particles in extragalactic magnetic fields has been 
performed in Ref.\ \cite{aloisio} by solving the diffusion equation 
analytically with the energy losses that are to be taken into account. In 
another study \cite{berezinsky_modif}, the authors obtained the ankle, 
instep and GZK cutoff in terms of the modification factor, which arises due 
to various energy losses suffered by CR particles while propagating through 
the complex galactic or intergalactic space \cite{berezinskyGK}. Similarly, in 
Ref.\ \cite{berezinski_four_feat}, authors obtained four features in the CR 
proton spectrum, viz.\ the ankle, instep, second ankle, and the GZK 
cutoff taking into consideration of extragalactic proton's interaction with 
CMB and assuming of resulting power-law spectrum.\\
\indent General relativity (GR) developed by Albert Einstein in 1915 to 
describe the ubiquitous gravitational interaction is the most beautiful, 
well-tested, and successful theory in this regard. The discovery of 
gravitational waves (GWs) by LIGO detectors in 2015 \cite{ligo} after almost 
a hundred years of their prediction by Einstein himself and the release of 
the first image of the supermassive black hole at the center of the elliptical 
supergiant galaxy M$87$ by the Event Horizon Telescope (EHT) in 2019 
\cite{m87a, m87b, m87c, m87d, m87e, m87f} are the robust supports amongst 
others in a century to GR. Even though the GR has been 
suffering from many drawbacks from the theoretical as well as observational 
fronts. For example, the complete quantum theory of gravity has 
remained elusive till now. The most important limitations of GR from the 
observational point of view are that it can not explain the observed current 
accelerated expansion \cite{reiss, perlmutter, spergel, astier} of the 
Universe, and the rotational dynamics of galaxies indicating the missing mass 
\cite{naselskii} in the Universe. Consequently, the modified theories of 
gravity (MTGs) have been developed as one of the ways to explain these 
observed cosmic phenomena, wherein these phenomena are looked at as some 
special geometrical manifestations of spacetime, which remain to be 
taken into account in GR. The most simplest but remarkable and widely used 
MTG is the $f(R)$ \cite{Sotiriou} theory of gravity, where the Ricci scalar $R$ in the Einstein-Hilbert (E-H) action is replaced by a function $f(R)$ of $R$. 
Various models of $f(R)$ gravity theory have been developed so far from 
different perspectives. Some of the viable as well as famous or popular models 
of $f(R)$ gravity are the Starobinsky model \cite{starobinsky, staro}, Hu-Sawicki model \cite{husawicki}, Tsujikawa model \cite{tsujikawa}, power-law model 
\cite{powerlaw} etc.\\    
\indent Till now several authors have studied the propagation of CRs 
in the domain of GR \cite{harari, molerach, berezinskyGK,berezinskyGG,nagano, 
Bhattacharjee, Olinto, s.mollerach, Supanitsky, aloisio,berezinsky_modif, 
prosekin}. The enhancement of the flux of CRs is obtained in the framework of 
the $\Lambda$CDM model by a variety of authors \cite{molerach,molerach_new}. 
Besides these, differential flux as well as the modification factor have also 
been studied \cite{berezinskyGK,berezinskyGG, Supanitsky,aloisio, 
berezinsky_modif, berezinski_four_feat}. Since MTGs have made significant 
contributions to the understanding of cosmological 
\cite{psarmah, gogoi_model} and astrophysical \cite{jbora} issues in recent 
times, it would be wise to apply the MTGs in the field of CRs to study the 
existing issues in this field. Keeping this point in mind, in this work, we 
study for the very first time the propagation of UHECRs and their consequent 
flux in the light of an MTG, the $f(R)$ theory of gravity. For this 
purpose, we consider two $f(R)$ gravity models, viz.\ the power-law model 
\cite{powerlaw} and the Starobinsky model \cite{staro}. 
%For the power law model, we consider $f(R) =\lambda R^n$ where 
%$\lambda$ and $n$ are model parameters while for the Starobinsky model 
%$f(R)=\alpha R + \beta R^2$, where $\alpha $ and $\beta$ are model 
%parameters. 
Considering these two models, we calculate the expression for the number 
density of particles. From the number density, we calculate the 
enhancement factor as well as the differential flux and modification 
factor for the UHECRs.\\
\indent The remaining part of the paper is arranged as follows. In Section 
\ref{secII}, we discuss the turbulent magnetic field and diffusive propagation 
mechanism. The basic cosmological equations that are used to calculate 
the cosmological parameters are introduced in Section \ref{secIII}. In 
Section \ref{secIV}, we define $f(R)$ gravity models of our interest and 
calculate the required cosmological parameters for those models. The fittings 
of predicted Hubble parameter values at different cosmological redshifts by 
those models to the observational Hubble parameter data are also shown in 
this section. In Section \ref{secV}, we calculate the number density of 
particles and hence the enhancement factor. 
Then the differential fluxes for both models along with the 
$\Lambda$CDM model were calculated and compared these results 
with the data of the Telescope Array (TA) experiment 
\cite{ta2019} and Pierre Auger Observatory (PAO) \cite{augerprd2020}. 
We also compare the calculated modification factors for 
those two models with the observational data of TA and PAO. 
Finally, we compare the results of all three models including 
the $\Lambda$CDM model, and then conclude our paper with a fruitful 
discussion incorporating the Chi-square test in Section \ref{secVI}.

\section{Propagation of Cosmic Rays in Turbulent Magnetic Fields}\label{secII}
It is a challenging task to build a model for the extragalactic magnetic 
fields since there are few observable constraints on them \cite{han}. Their 
exact amplitude values are unknown, and they probably change depending on the 
region of space being considered. In the cluster center 
regions, the large-scale magnetic fields have recorded amplitudes that vary 
from a few to tens of $\mu$G \cite{feretti}. Smaller strengths are anticipated 
in the vacuum regions, with the typical boundaries in unclustered regions 
being $1$ to $10$ nG. This means that considerable large-scale magnetic fields 
should also be present in the filaments and sheets of cosmic structures. The 
evolution of primordial seeds impacted by the process of structure building 
may result in TMFs in the Universe \cite{harari}. As a result, magnetic fields 
are often connected with the matter density and are therefore stronger in 
dense areas like superclusters and weaker in voids ($\leq\; \sim\! 10^{-15}$ G). In the local supercluster region, a pragmatic estimation places 
the coherence lengths of magnetic fields in between $10$ kpc and $1$ 
Mpc, while their root mean square (RMS) strengths lie in the range 
of $1$ to $100$ nG \cite{feretti,Valle,Vazza}. The regular component of the 
galactic magnetic field (GMF), which typically has a strength of only a few 
$\mu$G, may have an impact on the CRs' arrival directions, but due to 
its much lesser spatial extent, it is anticipated to have a 
subdominant impact on the CRs spectrum.

%The study of the CRs' propagation is what we will be most interested in here. 
In the local supercluster region, the rotation measures of polarised 
background sources have suggested the presence of a strong magnetic 
field, with a potential strength of $0.3$ to $2$ $\mu$G \cite{Valle}. It is 
the magnetic field within the local supercluster that is most 
relevant since the impacts of the magnetic horizon become noticeable when the 
CRs from the closest sources reach the Earth. Thus we will not consider here 
the larger-scale inhomogeneities from filaments and voids. The propagation of 
CRs in an isotropic, homogenous, turbulent extragalactic magnetic field will 
then be simplified. The rms amplitude of magnetic fields $B$, and the coherence 
length $l_\text{c}$ which depicts the maximum distance between any two points 
up to which the magnetic fields correlate with each other, can be used to 
characterize such magnetic fields. The RMS strength of magnetic fields 
can be defined as $B = \sqrt{\langle B^2(x)\rangle}$, which can take values 
from $1$ nG up to $100$ nG and the strength of the coherence length 
$l_\text{c}$ can take the values from $0.01$ Mpc to $1$ Mpc.

An effective Larmor radius for charged particles of charge $Ze$ moving with
energy $E$ through a TMF of strength $B$ may be defined as
\begin{equation}\label{larmor}
r_\text{L} = \frac{E}{ZeB} \simeq 1.1 \frac{E/\text{EeV}}{ZB/\text{nG}}\;\text{Mpc}.
\end{equation}
A pertinent quantity in the study of diffusion of charged particles in magnetic
fields is the critical energy of the particles. This energy can be defined as 
the energy at which the coherence length of a particle with charge $Ze$ is 
equal to its Larmor radius i.e., $r_\text{L}(E_{\text{c}}) = 
l_\text{c} $ and it is given by
\begin{equation}\label{cri_energy}
E_\text{c} = ZeBl_\text{c} \simeq 0.9 Z\, \frac{B}{\text{nG}}\, \frac{l_\text{c}}{\text{Mpc}}\;\text{EeV}.
\end{equation}
This energy distinguishes between the regime of resonant diffusion that occurs 
at low energies ($<E_\text{c}$) and the non-resonant regime at higher 
energies ($>E_\text{c}$). In the resonant diffusion regime, particles 
suffer large deflections due to the interaction with magnetic field $B$ with 
scales that are comparable to $l_\text{c}$, whereas in the latter 
scenario, deflections are small and can only take place across travel lengths 
that are greater than $l_\text{c}$. Extensive numerical simulations of 
proton's propagation yielded a fit to the diffusion coefficient $D$ as a 
function of energy \cite{harari}, which is given by
\begin{equation}\label{diff_coeff}
D(E) \simeq \frac{c\,l_\text{c}}{3}\left[4 \left(\frac{E}{E_\text{c}} \right)^2 + a_\text{I} \left(\frac{E}{E_\text{c}} \right) + a_\text{L} \left(\frac{E}{E_\text{c}} \right)^{2-m}   \right],
\end{equation}
where $m$ is the index parameter, $a_\text{I}$ and $a_\text{L}$ 
are two coefficients. For the case of TMF with Kolmogorov 
spectrum{, $m=5/3$ and} the coefficients are $a_\text{L} 
\approx 0.23 $ and $a_\text{I} \approx 0.9$, while that for Kraichnan 
spectrum one will have 
$m = 3/2$, $a_\text{L} \approx 0.42 $ and $a_\text{I} \approx 0.65$. 
The diffusion length $l_\text{D}$ relates to the distance after which 
overall deflection {of particles} is nearly one radian and is given by 
$l_\text{D} = 3D/c$. From Eq.\ \eqref{diff_coeff}, it is seen that for 
$E/E_\text{c}\ll 0.1$ the diffusion length, $l_\text{D} 
\simeq a_\text{L} l_\text{c} (E/E_\text{c})^{2-m}$ while for 
$E/E_\text{c}\gg 0.2$, the diffusion length will be $l_\text{D} 
\simeq 4\, l_\text{c} (E/E_\text{c})^{2}$.

\section{Basic Cosmological Equations}\label{secIII}
On a large scale, the Universe appears to be isotropic and homogeneous 
everywhere. In light of this, the simplest model to be considered is a 
spatially flat Universe, which is described by the 
Friedmann-Lema\^itre-Robertson-Walker (FLRW) metric and is defined as
\begin{equation}\label{flrw}
ds^2 = -\,dt^2 + a^2(t)\delta _{ij}dx^i dx^j,
\end{equation}
where $a(t)$ is the scale factor, $\delta_{ij}$ is the Kronecker delta 
function with $i,j =  \{1,2,3\}$ and $x^\mu = \{x^0,x^1,x^2,x^3\}$ are 
comoving coordinates with $x^0 = t$. Moreover, as a source of curvature, we
consider the perfect fluid model of the Universe with energy density $\rho$ and
pressure $p$ which is specified by the energy-momentum tensor 
$T^{\mu}_{\nu} = \operatorname{diag}(-\,\rho, p,p,p)$. At this 
stage, we are interested in the basic cosmological evolution equation to be 
used in our study and this equation is the Friedmann equation. The 
Friedmann equation in $f(R)$ gravity theory is derived by following the 
Palatini variational approach of the theory. In this approach both the 
metric $g_{\mu\nu}$ and the torsion-free connection $\Gamma_{\mu\nu}^\lambda$ 
are considered as independent variables. In our present case the metric is 
$g_{\mu\nu} = \operatorname{diag}(-\,1,a^2,a^2,a^2)$ and the 
connection can be obtained from the $f(R)$ gravity field equations in the 
Palatini formalism \cite{Sotiriou}. Following the Palatini formalism the 
generalized Friedmann equation for our Universe in terms of redshift in $f(R)$ 
gravity theory can be expressed as \cite{Santos}
\begin{equation}\label{fredmann}
\frac{H^2}{H_\text{0}^2}=\frac{3\,\Omega_{m0}(1+z)^3 + 6\,\Omega_\text{r0}(1+z)^4 + \frac{f(R)}{H_\text{0}^2}}{6 f'(R)\zeta^2 },
\end{equation}
where 
\begin{equation}\label{zeta}
\zeta = 1+ \frac{9f''(R)}{2f'(R)}\frac{H_\text{0}^2\, \Omega_{m0}(1+z)^3}{Rf''(R)-f'(R)}.
\end{equation}
In Eqs.\ \eqref{fredmann} and \eqref{zeta}, $H_\text{0} \approx 67.4$ 
km s$^{-1}$ Mpc$^{-1}$ \cite{planck2018} is the Hubble constant, 
$\Omega_{m0} \approx 0.315$ \cite{planck2018} is the present value of the 
matter density parameter and $\Omega_\text{r0} \approx 5.373 \times 10^{-5}$ 
\cite{nakamura} is the present value of the radiation density parameter. 
$f'(R)$ and $f''(R)$ are the first and second-order 
derivatives of the function $f(R)$ with respect to $R$. It is seen that 
Eqs.\ \eqref{fredmann} and \eqref{zeta} are $f(R)$ gravity model dependent.

Secondly, in our study, it is important to know how the cosmological redshift 
is related to the cosmological time evolution. This can be studied from 
the connection between the redshift and cosmological time evolution, which is 
given by
\begin{equation}\label{dtdz}
\bigg | \frac{dt}{dz} \bigg |=\frac{1}{(1+z)H}.
\end{equation}
The expression of the Hubble parameter $H(z)$ for different models of $f(R)$ 
gravity will be derived using Eqs.\ \eqref{fredmann} and \eqref{zeta} in the 
next section \ref{secIV}.
\section{$f(R)$ Gravity Models and Cosmological Evolution}
\label{secIV}
In this section, we will introduce the power-law model \cite{powerlaw} and 
Starobinsky model \cite{staro} of $f(R)$ theory of gravity, and then will derive
the expressions for the Hubble parameter and evolution Eq.\ \eqref{dtdz} for
these two models. The least-square fits of the derived Hubble 
parameters for the models to the recent observational data will also be done 
here to constrain the parameters of the models.
%Moreover, the likelihood fit method will be used here to further constrain 
%different model parameters with the observed cosmological data.  
\subsection{Power-law model and cosmological equations}
The general $f(R)$ gravity power-law model is given by \cite{udg, d_gogoi}
\begin{equation}\label{powerlaw}
f(R)=\lambda\, R^n,
\end{equation}
where $\lambda$ and $n$ are two model parameters. Here the parameter $n$ is 
apparently a constant quantity, but the parameter $\lambda$ depends on the 
value of $n$ as well as on the cosmological parameters $H_\text{0}$, $\Omega_{m0}$ and $R_0$ as given by \cite{d_gogoi}
\begin{equation}\label{lambda}
\lambda = -\,\frac{3H_\text{0}^2\, \Omega_\text{m0}}{(n-2)R_0^n}.
\end{equation}
This expression of the parameter $\lambda$ implies that the power-law model
has effectively only one unknown parameter, which is the $n$. 
For this model, the expression of the present value of the Ricci scalar $R_0$ 
can be obtained as \cite{d_gogoi}
\begin{equation}\label{R0}
R_0 = -\, \frac{3 (3-n)^2 H_\text{0}^2\, \Omega_{m0}}{2n\left[(n-3)\Omega_{m0} + 2 (n-2) \Omega_\text{r0}\right]}.
\end{equation}
The expression of the Hubble parameter $H(z)$ for the power-law model can be 
obtained from Eq.\ \eqref{fredmann} together with Eq.\ \eqref{zeta} as 
\cite{d_gogoi}
\begin{equation}\label{powerlawhubble}
H(z) = \left[-\,\frac{2nR_0}{3 (3-n)^2\, \Omega_{m0}} \Bigl\{(n-3)\Omega_{m0}(1+z)^{\frac{3}{n}} + 2 (n-2)\,\Omega_\text{r0} (1+z)^{\frac{n+3}{n}} \Bigl\}\right]^\frac{1}{2}.
\end{equation}
In our study for the model parameter $n$, we use its value from the 
Ref.\ \cite{d_gogoi} where a detailed study has been made on this model in the 
cosmological perspective and the values of $n = 1.25$, $1.4$ and $1.9$ have 
been taken into account. Among these values the best-fitted value of $n$ is 
$1.4$ according to this Ref.\ \cite{d_gogoi}.

\begin{figure}[h!]
\centerline{
\includegraphics[scale=0.42]{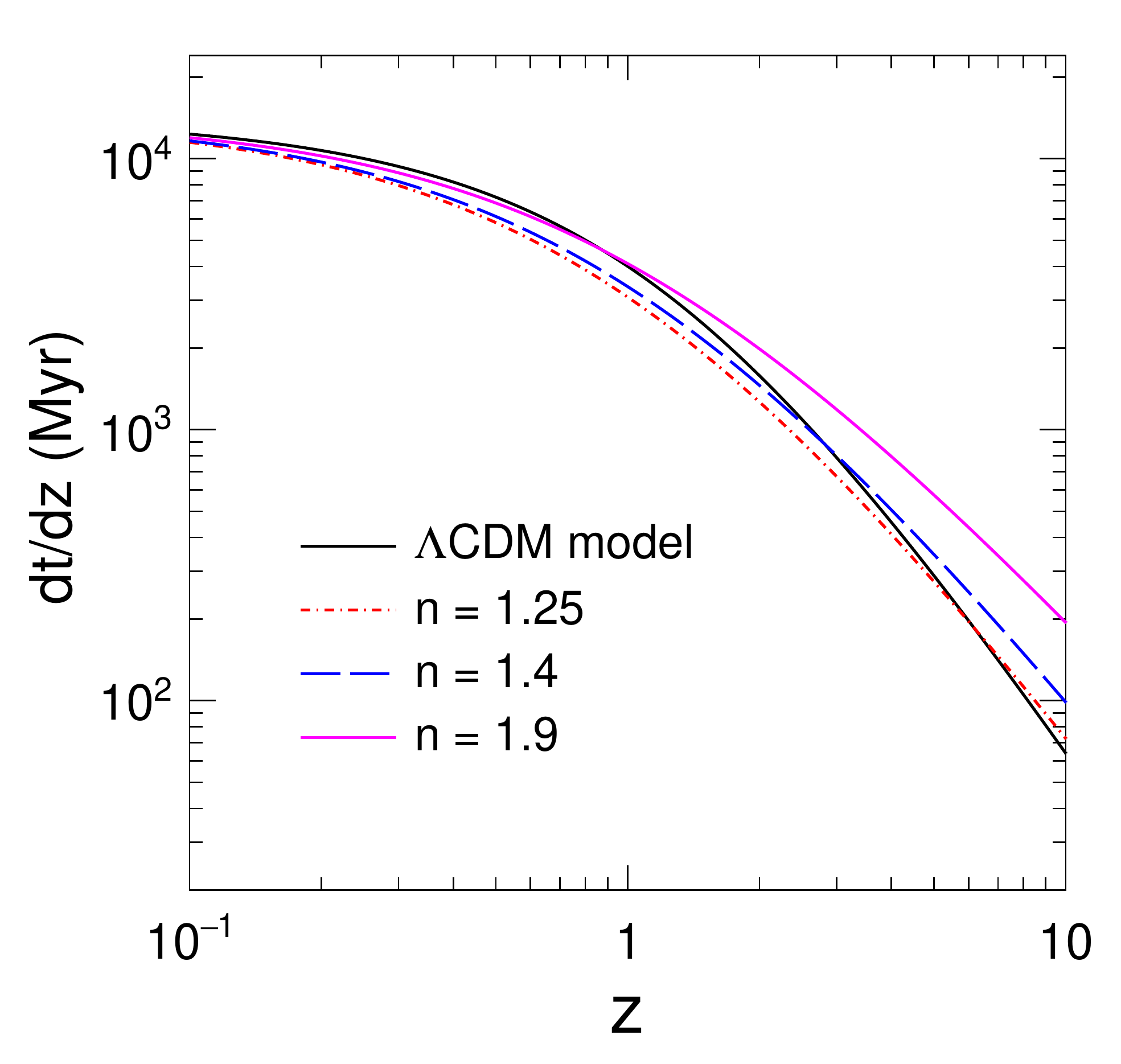}} 
\vspace{-0.2cm}
\caption{Variation of $dt/dz$ with the redshift $z$ for different values of 
the power-law model parameter $n$ along with the variation of the
same for the $\Lambda$CDM model.}
\label{fig2}
\end{figure}

The relation between the cosmological evolution time $t$ and redshift 
$z$ for the power-law model can be obtained by substituting 
Eq.\ \eqref{powerlawhubble} for $H(z)$ in Eq.\ \eqref{dtdz} as given by 
\begin{equation}\label{dtdz1}
\bigg | \frac{dt}{dz} \bigg | = (1+z)^{-1}  \left[-\,\frac{2nR_0}{3 (3-n)^2 \Omega_{m0}} \Bigl\{(n-3)\Omega_{m0}(1+z)^{\frac{3}{n}} + 2 (n-2)\Omega_\text{r0} (1+z)^{\frac{n+3}{n}} \Bigl\} \right]^{-\,\frac{1}{2}}\!\!.
\end{equation}
In Fig.\ \ref{fig2}, we plot the differential variation of cosmological time 
$t$ with respect to redshift $z$ i.e. the variation of $dt/dz$ with the 
redshift $z$ for different values for model parameter $n$ along with 
that for the $\Lambda$CDM model.
It is seen from Fig.\ \ref{fig2} that the difference of variation of 
$dt/dz$ for the power-law model from the $\Lambda$CDM model is both 
redshift $z$ and model parameter $n$ dependent. The difference appears to be 
less significant for all values of $n$ when values of $z< 0.2$, while for 
higher values of $z$, it has shown a notable deviation depending on the value 
of $n$. However, at around certain higher values $z$ the power-law model 
predicts the same values of $dt/dz$ as that of $\Lambda$CDM model depending 
on the parameter $n$. For example, at around $z=2.8$ the power-law model with 
$n=1.4$ and the $\Lambda$CDM model predict the same $dt/dz$. Beyond such 
values of $z$ corresponding to $n$ values the prediction of the power-law 
model deviates significantly from the $\Lambda$CDM model.  
%Therefore for the rest of the paper, a possible higher value of redshift 
%will have to be taken into account. 
It should be mentioned that although $n = 1.4$ is found as the most 
suitable value of the parameter of the power-law model as informed earlier, 
we use other two values of $n$ in this plot to see how the model prediction 
varies from that of the $\Lambda$CDM model with different values of $n$. It 
is clear that the higher values of $n$ obviously show more deviation from 
the $\Lambda$CDM model prediction for all appropriate values of $z$ and hence 
the most favorable value $n = 1.4$ shows appreciable behavior in this regard.  
\begin{center}
\begin{table}[h!]
\caption{Currently available observational Hubble parameter data 
($H^\text{obs}(z)$ [km s$^{-1}$ Mpc$^{-1}$]).}
\vspace{0.2cm}
\begin{tabular}{ccc|ccc}
\hline 
\rule[-1ex]{0pt}{2.5ex} \hspace{0.5cm} $z$ \hspace{0.5cm}  & \hspace{0.0cm} $ H^\text{obs}(z) $ \hspace{0.0cm} & \hspace{0.0cm} Reference \hspace{0.0cm} & \hspace{0.5cm}  $z$ \hspace{0.5cm}  & \hspace{0.0cm} $ H^\text{obs}(z)$ \hspace{0.0cm} & \hspace{0.0cm} Reference \hspace{0.0cm} \\ 
\hline
\rule[-1ex]{0pt}{2.5ex} 0.0708 & $69.0 \pm 19.68$ & \cite{Zhang_2014} & 0.48 & $97.0 \pm 62.0$ & \cite{Ratsimbazafy_2017}\\
\rule[-1ex]{0pt}{2.5ex} 0.09 & $69.0 \pm 12.0$ & \cite{Simon_2005} & 0.51 & $90.8 \pm 1.9$ & \cite{Alam_2017}\\
\rule[-1ex]{0pt}{2.5ex} 0.12 & $68.6 \pm 26.2$ & \cite{Zhang_2014} & 0.57 & $92.4 \pm 4.5$ & \cite{Samushia_2013}\\
\rule[-1ex]{0pt}{2.5ex} 0.17 & $83.0 \pm 8.0$  & \cite{Simon_2005} & 0.593 & $104.0 \pm 13.0$ & \cite{Moresco_2012}\\
\rule[-1ex]{0pt}{2.5ex} 0.179 & $75.0 \pm 4.0$ & \cite{Moresco_2012} & 0.60 & $87.9 \pm 6.1$ & \cite{Blake_2012}\\
\rule[-1ex]{0pt}{2.5ex} 0.199 & $75.0 \pm 5.0$ & \cite{Moresco_2012}& 0.61 & $97.8 \pm 2.1$  & \cite{Alam_2017}\\
\rule[-1ex]{0pt}{2.5ex} 0.2 & $72.9 \pm 29.6$  & \cite{Zhang_2014}& 0.68 & $92.0 \pm 8.0$  & \cite{Moresco_2012}\\
\rule[-1ex]{0pt}{2.5ex} 0.24 & $79.69 \pm 2.65$ & \cite{Gazetanga_2009}& 0.73 & $97.3 \pm 7.0$  & \cite{Blake_2012}\\
\rule[-1ex]{0pt}{2.5ex} 0.27 & $77.0 \pm 14.0$ & \cite{Simon_2005}& 0.781 & $105.0 \pm 12.0$ & \cite{Moresco_2012}\\
\rule[-1ex]{0pt}{2.5ex} 0.28 & $88.8 \pm 36.6$ & \cite{Zhang_2014}& 0.875 & $125.0 \pm 17.0$  & \cite{Moresco_2012}\\
\rule[-1ex]{0pt}{2.5ex} 0.35 & $84.4 \pm 7.0$ & \cite{XXu_2013}& 0.88 & $90.0 \pm 40.0$ & \cite{Ratsimbazafy_2017}\\
\rule[-1ex]{0pt}{2.5ex} 0.352 & $83.0 \pm 14.0$ & \cite{Moresco_2012}& 0.9 & $117.0 \pm 23.0$  & \cite{Simon_2005}\\
\rule[-1ex]{0pt}{2.5ex} 0.38 & $81.9 \pm 1.9$ & \cite{Alam_2017}& 1.037 & $154.0 \pm 20.0$  & \cite{Moresco_2012}\\
\rule[-1ex]{0pt}{2.5ex} 0.3802 & $83.0 \pm 13.5$  & \cite{Moresco_2016}& 1.3 & $168.0 \pm 17.0$  & \cite{Simon_2005}\\
\rule[-1ex]{0pt}{2.5ex} 0.40 & $95.0 \pm 17.0$ & \cite{Simon_2005}& 1.363 & $160.0 \pm 33.6$ & \cite{Moresco_2015}\\
\rule[-1ex]{0pt}{2.5ex} 0.4004 & $77.0 \pm 10.2$ & \cite{Moresco_2016}& 1.43 & $177.0 \pm 18.0$ & \cite{Simon_2005}\\
\rule[-1ex]{0pt}{2.5ex} 0.4247 & $87.1 \pm 11.2$ & \cite{Moresco_2016}& 1.53 & $140.0 \pm 14.0$ & \cite{Simon_2005}\\
\rule[-1ex]{0pt}{2.5ex} 0.43 & $86.45 \pm 3.68$ & \cite{Gazetanga_2009}& 1.75 & $202.0 \pm 40.0$ & \cite{Simon_2005}\\
\rule[-1ex]{0pt}{2.5ex} 0.44 & $82.6 \pm 7.8$ & \cite{Blake_2012}& 1.965 & $186.5 \pm 50.4$ & \cite{Moresco_2015}\\
\rule[-1ex]{0pt}{2.5ex} 0.4497 & $92.8 \pm 12.9$ & \cite{Moresco_2016} & 2.34 & $223.0 \pm 7.0$ & \cite{Delubac_2015} \\
\rule[-1ex]{0pt}{2.5ex} 0.47 & $89.0 \pm 50.0$ & \cite{Ratsimbazafy_2017}& 2.36 & $227.0 \pm 8.0$ & \cite{Ribera_2014} \\
\rule[-1ex]{0pt}{2.5ex} 0.4783 & $80.9 \pm 9.0$ & \cite{Moresco_2016} &  &  & \\
\hline
\end{tabular}
\label{table1}
\end{table}
\end{center}
\subsection{Starobinsky Model and cosmological equations}
The Starobinsky model of $f(R)$ gravity considered here is of the form \cite{staro}:
\begin{equation}\label{starobinsky}
f(R) = \alpha R + \beta R^2,
\end{equation}
where $\alpha$ and $\beta$ are two free model parameters to be constrained by 
using observational data associated with a particular problem of study. 
Similar to the previous case the expression of the Hubble parameter 
$H(z)$ for the Starobinsky model can be obtained from 
Eq.\ \eqref{fredmann} along with Eq.\ \eqref{zeta} as
\begin{equation}\label{hubble_staro}
H(z) = H_\text{0} \left[ \frac{3\, \Omega_{m0} (1+z)^3 + 6\, \Omega_\text{r0} (1+z)^4 + \left(\alpha R + \beta R^2\right)\!H_\text{0}^{-2}}{6(\alpha + 2 \beta R)\Bigl\{ 1-\frac{9\,\beta H_\text{0}^2\, \Omega_{m0} (1+z)^3}{\alpha(\alpha+2\beta R)} \Bigl\}^2 }\right]^{\frac{1}{2}}.
\end{equation}
\begin{figure}[h!]
\centerline{
\includegraphics[scale=0.42]{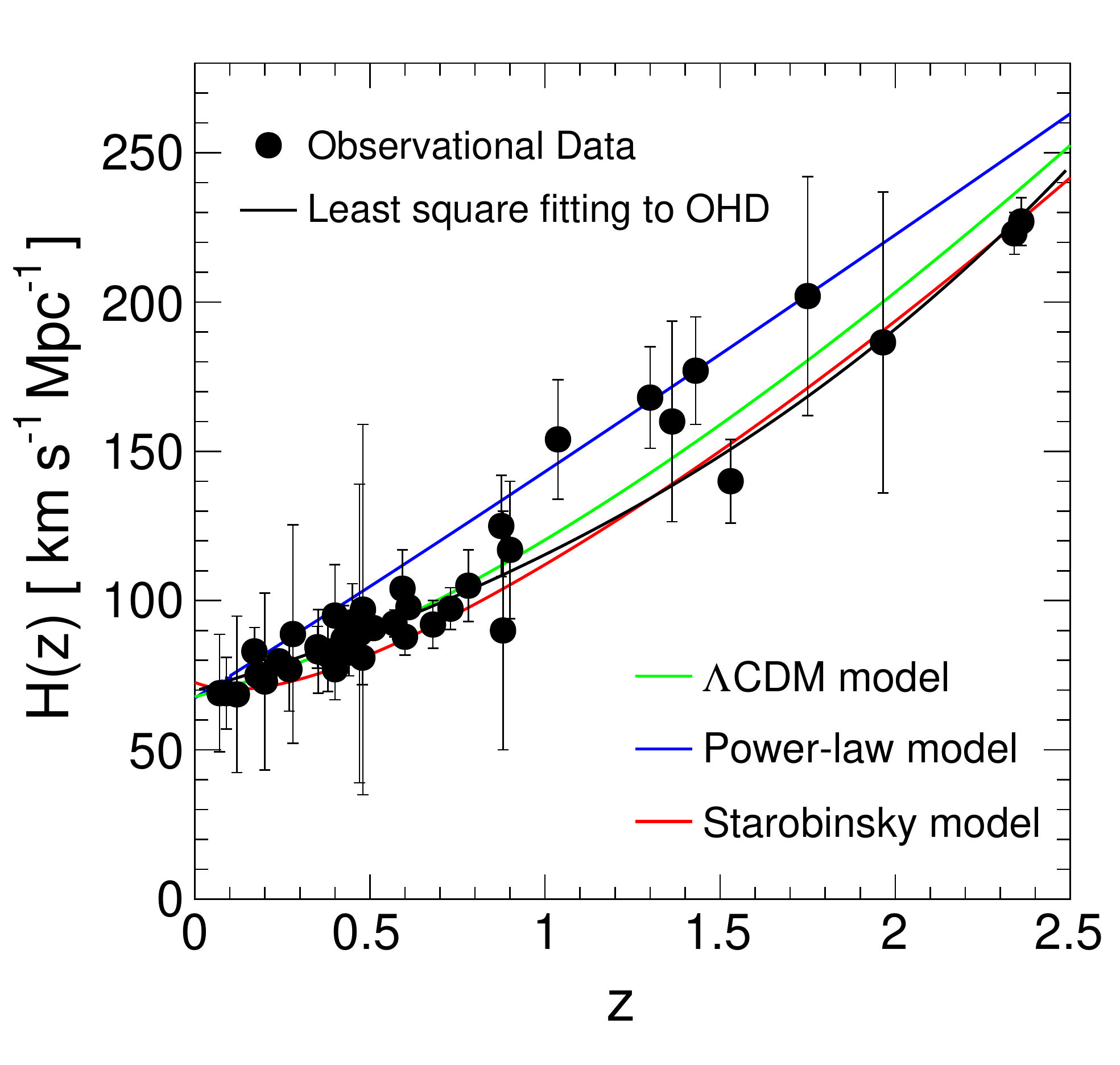}} 
\vspace{-0.3cm}
\caption{Least-square fitting to the observational Hubble 
data (OHD) as shown in Table \ref{table1} and the best-fitted curve 
for the Starobinsky model with parameters $\alpha = 1.07$ and 
$\beta = 0.00086$. Also, a curve for the power-law model with the 
model parameter $n=1.4$ is shown here along with the curve for the 
$\Lambda$CDM model.}
\label{fig3}
\end{figure}
To use this expression of $H(z)$ for further study we have to 
constrain the values of the model parameters $\alpha$ and $\beta$ 
within their realistic values as the behaviour of $H(z)$ depends significantly 
on these two model parameters. For this, we use the currently available 
observational Hubble parameter ($H^\text{obs}(z)$) data set 
\cite{p_sarmah} as shown in Table \ref{table1}. 
%The best fitted curve of $H(z)$ for the values of $\alpha$ and $\beta$ as 
%$1.07$ and $0.00086$ respectively has obtained from the $H^{obs}(z)$ data 
%set as shown in Table \ref{table1}.
Here we consider the combination of 43 observational Hubble 
parameter data against 43 distinct values of redshift $z$ (as they are 
available in the references mentioned) to obtain the precise values 
of the aforementioned free model parameters, so that the predicted 
$H(z)$ should be consistent with the $\Lambda$CDM model value at 
least around the current epoch i.e.\ at $z\sim0$.
\begin{figure}[h!]
\centerline{
\includegraphics[scale=0.42]{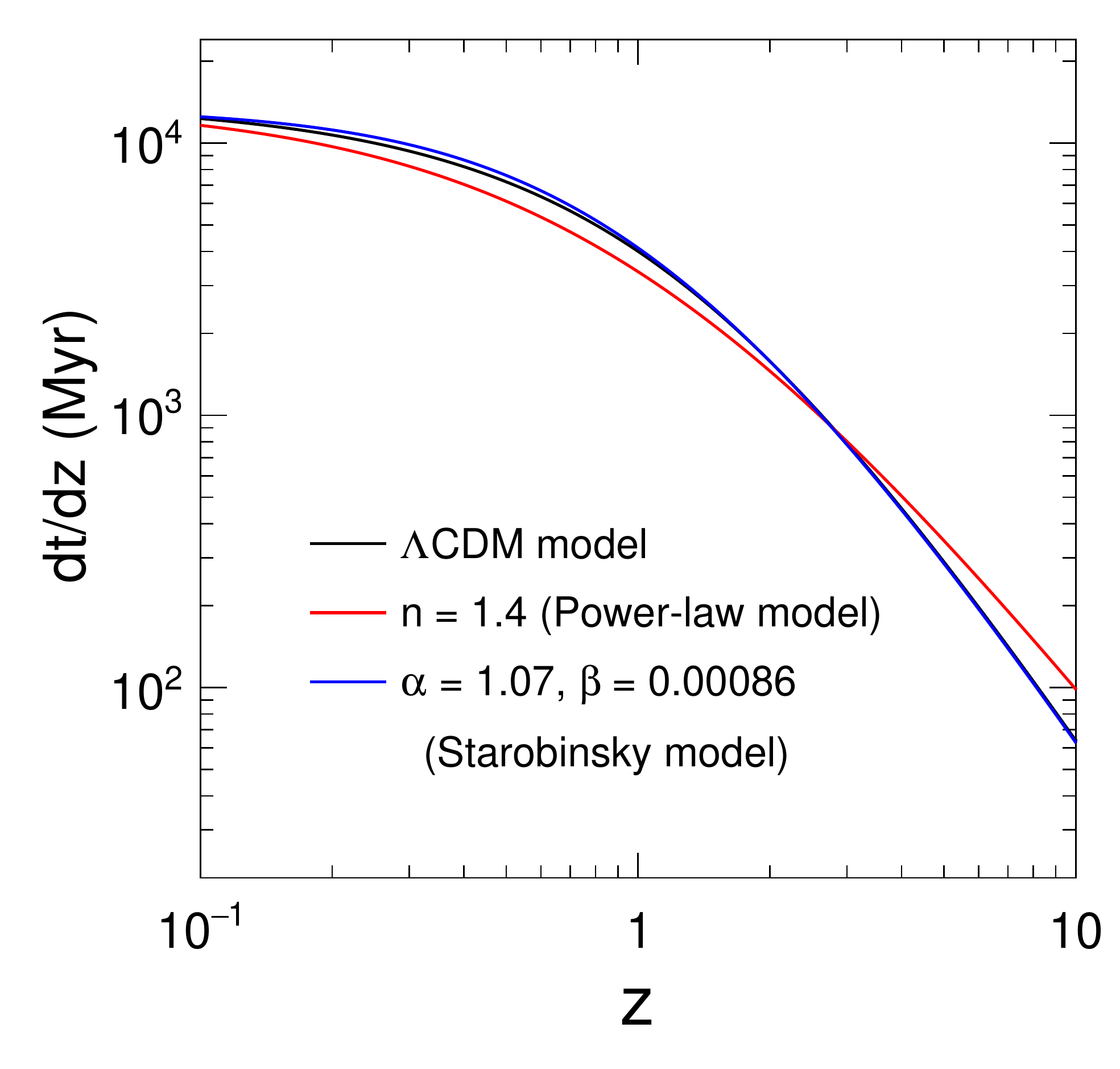}}
\vspace{-0.3cm} 
\caption{Variation of $dt/dz$ with respect to redshift $z$ for 
both $f(R)$ gravity models (power-law model and Starobinsky model) in 
comparison with the variation of the same for the $\Lambda$CDM model. Here 
the constrained parameter(s) is(are) used for the associated $f(R)$ gravity 
model.}
\label{fig4}
\end{figure} 
Using the least square fitting technique in ROOT software \cite{root}, we plot 
the best-fitted curve to this set of Hubble parameter data with respect to 
redshift as shown in Fig\ \ref{fig3}. For this least-square fitting, we 
use an exponential function of the form: $a \exp (bz)$, where $a$ and $b$ are 
two constants whose values are found after the fitting as 
$a = 69.750 \pm 0.927$ and $b = 0.503 \pm 0.012$. Using this fitting we infer 
values of $\alpha$ and $\beta$ as $1.07$ and $0.00086$ respectively 
by using the chi-square minimization method (as discussed in 
\cite{d_gogoi}). The value of $\chi^2$ is 29.38 with the critical value 
(in 95 \% confidence level) 56.94.

Now, we are in a position to write the expression for $dt/dz$ for this model 
and it can be expressed as
\begin{equation}\label{dtdz_staro}
\bigg | \frac{dt}{dz} \bigg | = \big[(1+z) H_\text{0}\big]^{-1} \left[ \frac{3\, \Omega_{m0} (1+z)^3 + 6\, \Omega_\text{r0} (1+z)^4 + \frac{\alpha R + \beta R^2}{H_\text{0}^2}}{6(\alpha + 2 \beta R)\Bigl\{ 1-\frac{9\beta H_\text{0}^2 \Omega_{m0} (1+z)^3}{\alpha(\alpha+2\beta R)} \Bigl\}^2 }\right]^{-\frac{1}{2}}.
\end{equation}
%where $H_0 \approx 67.4 \ kms^{-1}Mpc^{-1}$ \cite{planck2018} is the Hubble 
%constant, the matter density parameter $\Omega_{m0} \approx 0.315$ 
%\cite{planck2018} and $\Omega_\text{r0} \approx 5.373 \times 10^{-5}$ 
%\cite{nakamura}. 
In Fig.\ \ref{fig4}, variations of $dt/dz$ with respect 
to redshift $z$ are shown for both $f(R)$ gravity models, 
i.e.\ for the power-law model and the Starobinsky model in the
comparison with the prediction of the $\Lambda$CDM model. It can be observed 
that the Starobinsky model with the constrained set of parameters 
predicts the values of $dt/dz$ which are almost comparable to 
the values of the same predicted by the $\Lambda$CDM model over the considered 
range of $z$ (especially for $z>1$). Whereas, except at $z\sim 2.8$ there is a noticeable difference in the prediction of the power-law model from that of
the $\Lambda$CDM model, although the difference is small at $z<0.2$ as 
mentioned already. Power-law model predicts lower values of $dt/dz$ than that
for the other two models from $z = 0.1$ to $z \sim 2.8$ and above this range 
the trend becomes reversed. Moreover, it is found that at $z=0$, i.e.\ at the 
present epoch the $\Lambda$CDM model predicts the highest value and the 
power-law model predicts the lowest value of $dt/dz$.

In the next section, we will employ the results of this section to 
calculate the density and differential flux of CRs for the power-law and 
Starobinsky models.
\section{Cosmic Rays Density and Flux in the Domain of $f(R)$ 
Gravity}  
\label{secV}
The first thing that piques our curiosity is how the density of CRs is being 
modulated at a certain distance from the originating source in a TMF. For this, 
it is necessary to calculate the density enhancement of CRs at a certain 
distance $r_\text{s}$ from the originating source while being surrounded 
by a TMF. Specifically, we wish to investigate the reliance 
of density enhancement on different CR parameters taking into account the 
diffusive propagation of CRs in the light of $f(R)$ gravity theory.

In the diffusive regime, the diffusion equation for UHE particles 
propagating in an expanding Universe from a source which is located at a 
position ${x}_\text{s}$ can be expressed as \cite{berezinkyGre}
\begin{equation}\label{diff_eqn}
\frac{\partial \rho}{\partial t} + 3 H(t)\, \rho - b(E,t)\,\frac{\partial  \rho}{\partial E}- \rho\, \frac{\partial  \rho}{\partial E}-\frac{D(E,t)}{a^2(t)}\,\nabla^2  \rho = \frac{Q_\text{s}(E,t)}{a^3(t)}\,\delta^3({x}-{\bf{x}_\text{s}}),
\end{equation}
where $H(t)= \dot{a}(t)/a(t)$ is the Hubble parameter as a function of 
cosmological time $t$, $\dot{a}(t)$ is the time derivative of the 
scale factor $a(t)$, ${x}$ denotes the comoving coordinates, 
$\rho$ is 
the density of particle at time $t$ and position ${x}$, $Q_\text{s}(E)$ 
is the source function that depicts the number of emitted particles with 
energy $E$ per unit time. Thus, at time $t$, which corresponds to redshift $z$, 
$r_\text{s} = {x}-{\bf{x}_\text{s}}$. The energy losses of
particles due to expansion of the Universe and interaction with CMB 
are described by
\begin{equation}
\frac{dE}{dt} = -\, b(E,t),\;\; b(E,t) = H(t)E + b_\text{int}(E).
\end{equation}
Here $H(t)E$ represents the adiabatic energy losses due to expansion and 
$b_\text{int}(E)$ denotes the interaction energy losses. The interaction 
energy losses with CMB include energy losses due to pair production and 
photopion production (for details see \cite{harari}). The general solution of 
Eq.\ \eqref{diff_eqn} was obtained in Ref.~\cite{berezinkyGre} considering the
particles as protons and it is given as
\begin{equation}\label{density}
\rho(E,r_\text{s})= \int_{0}^{z_{i}} dz\, \bigg | \frac{dt}{dz} \bigg |\, Q(E_\text{g},z)\, \frac{\textrm{exp}\left[-r_\text{s}^2/4 \lambda^2\right]}{(4\pi \lambda^2)^{3/2}}\, \frac{dE_\text{g}}{dE},
\end{equation}
where $z_i$ is the redshift at the initial time when a particle was 
just emitted by a source and $E_\text{g}$ is the generation 
energy at redshift $z$ of a particle whose energy is $E$ at $z=0$, i.e.\ at 
present time. The source function $Q(E_\text{g},z)$ is considered to 
follow a power-law spectrum, 
$Q \varpropto E_\text{g}^{-\gamma_\text{g}}$ with 
$\gamma_\text{g}$ as the spectral index of generation at the source. 
$\lambda$ is the Syrovatsky variable \cite{syrovatsy_1959, mollerach2013} and 
is given by
\begin{equation}\label{syrovatsky}
\lambda^2(E,z)=\int_{0}^{z}dz\, \bigg | \frac{dt}{dz} \bigg |\,(1+z)^2 D(E_\text{g},z).
\end{equation}
Here $\lambda(E,z)$ refers to the usual distance that CRs travel from the 
location of their production at redshift $z$ with energy $E_\text{g}$, to 
the present time at which they are degraded to energy $E$. The expression of 
the rate of degradation of energy of particles at the source 
with respect to their energy at $z=0$, i.e.\ $dE_\text{g}/dE$ is given by 
\cite{berezinkyGre, berezinski_four_feat}
\begin{equation}\label{degde}
\frac{dE_\text{g}}{dE}= (1+z)\, \exp \left[\int_{0}^{z} dz\, \bigg | \frac{dt}{dz} \bigg | \left(\frac{\partial\, b_\text{int}}{\partial E} \right) \right].
\end{equation} 
The detailed derivation of this expression was nicely performed by 
Berezinsky et al.\ in Appendix B of Ref.~\cite{berezinski_four_feat}. It is 
clear that using Eqs.\ \eqref{dtdz1} and \eqref{dtdz_staro} in Eqs.\
\eqref{syrovatsky} and \eqref{degde} the density of UHE protons in the 
diffusive medium at any cosmological time $t$ with energy $E$ and at a distance
$r_\text{s}$ from the source can be obtained for the $f(R)$ 
gravity power-law model and the Starobinsky model respectively, as given by 
Eq.\ \eqref{density}. So, in the following, we will implement the 
results of the power-law and Starobinsky models from 
Section \ref{secIV} to obtain the CR protons density enhancement 
factor, and subsequently their flux and energy spectrum as 
predicted by these two $f(R)$ gravity models.

\subsection{Projections of \boldmath{f(R)} power-law model}
To calculate the CR protons density from Eq.~\eqref{density} 
and hence its enhancement factor in the TMF of extragalactic space 
projected by the power-law model of $f(R)$ gravity, as a prerequisite we 
calculate first the Syrovatsky variable $\lambda$ for this model from 
Eq.\ \eqref{syrovatsky} using Eq.\ \eqref{dtdz1}. In this calculation 
we use different values of the model parameters $n$ taking the feasible values 
of field parameters as $l_\text{c} = 0.1$ Mpc and $B = 50$ nG with the 
corresponding critical energy of protons as $E_\text{c} = 4.5$ EeV. 
Then we study the behaviour of the variable $\lambda$ for the both Kolmogorov 
spectrum and the Kraichnan spectrum. We also calculate this variable for 
the $\Lambda$CDM model for those two spectra for comparison. 
Here and rest of calculations we use the values of $z=0 - 5$ keeping 
in view of possible source locations of CRs as well as the present and 
probable future cosmological observable range.  
\begin{figure}[h!]
\centerline{
\includegraphics[scale=0.27]{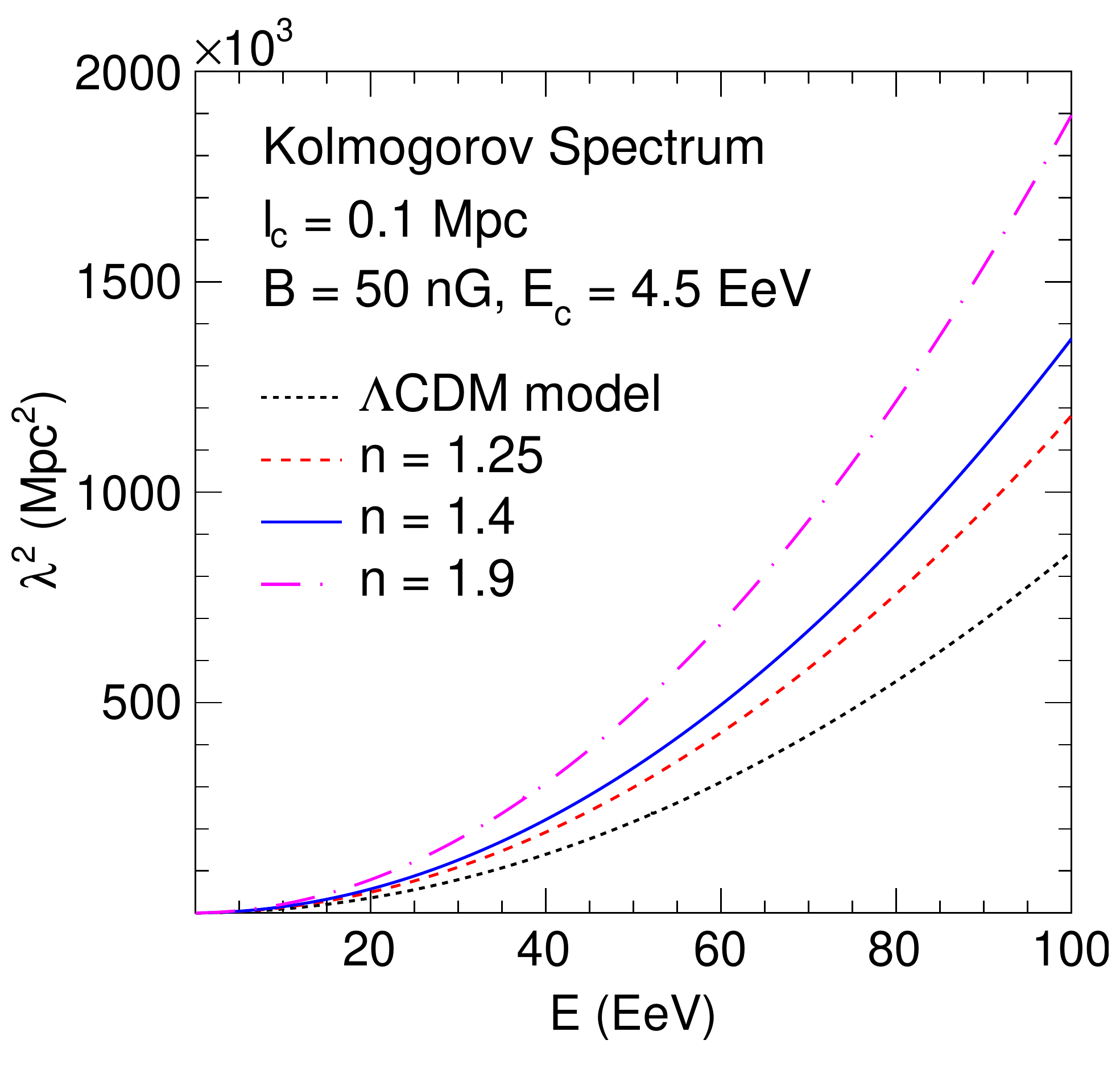}\hspace{0.3cm} 
\includegraphics[scale=0.27]{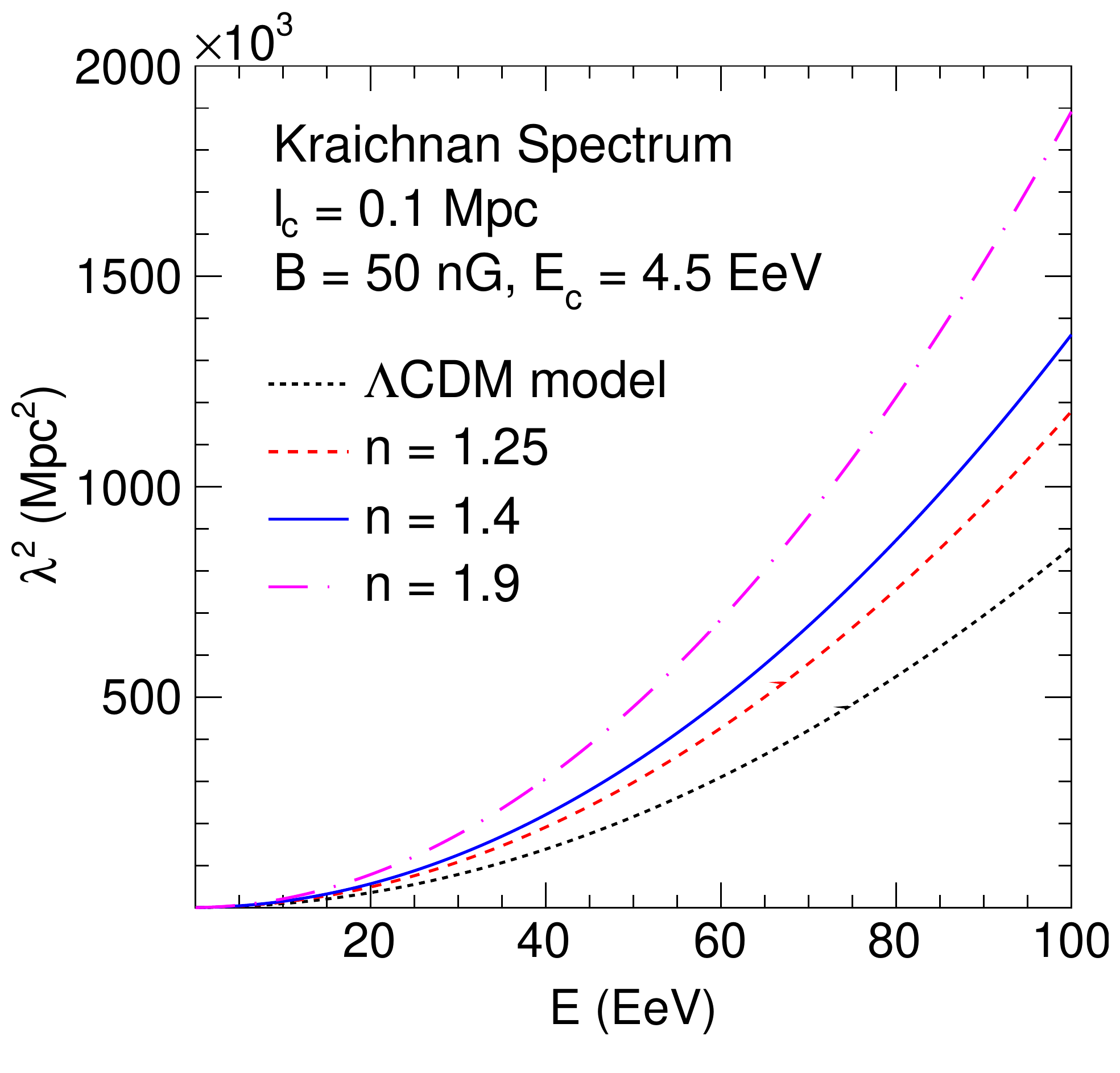}\hspace{0.3cm}
\includegraphics[scale=0.27]{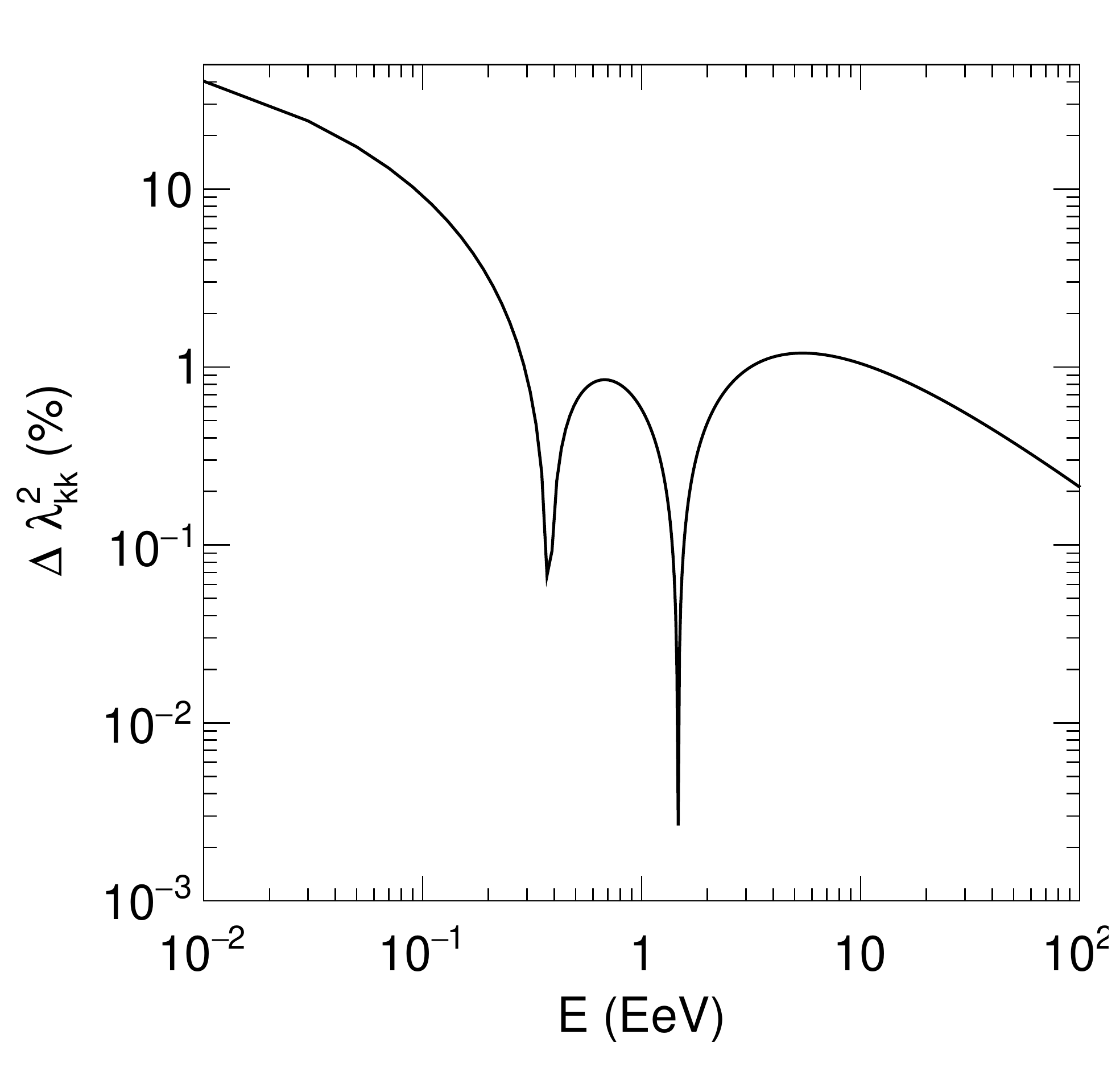}
}
\vspace{-0.3cm}
\caption{Variations of $\lambda^2$ {concerning} energy $E$ for the 
Kolmogorov spectrum (left panel) and {the} Kraichnan spectrum 
(middle panel) according to the $f(R)$ gravity power-law model and 
the standard $\Lambda$CDM model. {These plots are} obtained by 
{considering different values of the power-law model parameter $n$ with} 
$l_\text{c}=0.1$ Mpc, $B=50$ nG and $E_\text{c}=4.5$ EeV. The right 
panel shows the percentage {of per average bin} difference between 
$\lambda^2$ values for the Kolmogorov spectrum and {the} Kraichnan 
spectrum in each energy bin as per the power-law model with $n=1.4$. Here 
and {in the rest of the corresponding plots} we use $z=0-5$.}
\label{fig5}
\end{figure}
The results of these calculations are shown in Fig.\ \ref{fig5} with respect 
to energy $E$ for the Kolmogorov spectrum ($m=5/3$, $a_\text{L} \approx 0.23 $ 
and $a_\text{I} \approx 0.9$) (left panel) and the Kraichnan spectrum 
($m = 3/2$, $a_\text{L} \approx 0.42 $ and $a_\text{I} \approx 0.65$) (middle 
panel). It is seen from the figure that the value of $\lambda^2$ increases 
{substantially} with increasing energy of particles. The power-law model 
predicts higher values of $\lambda^2$ for all values of $n$ in comparison to 
that of the $\Lambda$CDM models for both spectra and this difference 
increases {significantly} with the increasing energy $E$. Similarly higher 
values of the parameter $n$ give increasingly higher values $\lambda^2$ in 
comparison to the smaller values of $n$. No difference can be 
observed between the values of $\lambda^2$ obtained for the Kolmogorov 
spectrum and the Kraichnan spectrum from the respective plots. So, to quantify 
the difference of values of $\lambda^2$ for these two spectra to a visible 
one we calculate the percentage {of per average bin} difference 
between $\lambda^2$ values obtained for the Kolmogorov spectrum and {the} 
Kraichnan spectrum in each energy bin ($\Delta \lambda_{kk}^2 (\%)$) for the 
power-law model with $n=1.4$, which is shown in the right panel of the figure. 
A peculiar behaviour of the variation of $\Delta \lambda_{kk}^2 (\%)$ with 
energy is seen from the plot. The $\Delta \lambda_{kk}^2 (\%)$ is energy 
dependent, it decreases rapidly with $E$ up to $\sim 0.4$ EeV, after which it
shows oscillatory behaviour with the lowest minimum at $\sim 1.55$ EeV. At
energies above $0.1$ EeV, the values of $\Delta \lambda_{kk}^2 (\%)$ are seen to
be mostly below the $1\%$. Thus at these {UHEs} differences of 
$\lambda^2$ values for the Kolmogorov spectrum and the Kraichnan spectrum are 
not so significant.     

In the diffusive regime, the density of particles has been enhanced by a 
factor depending on the energy, distance of the particles from the 
source and TMF properties. The density enhancement factor can be defined as 
the ratio of actual density to the density of particles that would lead
to their rectilinear propagation, which is given by \cite{molerach}
\begin{equation}\label{enhancement}
\xi(E,r_\text{s})=\frac{4\pi r_\text{s}^2c\, \rho(E, r_\text{s})}{\mathcal{L}(E)},
\end{equation}
where $\mathcal{L}(E)$ is the spectral emissivity of the source, which has 
a power-law dependency on the energy of the particles.
%The enhancement of the density of 
%particles for different values of parameters are shown in Fig.\ref{fig6}.\\
\begin{figure}[h!]
\centerline{
\includegraphics[scale = 0.4]{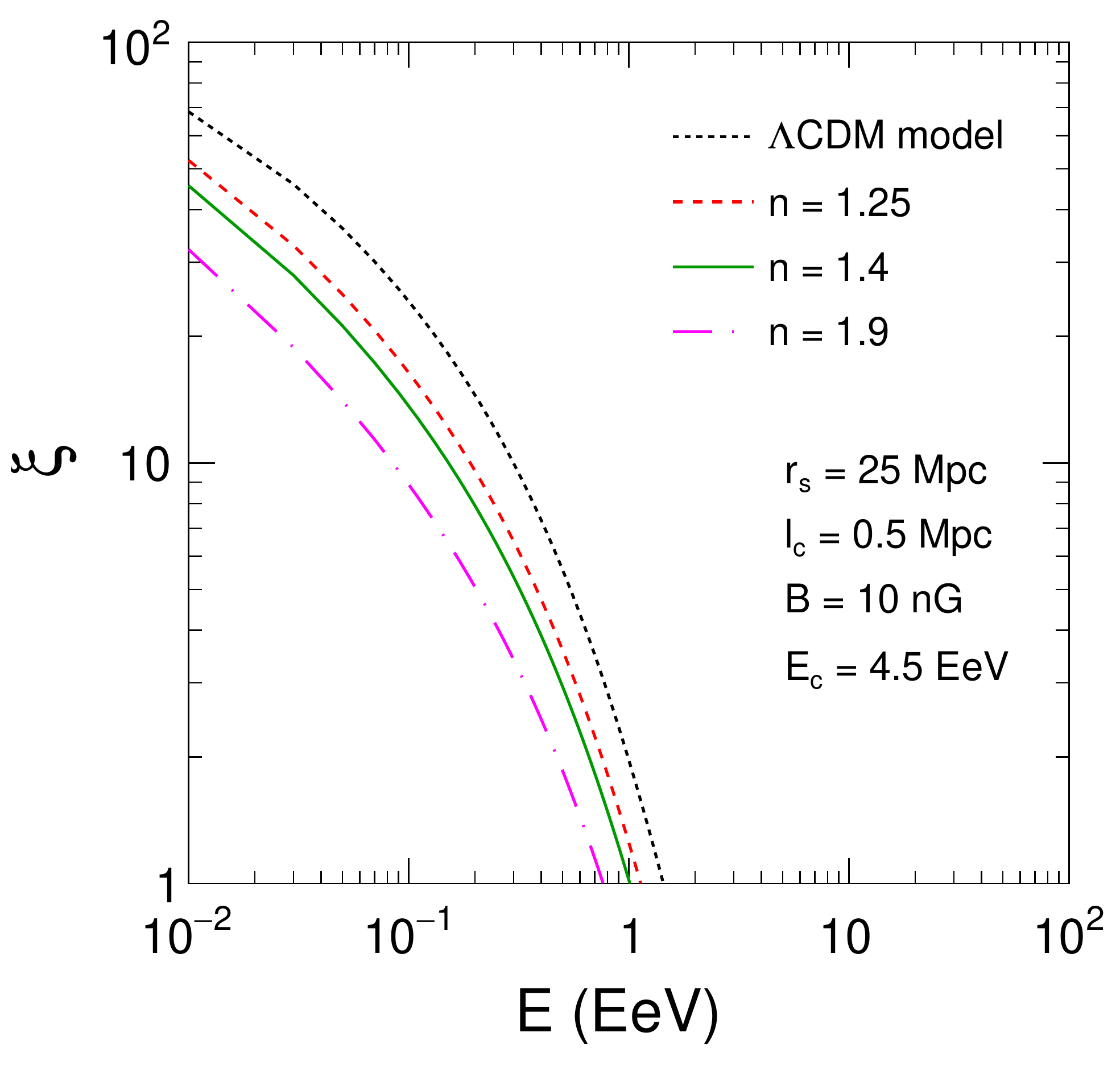}\hspace{0.5cm}
\includegraphics[scale = 0.4]{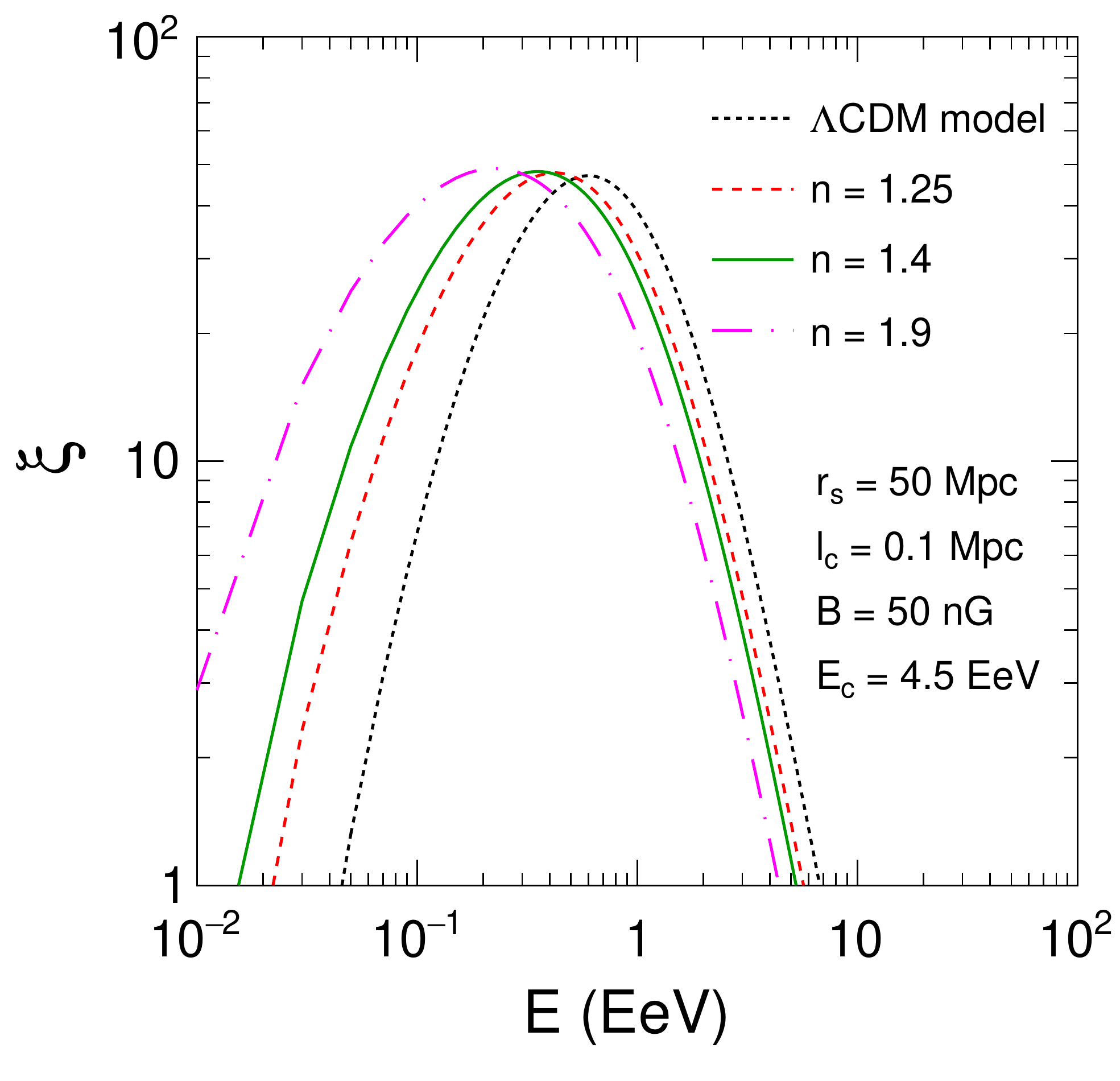}}
\centerline{
\includegraphics[scale = 0.4]{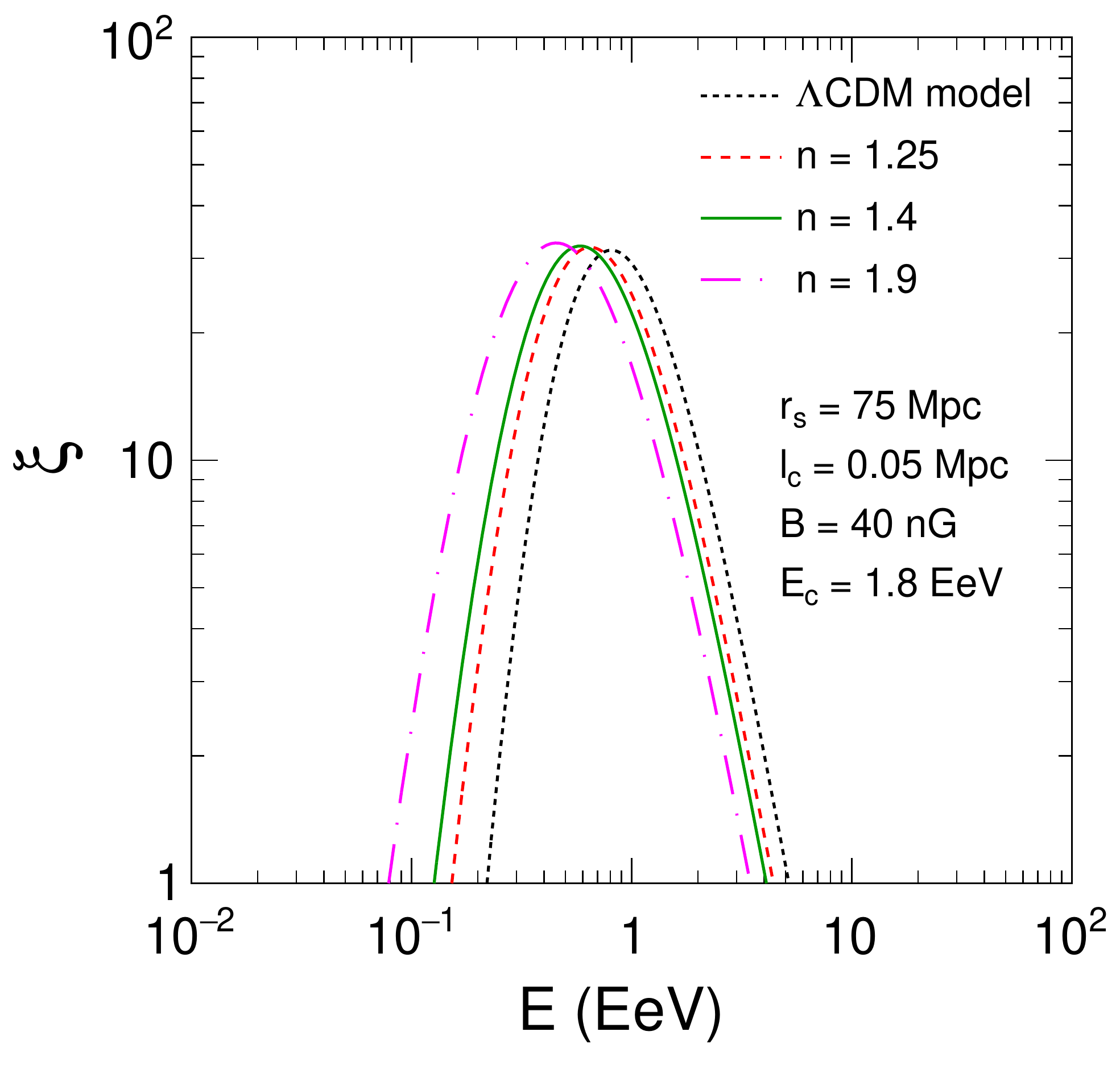}\hspace{0.5cm}
\includegraphics[scale = 0.4]{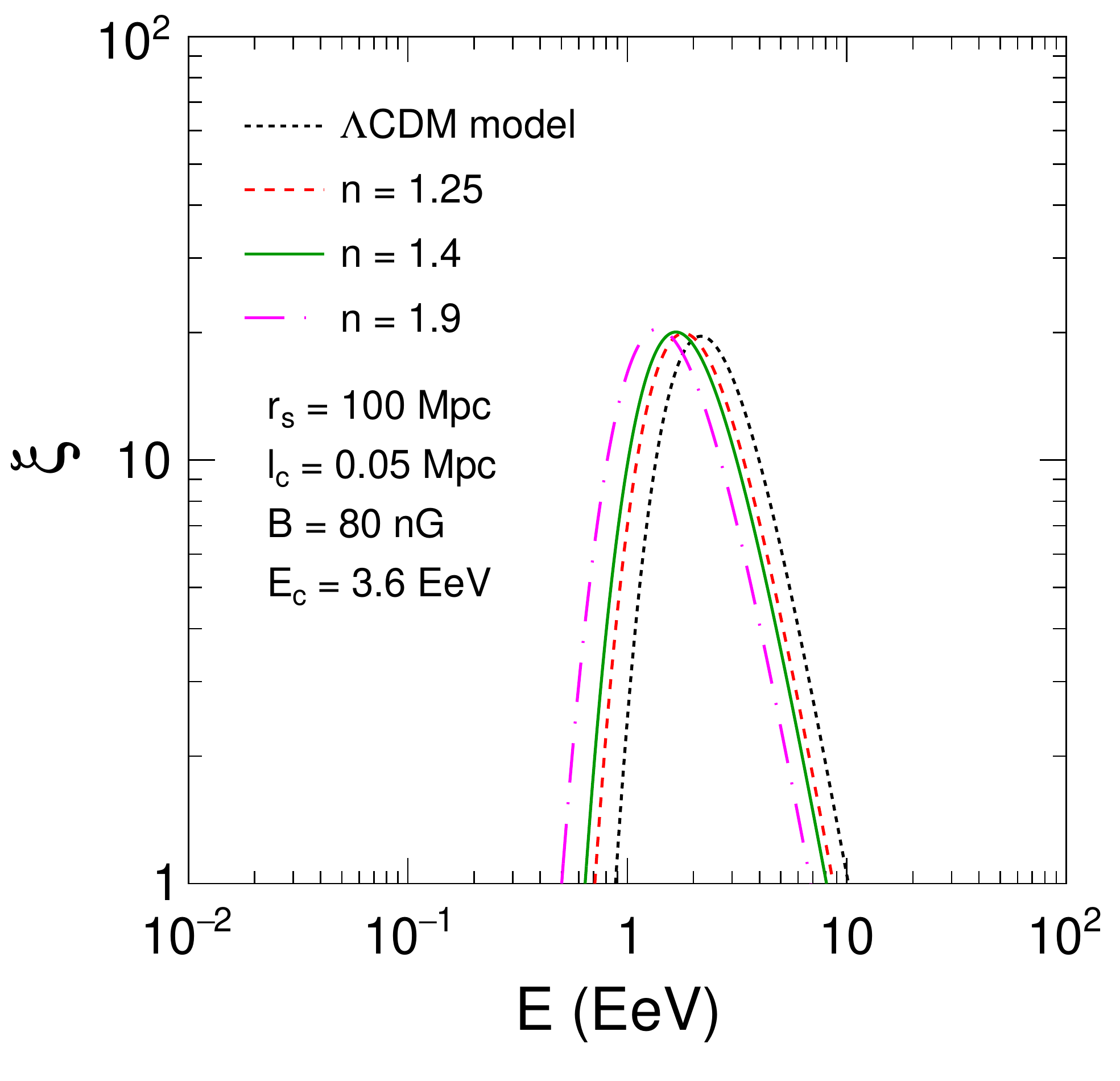}}
\vspace{-0.3cm}
\caption{Variation of density enhancement factor $\xi$ with energy $E$ for 
the $f(R)$ gravity power-law model and the $\Lambda$CDM model obtained by 
considering $r_\text{s}=25-100$ Mpc, $l_\text{c}=0.05-0.5$ Mpc, 
$B=10-80$ nG and $E_\text{c}=1.8-4.5$ EeV.}
\label{fig6}
\end{figure}
The results of the enhancement of the density for a proton source and 
for various parameters values obtained from 
Eq.~\eqref{enhancement} by numerically integrating Eq.\ \eqref{density} 
are displayed in Fig.\ \ref{fig6}. The distance to the source $r_\text{s}$, 
the magnetic field amplitude $B$, and its coherence length $l_\text{c}$ 
are the major factors that determine the lower-energy suppression of the 
density enhancement factor. For $r_\text{s}=25$ Mpc, $l_\text{c}=0.5$ 
Mpc, $B=10$ nG, and $E_\text{c}=4.5$ EeV (upper left panel), the 
enhancement has become noticeable for different gravity models in 
the energy range $E < 1$ EeV. For the energy range $0.01 < E < 10$ EeV, 
$r_\text{s}=50$ Mpc, $l_\text{c}=0.1$ Mpc, $B=50$ nG and 
$E_\text{c}=4.5$ EeV (upper right panel) are taken into account. In this 
case, below $1$ EeV the variation of enhancement for different gravity models 
is more distinguished compared to $E > 1$ EeV. In the lower left panel, 
$r_\text{s}=75$ Mpc, $l_\text{c}=0.05$ Mpc, $B=40$ nG and 
$E_\text{c}=1.8$ EeV are used to plot the enhancement factor for the 
$\Lambda$CDM and $f(R)$ power-law models, while this is done for 
$r_\text{s}=100$ Mpc, $l_\text{c}=0.05$ Mpc, $B=80$ nG and 
$E_\text{c}=3.6$ EeV in the lower 
right panel. In the lower panels, the enhancement energy range is less as 
compared to the upper panels, which is lowest in the case of the lower right 
panel. As the distance from the source is far away, the enhancement of density 
is limited to a smaller range of energies but shifted towards the higher 
energy side. The final verdict from Fig.~\ref{fig6} is that as the distance 
from the source $r_\text{s}$ increases, the enhancement becomes gradually 
model independent. Also one can appreciate that the $f(R)$ gravity power-law 
model has done a perfect job by enhancing density in a wider range 
of energies as compared to the $\Lambda$CDM model.

\begin{figure}[h!]
\centerline{
\includegraphics[scale=0.4]{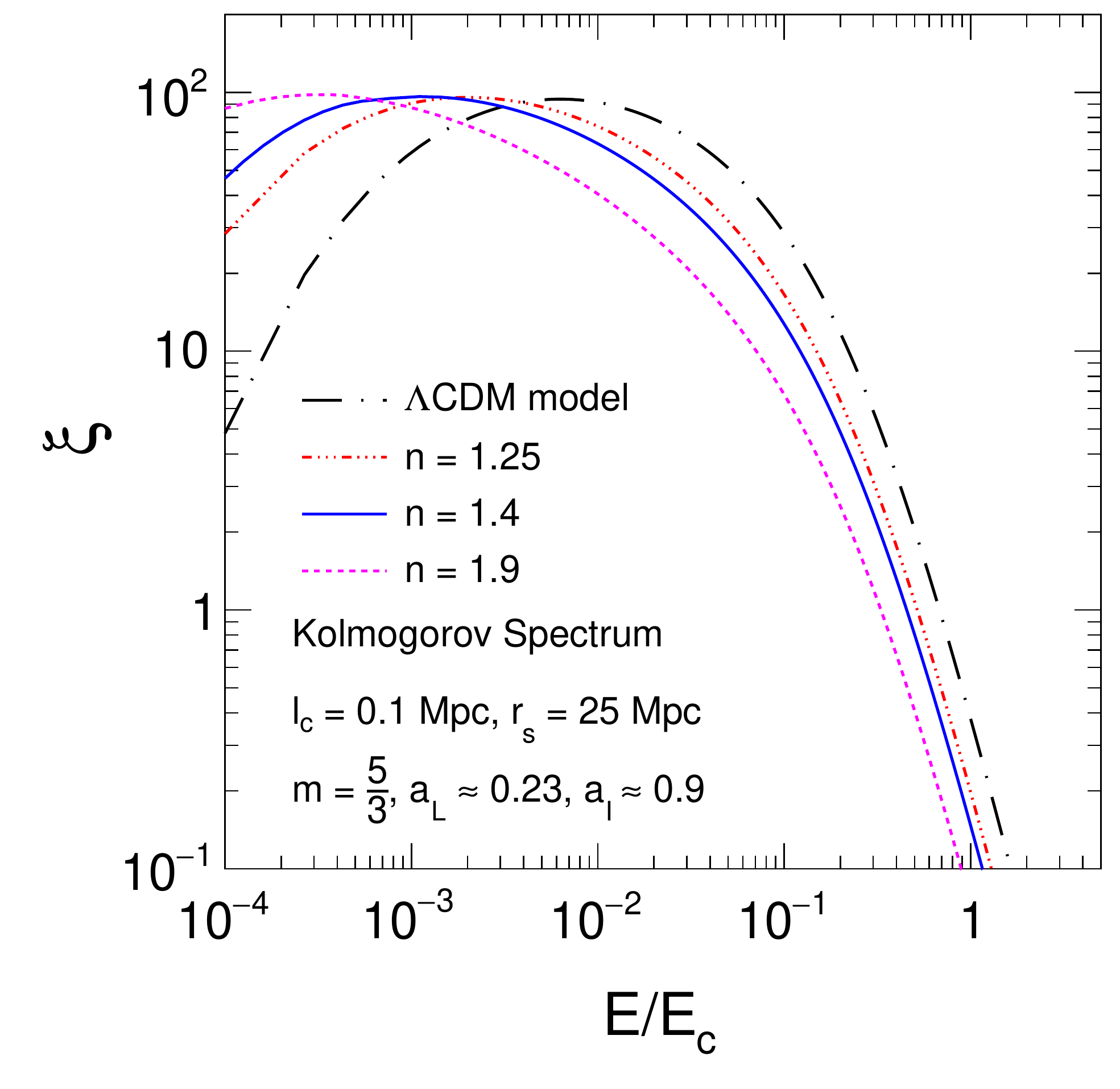}\hspace{0.5cm}
\includegraphics[scale=0.4]{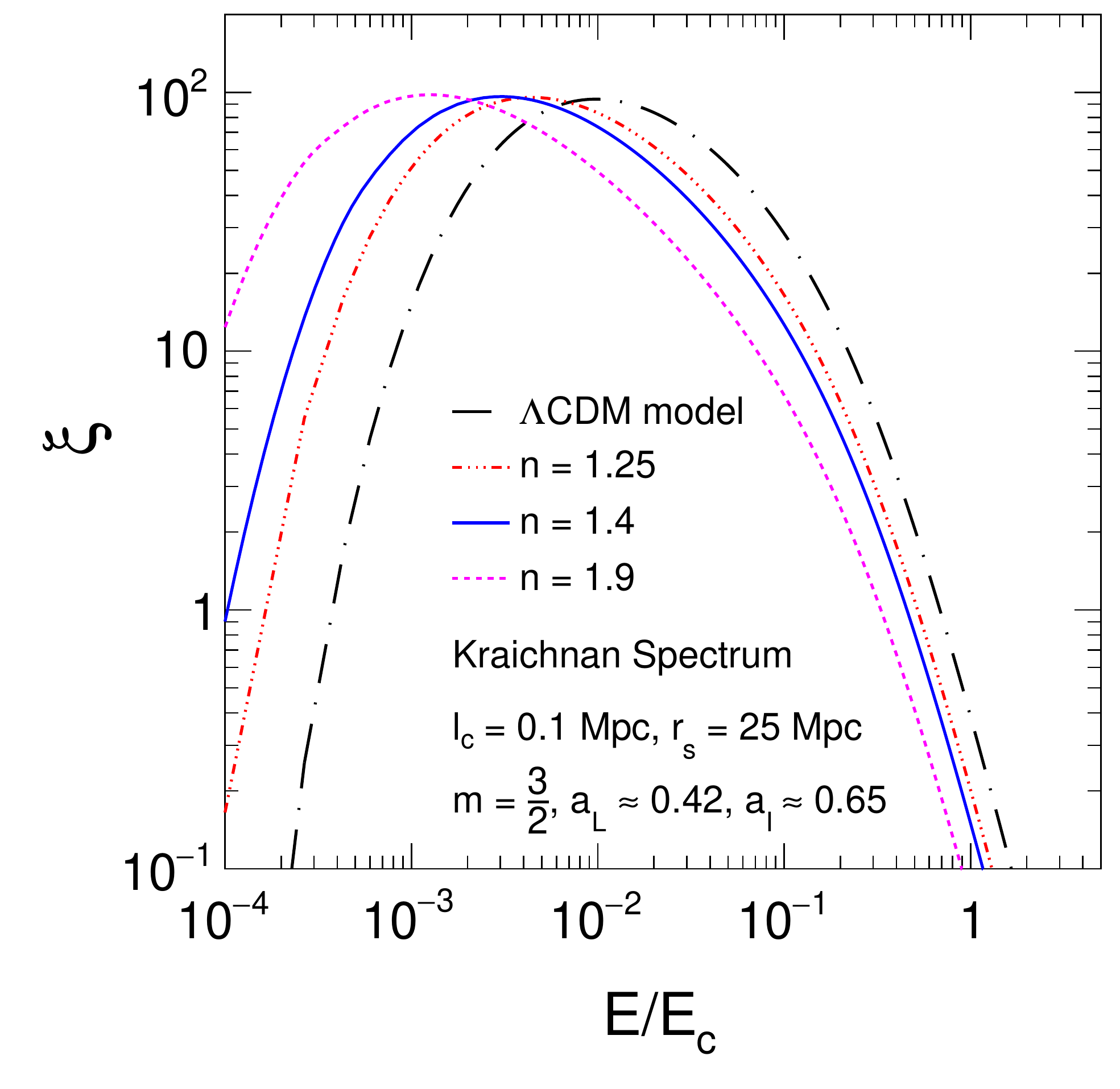}}
\vspace{-0.2cm}
\caption{Variation of density enhancement $\xi$ with 
$E/E_\text{c}$. The left panel is for the Kolmogorov spectrum 
while the right panel is for the Kraichnan spectrum obtained by considering 
the $\Lambda$CDM and $f(R)$ gravity power-law models with $l_\text{c}=0.1$ 
Mpc and $r_\text{s} = 25$ Mpc.}
\label{fig7}
\end{figure}
For a given source distance of $25$ Mpc and coherence length of $0.1$ Mpc, we 
depict the enhancement factor $\xi$ as a function of $E/E_\text{c}$ in 
Fig.\ \ref{fig7} to better highlight the fact that for 
$E/E_\text{c} < 0.01$ the Kolmogorov spectrum (left panel) and Kraichnan 
spectrum (right panel) have shown different behaviours while for 
$E/E_\text{c} > 0.01$ both Kolmogorov and Kraichnan spectra have shown 
similar patterns. In this case, the $f(R)$ power-law model is more suitable 
as it gives the enhancement in the higher as 
well as lower values of $E/E_\text{c}$, while in the case of the 
$\Lambda$CDM the range it gives the enhancement is less {wide} than the 
power-law model. From this Fig.\ \ref{fig7} it is seen that the 
Kolmogorov spectrum has given a better range of $E/E_\text{c}$ than 
{the} Kraichnan spectrum for both $\Lambda$CDM and $f(R)$ power-law model.

\begin{figure}[h!]
\centerline{
\includegraphics[scale=0.4]{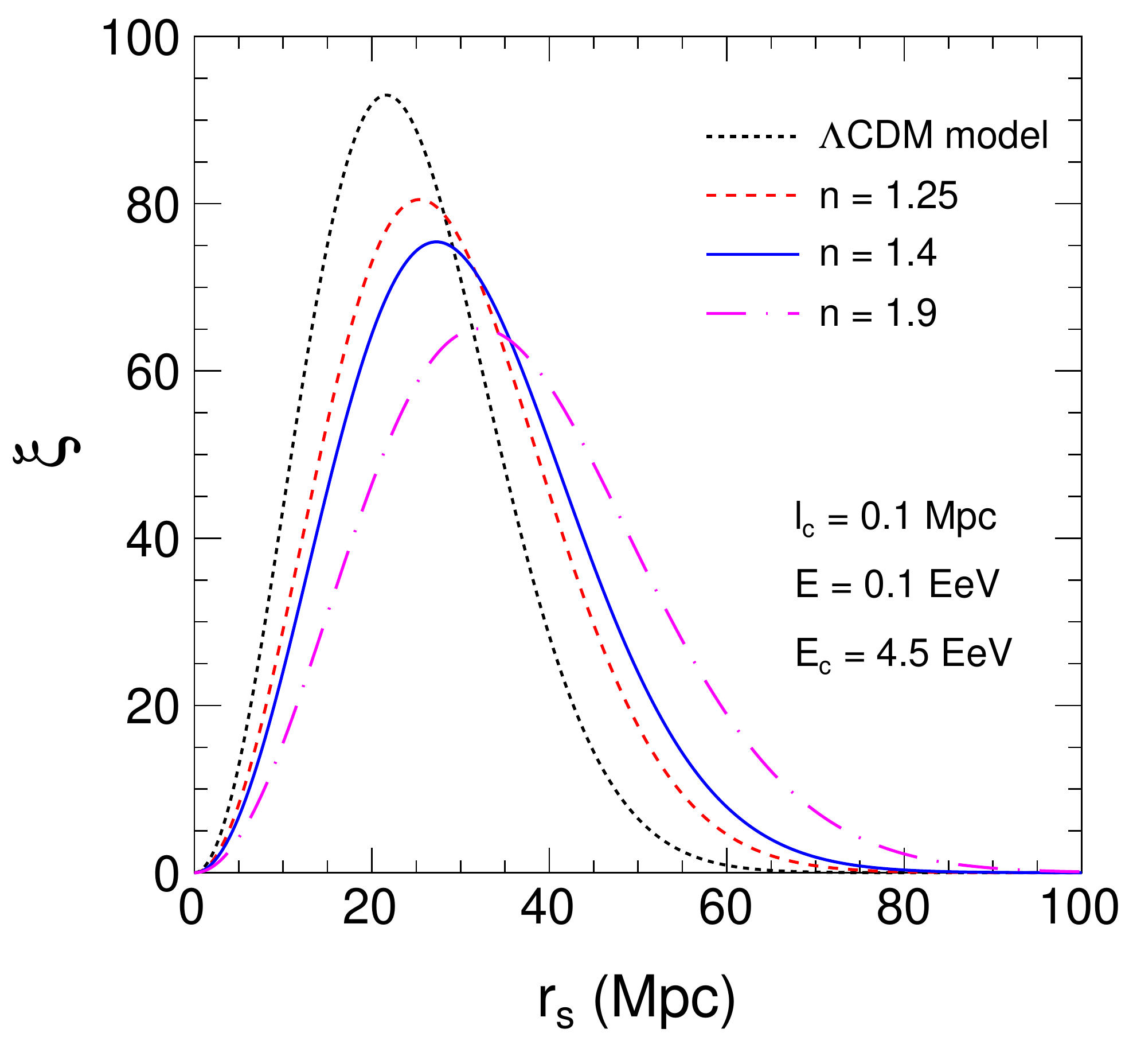}\hspace{0.5cm}
\includegraphics[scale=0.4]{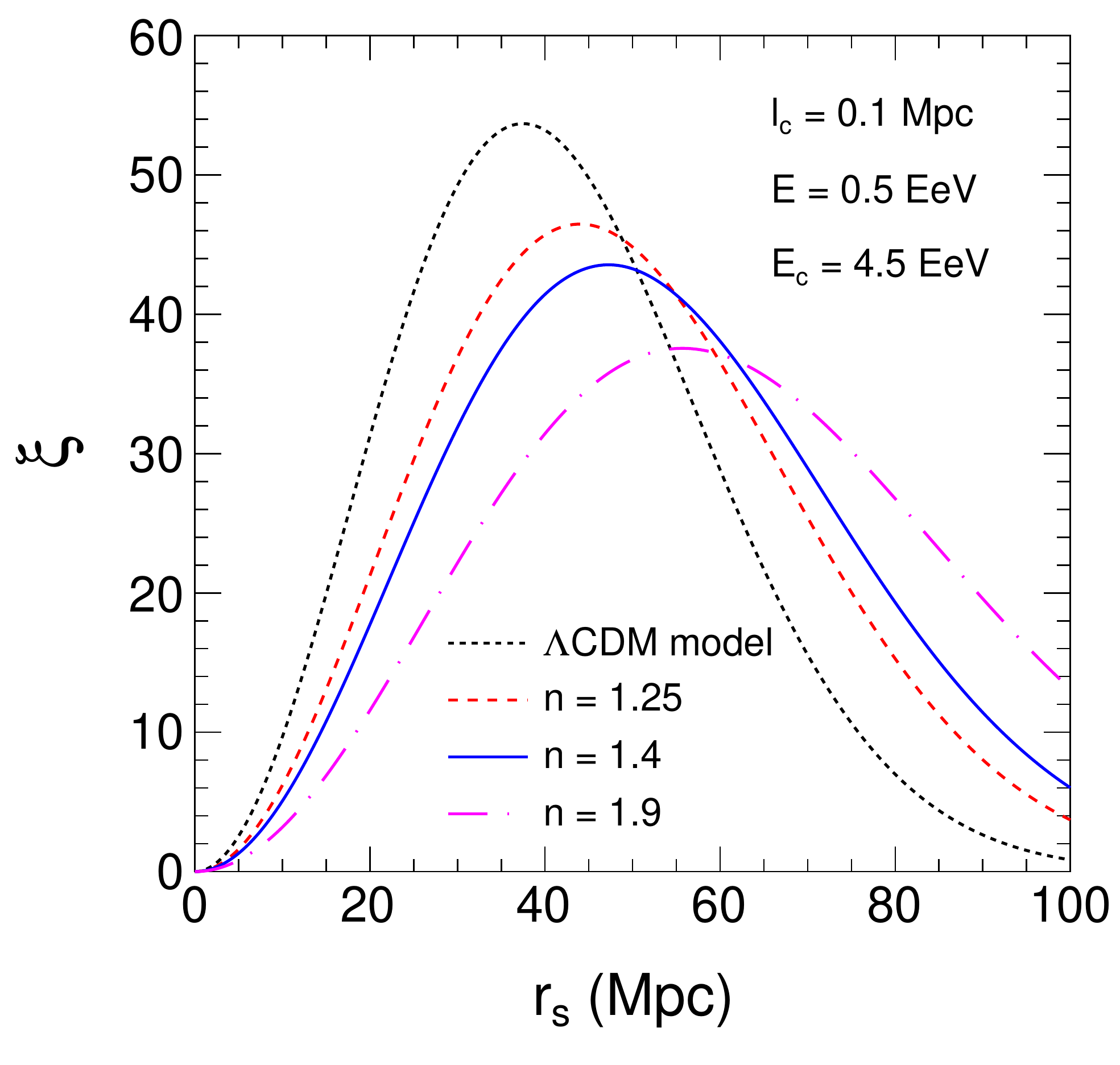}}
\centerline{
\includegraphics[scale=0.4]{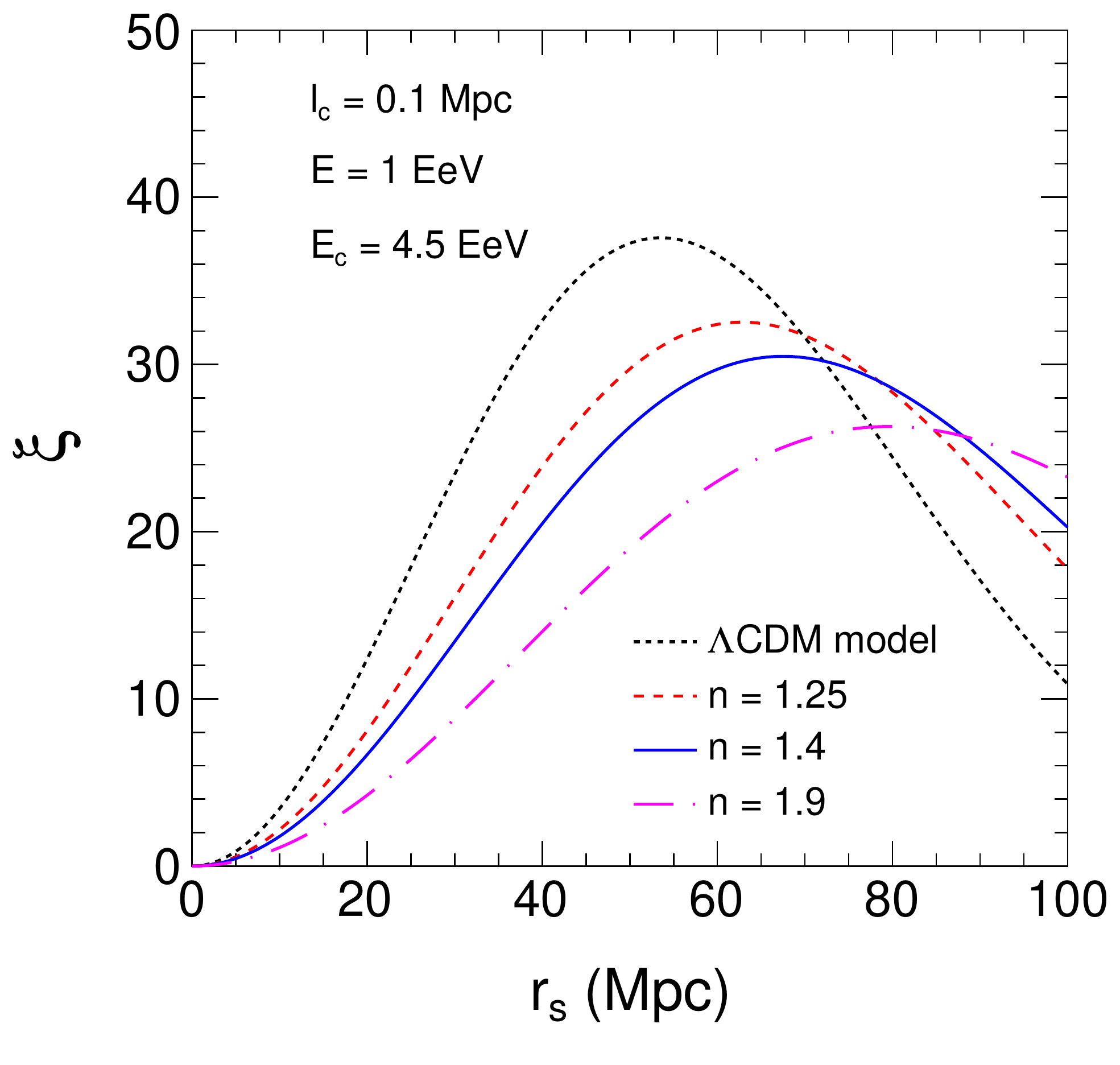}\hspace{0.5cm}
\includegraphics[scale=0.4]{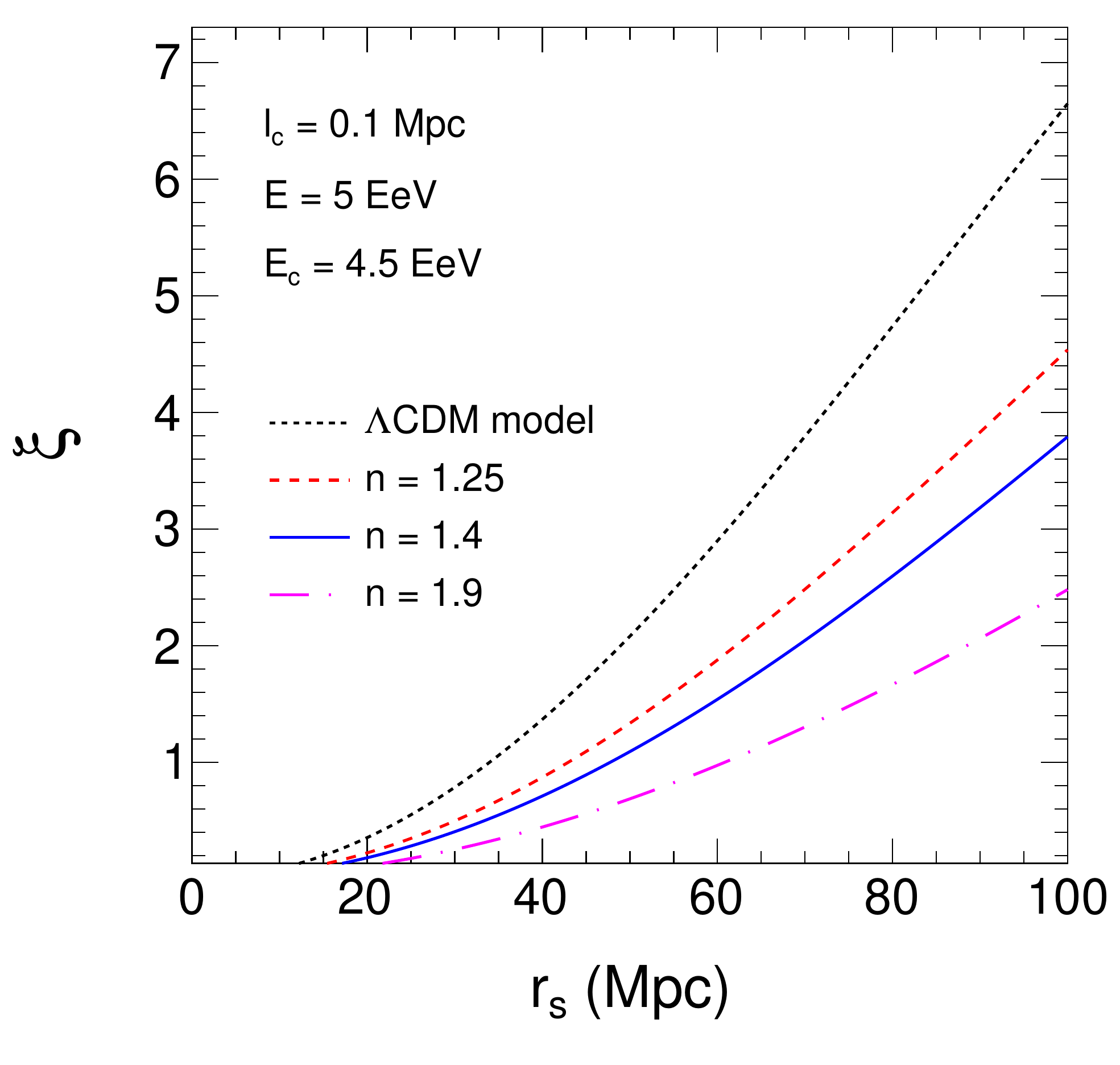}}
\vspace{-0.3cm}
\caption{Variation of $\xi$ with source distance $r_\text{s}$ for the 
$\Lambda$CDM model and $f(R)$ power-law model obtained by considering 
$l_\text{c}=0.1$ Mpc {with} E = 0.1 EeV (upper left panel), 0.5 EeV 
(upper right panel), 1 EeV (lower left panel) and 5 EeV (lower right panel).}
\label{fig8}
\end{figure}
\begin{figure}[h!]
\centerline{
\includegraphics[scale=0.8]{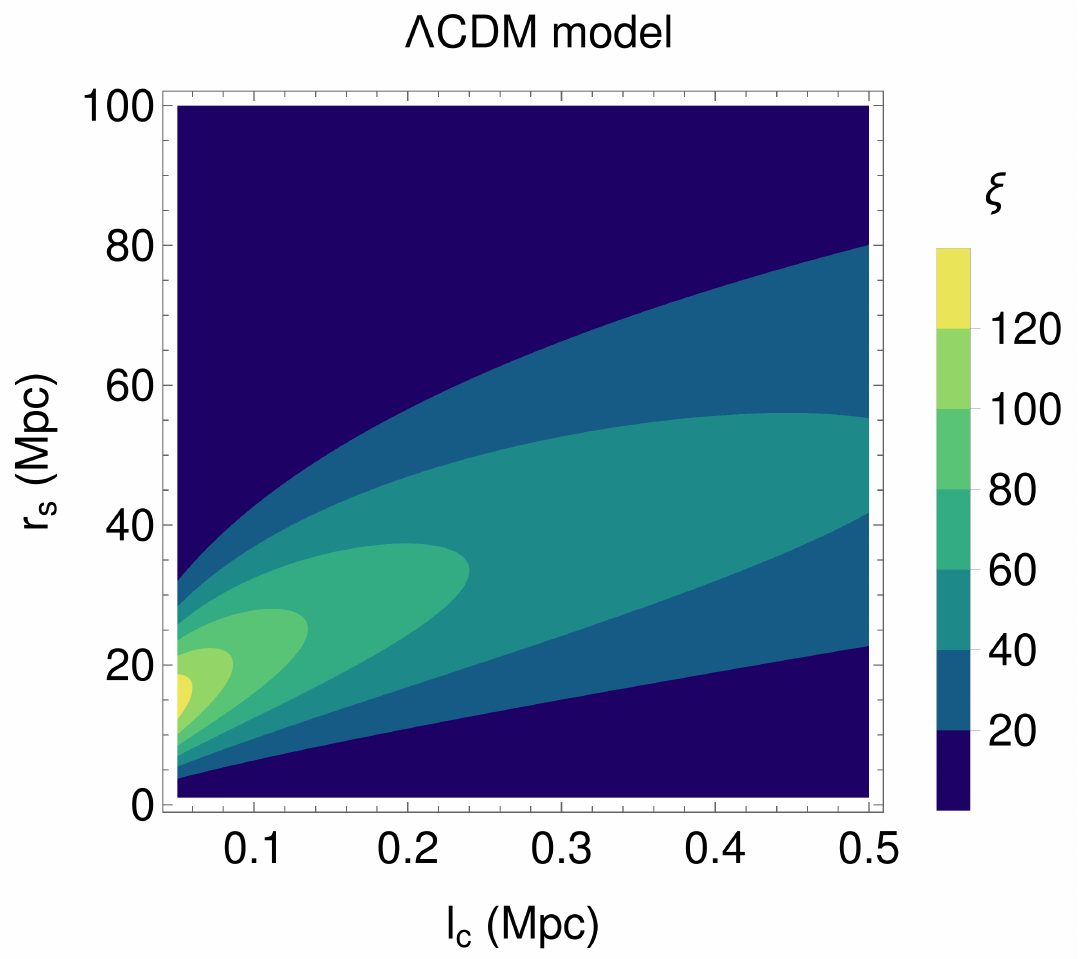}
\includegraphics[scale=0.8]{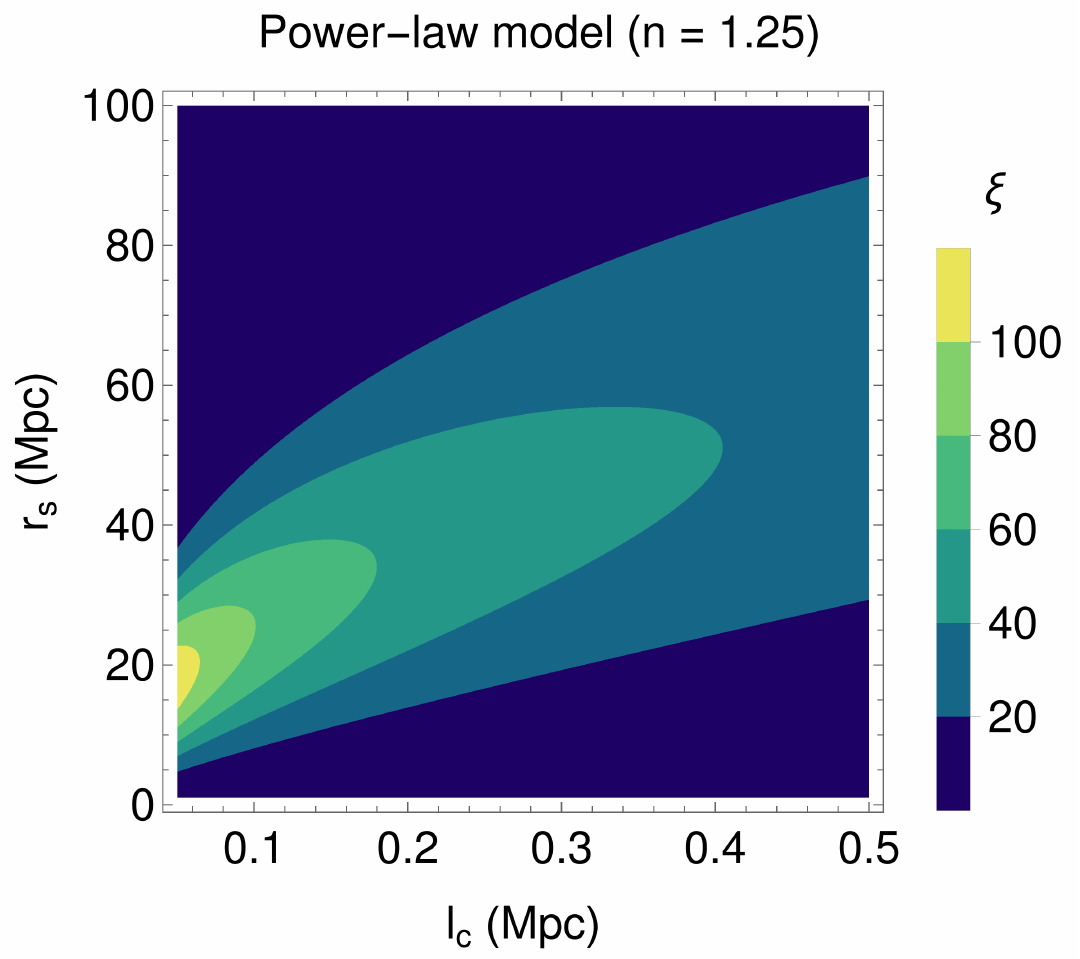}}
\centerline{
\includegraphics[scale=0.8]{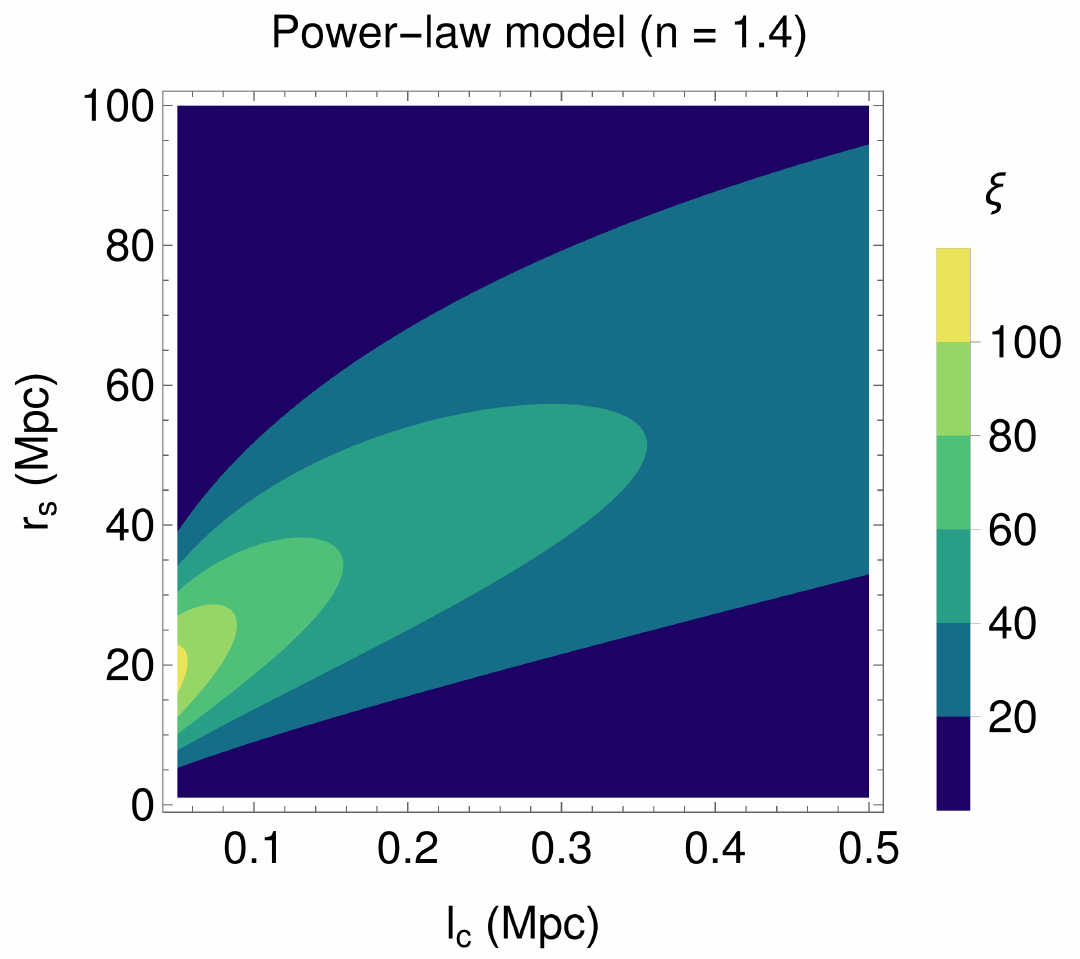}
\includegraphics[scale=0.8]{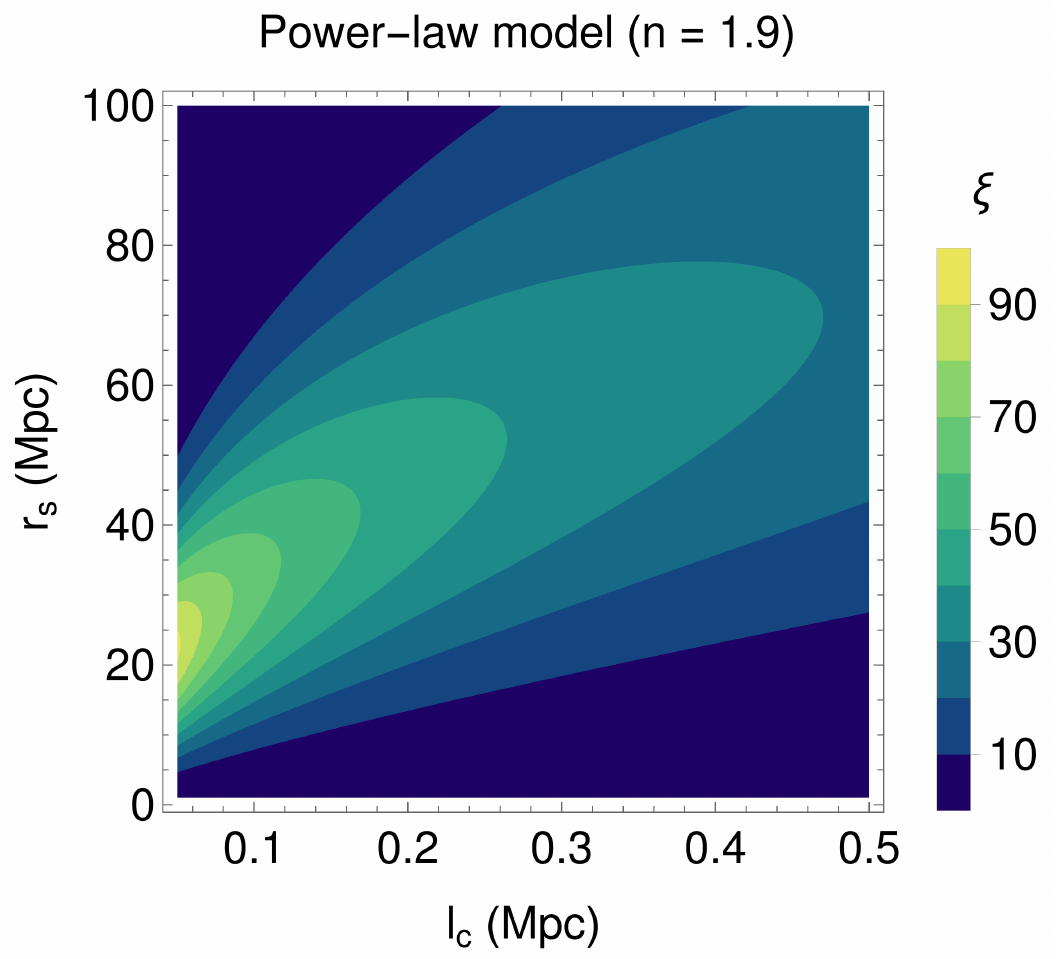}}
\caption{Contour plots of variation of density enhancement factor 
$\xi$ as a function of source distance $r_\text{s}$ and coherence length 
$l_\text{c}$ obtained by considering $E = 0.1$ EeV and $E_\text{c} = 4.5$ EeV 
for the $f(R)$ gravity power-law model and the $\Lambda$CDM model.}
\label{fig8a}
\end{figure}
The diffusive character of the propagation of UHE protons is shown in 
Fig.\ \ref{fig8}. Here we plot the density enhancement factor $\xi$ as a 
function of source distance $r_\text{s}$. In these plots, we fix the 
coherence length $l_\text{c}=0.1$ Mpc, while energy $E = 0.1$ to $5$ EeV 
have been taken into account. From these plots, we can say that 
the lower $E/E_\text{c}$ value results in a higher peak of the density 
enhancement with the peak position towards the smaller value of 
$r_\text{s}$, and also the enhancement peak lies in the diffusive region 
for smaller $E/E_\text{c}$ value. Again, the $\Lambda$CDM model 
shows the highest peak in the CRs density enhancement, while the $f(R)$ 
gravity power-law model depicts a better distribution of enhancement with the 
source distance. The power-law model with parameter 
values $n=1.25$ and $1.4$ results in a similar distribution, while 
for $n=1.9$, it shows a larger distribution. In the lower right panel, 
we consider a larger 
value of $E/E_\text{c}$ which results in a very poor peak for both 
$\Lambda$CDM and $f(R)$ power-law {models}. In this case, the enhancement 
peak is very far away from the diffusive regime. So from these results, we 
can finally say that for the suitable values of $l_\text{c}$ and 
$E/E_\text{c}$, the $\Lambda$CDM model depicts a better peak, while 
$f(R)$ power-law model depicts 
the enhancement in a much wider distribution. For a better illustration, we 
also draw contour plots of the density enhancement with 
the source distance $r_\text{s}$ ($0 - 100$ Mpc) and coherence length 
$l_\text{c}$ ($0.05 - 0.5$ Mpc) taking $E = 0.1$ EeV and 
$E_\text{c} = 4.5$ EeV for the $\Lambda$CDM and $f(R)$ power-law 
models as shown in Fig.\ \ref{fig8a}. One can see that the density
enhancement depends on the coherence length $l_\text{c}$ also. For the 
higher value of $l_\text{c}$ the density enhancement decreases, shifts 
away its maximum value from the source and takes place in the 
non-diffusive regime.

For reckoning the diffuse spectrum of UHE particles the separation between
sources plays a crucial role. If the sources are distributed 
uniformly with separations, which are much smaller than the propagation and 
interaction lengths, then the diffuse spectrum of UHE particles has a 
universal form, regardless of the mode of propagation of such particles  
\cite{aloisio}. To this end the explicit form of the source function $Q(E,z)$ 
for the power-law generation of the particles can be written as 
\cite{berezinski_four_feat}
\begin{equation}
Q(E,z)= \mathcal{L}_0(1+z)^{\delta} K q_{\text{gen}}(E_\text{g}),
\end{equation}
where $\mathcal{L}_0 = \int \mathcal{L}(E)\,dE$ is the total emissivity, 
$(1+z)^{\delta}$ represents the probable cosmological evolution 
of the sources with an index parameter $\delta$, $K$ 
is a normalisation constant with $K= \gamma_\text{g} -2$ for 
$\gamma_\text{g} >2$ and for $\gamma_\text{g} = 2$, 
$K = (\ln E_\text{max}/E_\text{min})^{-1}$, and 
$q_{\text{gen}}=E^{-\gamma_\text{g}}_\text{g}$ (see 
\hyperref[append]{Appendix A} for $E_\text{g}$). Utilizing the 
formalism of Ref.\ \cite{berezinkyGre}, it is possible 
to determine the spectrum of UHE protons in the model with a uniform source 
distribution and hence one can obtain the diffuse flux of UHE protons as
\begin{equation}\label{flux}
J_\text{p} (E)=\frac{c}{4\pi}\,\mathcal{L}_0 K \int_{0}^{z_\text{max}}\!\! dz\, \biggl | \frac{dt}{dz}\biggl |\, (1+z)^{\delta} q_\text{gen}(E_\text{g})\, \frac{dE_\text{g}}{dE}.
\end{equation}
Following Eq.\ \eqref{dtdz1} one can rewrite this diffuse flux 
Eq.\ \eqref{flux} as
\begin{align}\label{flux1}
J_\text{p} (E)= \frac{c}{4\pi}\, \mathcal{L}_0 K \displaystyle \int_{0}^{z_\text{max}}\!\! dz & \bigg[(1+z)^{-1} \Big[-\frac{2\,nR_0}{3 (3-n)^2\ \Omega_{\text{m}0}}\, \Bigl\{(n-3)\Omega_{\text{m}0}(1+z)^{\frac{3}{n}} \nonumber \\[5pt] & + 2 (n-2)\,\Omega_{\text{r}0} (1+z)^{\frac{n+3}{n}}\Bigl\}  \Big]^{-\frac{1}{2}} \bigg] (1+z)^{\delta} q_\text{gen}(E_\text{g})\, \frac{dE_\text{g}}{dE}.
\end{align}
\begin{figure}[h!]
\centerline{
\includegraphics[scale=0.43]{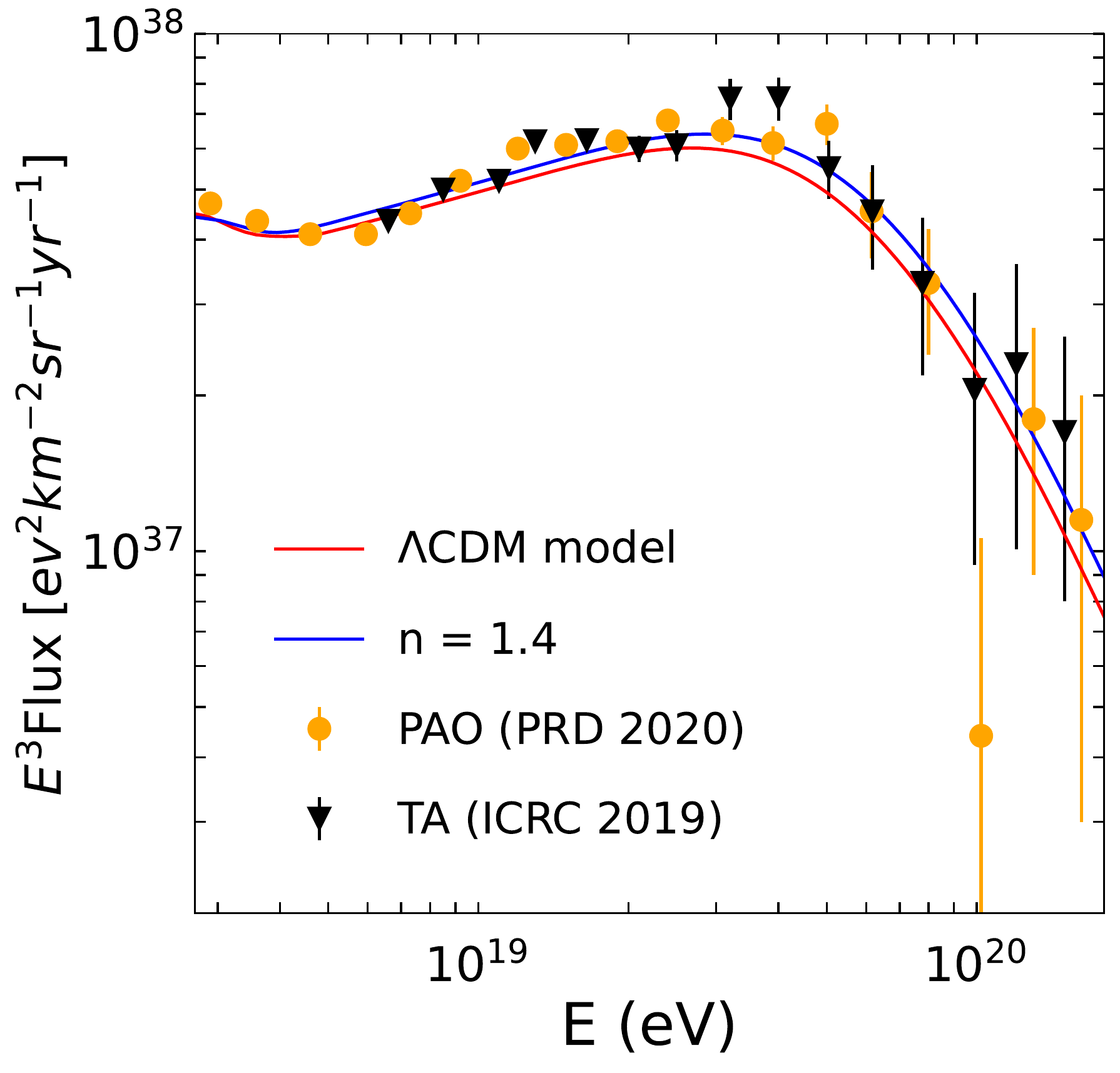}}
\vspace{-0.3cm}
\caption{UHECR protons flux is shown for the $f(R)$
gravity power-law model ($n = 1.4$) and the $\Lambda$CDM model in comparison with 
experimental data of the TA experiment \cite{ta2019} and PAO 
\cite{augerprd2020}.}
\label{fig9}
\end{figure} 
The spectrum given by Eq.\ \eqref{flux} is known as the universal spectrum as 
it is independent of the mode of propagation of particles which is
the consequence of the small separation of sources as mentioned earlier. 
The shape of the universal spectrum may theoretically be changed by a variety 
of effects, which include fluctuations in interaction, discreteness in 
the source distribution, large-scale inhomogeneous source 
distribution and local source overdensity or deficit. However, the 
aforementioned effects only slightly change the form of the universal spectrum, 
except for energies below $1$ EeV.
Numerical simulations demonstrate that the energy spectrum is changed by the 
propagation of UHE protons in the strong magnetic fields depending on the 
separation of sources. For small separation of sources with their uniform 
distribution the spectrum becomes the universal one as mentioned 
already \cite{Sigl_1999,Yoshiguchi_2003}. In Fig.\ \ref{fig9}, we plot the 
diffusive flux with no cosmological evolution ($\delta=0$) 
\cite{berezinski_four_feat, aloisio}. The emissivity $\mathcal{L}_0$ 
is taken to fit the curve with the available observational data 
\cite{berezinski_four_feat}. The energy-rescaling data of the 
TA experiment and PAO have been taken from
Ref.\ \cite{epjweb1}. It needs to be mentioned that the energy rescaling in the 
data of these two experiments is used to avoid the effect of the difference in 
the energy scales used by these two observatories. The uncertainty present in 
the energy scale contributes a significant impact on the uncertainty in the 
normalisation of the spectrum \cite{epjweb1}. The considered $f(R)$ gravity 
power-law model has shown a very good agreement with the observational data 
in predicting the energy spectrum of UHECRs and has also predicted 
similar result with that of the $\Lambda$CDM one. However, only a 
slightly higher 
flux is obtained for the power-law model in comparison to the $\Lambda$CDM 
model above $4$ EeV.  In data, a dip (the ankle) is seen at 
the energy around $4.5$ EeV, while at about $30$ EeV a bump 
(the instep) is observed. The ankle predicted by 
both power-law and $\Lambda$CDM models is at slightly lower energy around 
$3.5$ EeV, but the predicted position of the instep is the same as 
that of the data.

\begin{figure}[h!]
\centerline{
\includegraphics[scale=0.42]{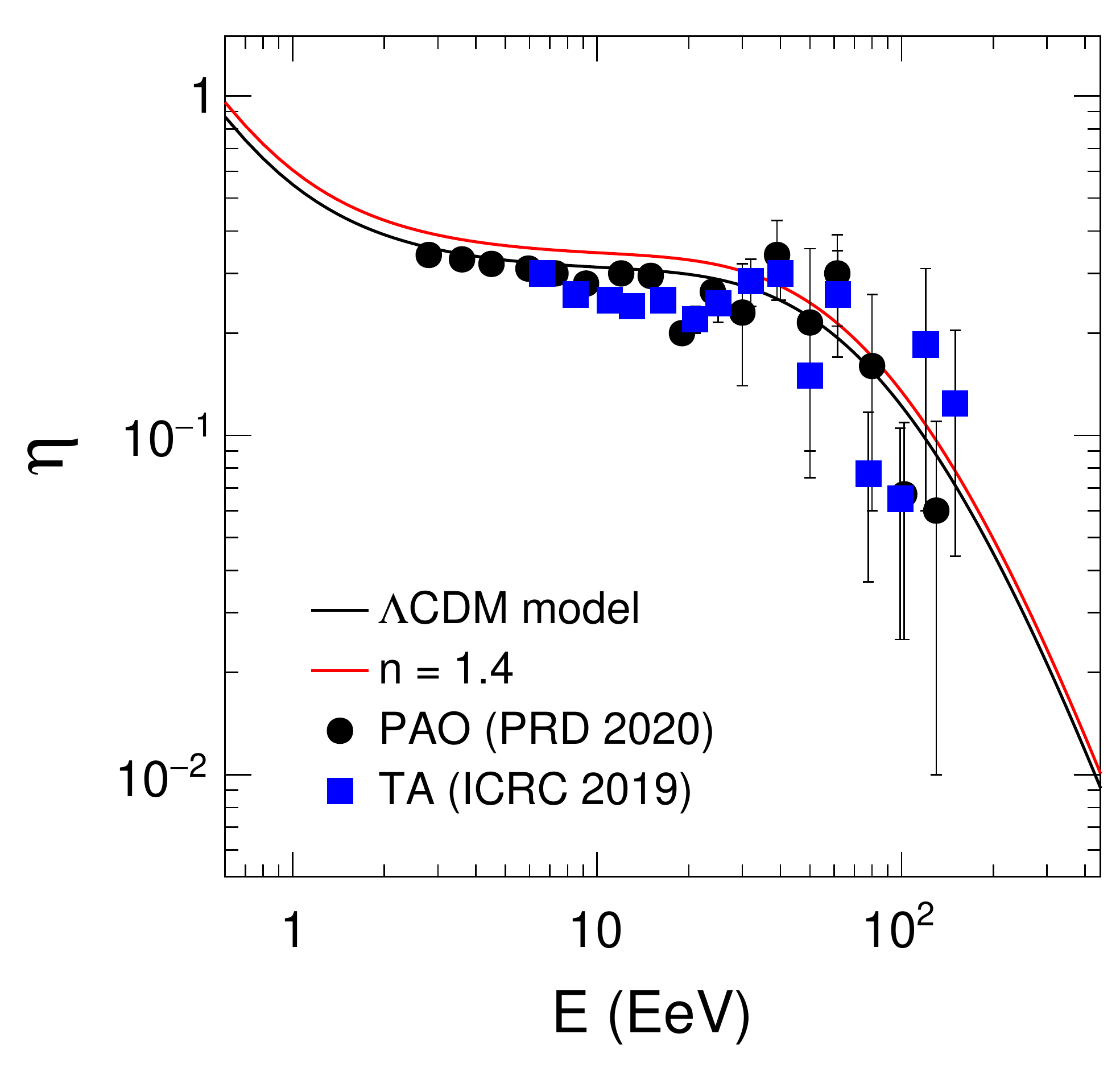}}
\vspace{-0.3cm}
\caption{Spectra of modification factor with $\gamma_g=2.7$ for the $f(R)$ 
gravity power-law model ($n = 1.4$) and the $\Lambda$CDM model in comparison with 
experimental data of the TA experiment \cite{ta2019} and PAO \cite{augerprd2020}.}
\label{fig10}
\end{figure}
These two signatures, the ankle and the instep are also 
observed in the modification factor of the energy spectrum plot as shown in 
Fig.\ \ref{fig10}. The modification factor of the energy spectrum is a 
convenient parameter for analysing the energy spectrum of UHECRs. 
This parameter corresponds to 
the enhancement factor of the density of UHECR particles discussed earlier. The 
modification factor of energy spectrum $\eta(E)$ is calculated as the ratio of 
the universal spectrum $J_\text{p}(E)$ which accounts for all 
energy losses to the unmodified spectrum $J_\text{p}^\text{unm}(E)$, 
in which only adiabatic energy losses due to the redshift are taken into 
consideration \cite{berezinski_four_feat}, i.e.
\begin{equation}\label{modification}
\eta (E) = \frac{J_\text{p}(E)}{J_\text{p}^\text{unm}(E)}
\end{equation}
Without any cosmological evolution, the unmodified spectrum can be written as
\begin{equation}
J_\text{p}^\text{unm}(E)=\frac{c}{4\pi}\,\mathcal{L}_0 (\gamma_g -2) E^{-\gamma_g} \int_{0}^{z_\text{max}} \!\! dz\, \biggl |\, \frac{dt}{dz}\biggl | (1+z)^{(1-\gamma_g)}.
\end{equation}
The modification factor as a function of energy with the spectral 
index $\gamma_g =2.7$ is shown in Fig.\ \ref{fig10} for the $f(R)$ gravity
power-law model and the $\Lambda$CDM model. At about $1$ EeV, the 
ankle is seen in the spectrum as predicted by both models in 
agreement with the observation of the TA experiment \cite{ta2019} and 
PAO \cite{augerprd2020} as well as a good agreement for the instep 
in the spectrum is also seen. From Fig.\ \ref{fig10}, it can also be said 
that the modification factor of the energy spectrum is a weak 
model-dependent parameter.

\subsection{Projections of Starobinsky \boldmath{f(R)} gravity model}
For this model of $f(R)$ gravity also we will follow the same procedure as we 
have already done in the case of the power-law model. So here also we 
have to calculate the Syrovatsky variable $\lambda^2$ and for this purpose, we 
express $\lambda^2(E,z)$ from Eq.\ \eqref{syrovatsky} using 
Eq.\ \eqref{dtdz_staro} for the Starobinsky model as
\begin{equation}\label{lambda_staro}
\lambda^2(E,z)= H_0^{-1}\int_{0}^{z}dz\, (1+z) \left[ \frac{3\, \Omega_{\text{m}0} (1+z)^3 + 6\, \Omega_{\text{r}0} (1+z)^4 + \frac{\alpha R + \beta R^2}{H_0^2}}{6(\alpha + 2 \beta R)\Bigl\{ 1-\frac{9\beta H_0^2 \Omega_{\text{m}0} (1+z)^3}{\alpha(\alpha+2\beta R)} \Bigl\}^2 }\right]^{-\frac{1}{2}}\!\!\!\!\!\!\! D(E_\text{g},z).
\end{equation}
\begin{figure}[h!]
\centerline{
\includegraphics[scale=0.42]{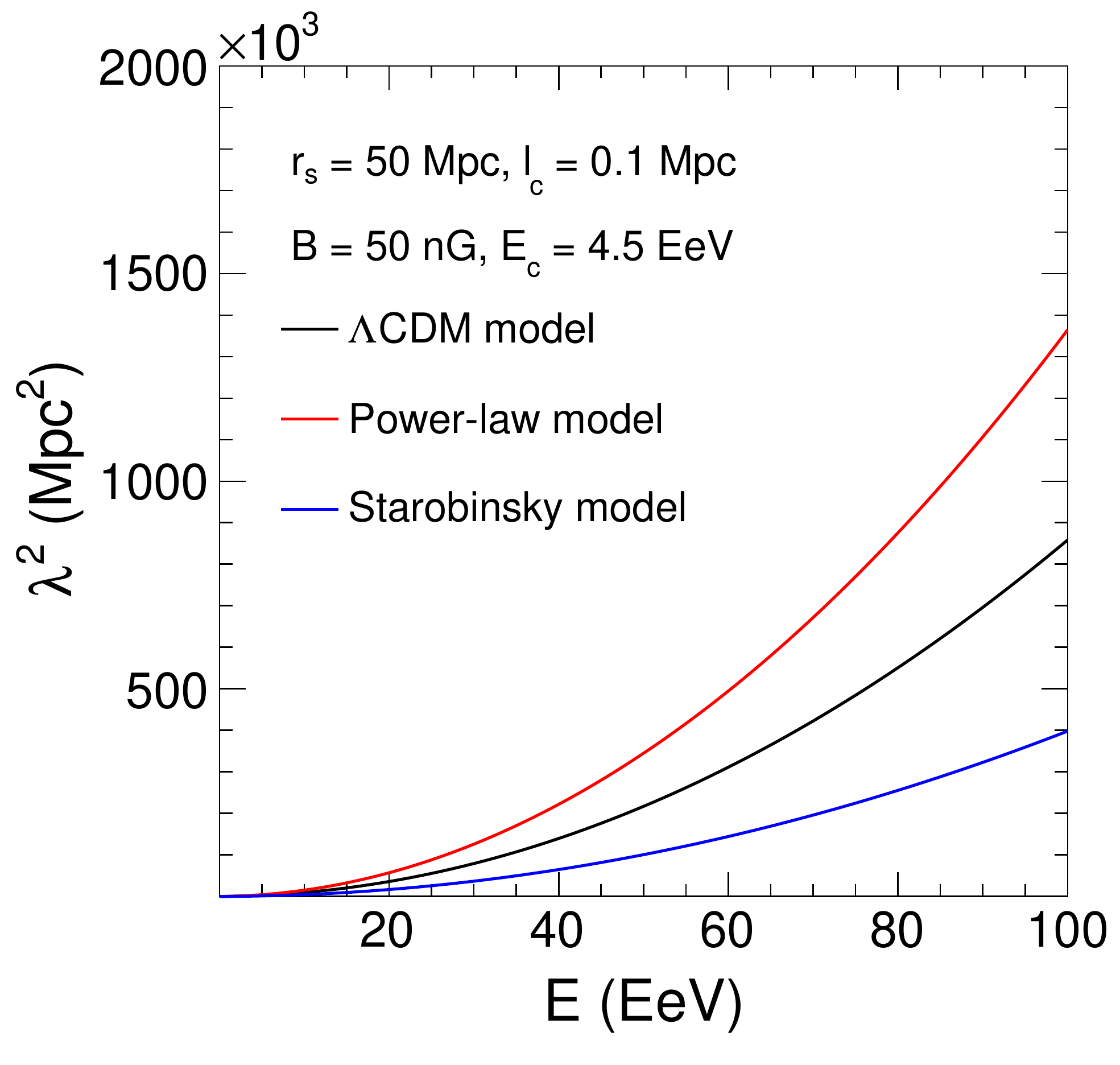}}
\vspace{-0.3cm}
\caption{Variation of $\lambda^2$ with energy $E$ for the $\Lambda$CDM, 
$f(R)$ gravity power-law ($n = 1.4$) and Starobinsky models obtained by 
considering $r_\text{s}=50$ Mpc, $l_\text{c}=0.1$ Mpc, $B=50$ nG and 
$E_\text{c}=4.5$ EeV.}
\label{fig11}
\end{figure}
In Fig.\ \ref{fig11} we plot the variation of $\lambda^2$ with respect to 
energy for the $f(R)$ gravity Starobinsky model and power-law model in 
comparison with the $\Lambda$CDM model. For this, we consider the source 
distance $r_\text{s} = 50$ Mpc, the coherence length 
$l_\text{c} = 0.1$ Mpc 
and the strength of the TMF, $B=50$ nG, and use only the Kolmogorov spectrum 
of the diffusion coefficient. A noticeable variation with respect to the 
energy is observed in  $\lambda^2$ values for all of the mentioned gravity 
models. Moreover, the $f(R)$ gravity Starobinsky model gives the lowest value 
of $\lambda^2$ although its pattern of variation with respect to energy is 
similar for all three models.    

\begin{figure}
\centerline{
\includegraphics[scale=0.4]{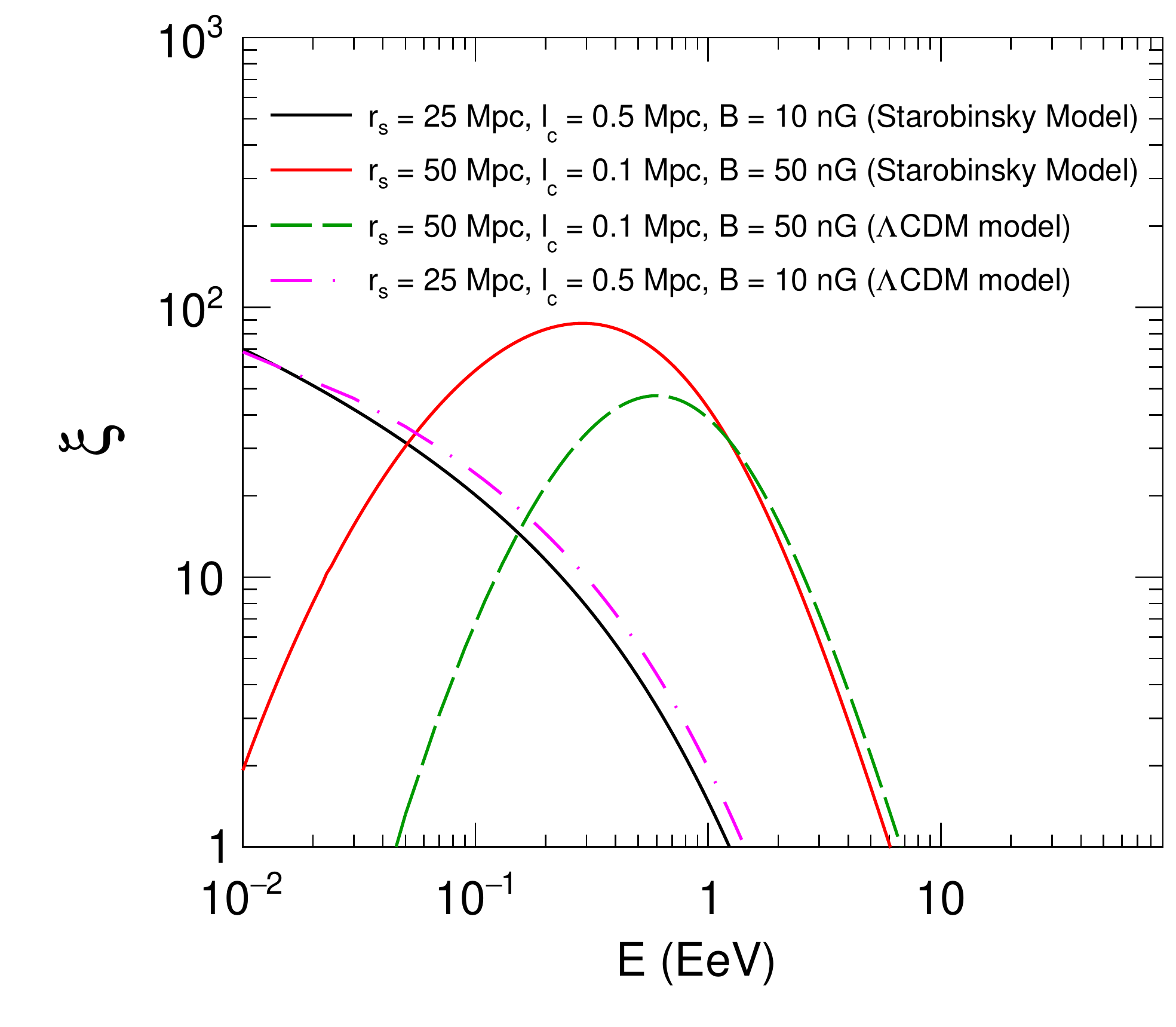}\hspace{0.5cm}
\includegraphics[scale=0.4]{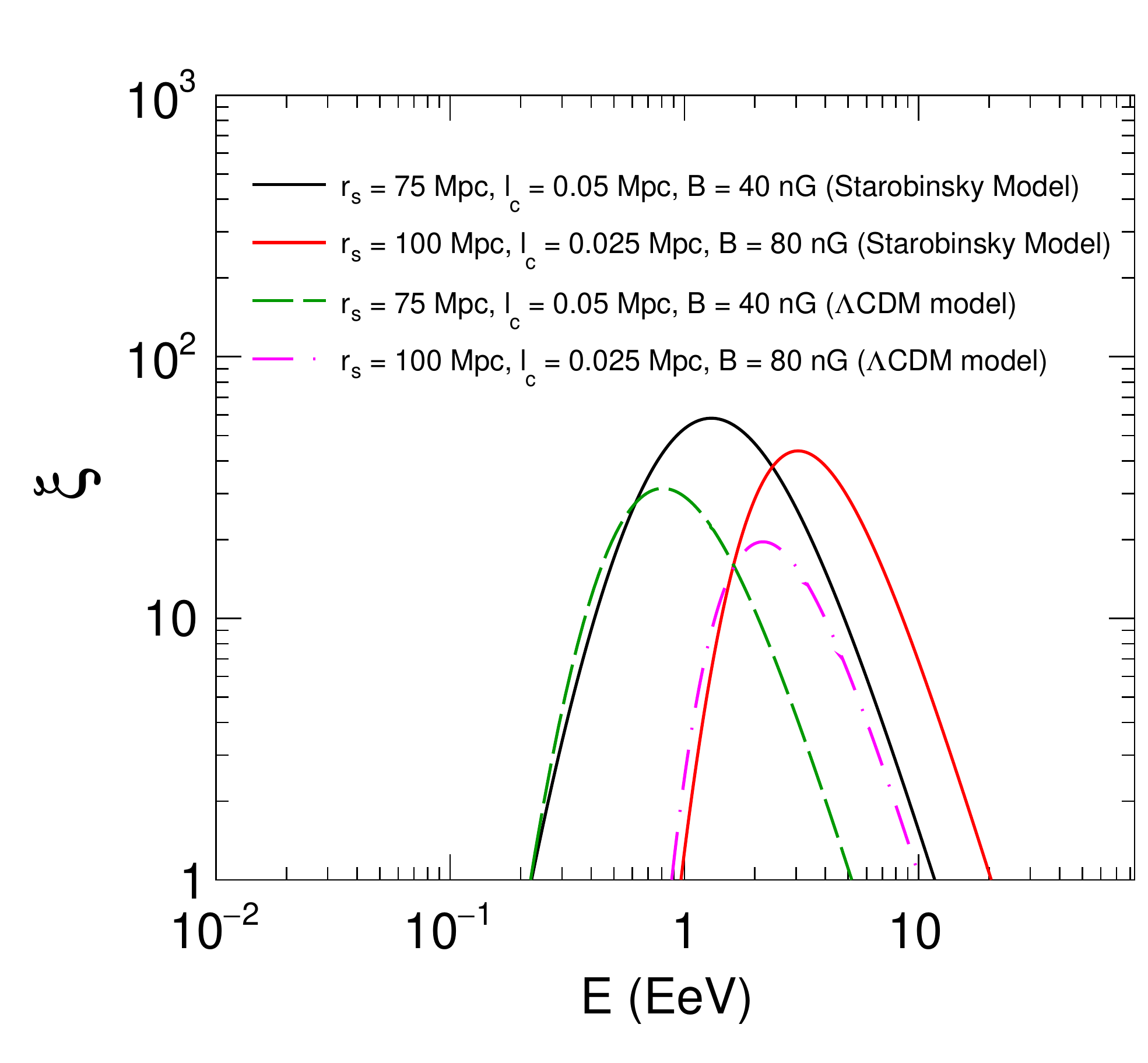}}
\vspace{-0.3cm}
\caption{Variation of density enhancement $\xi$ with respect to energy $E$ for 
the $f(R)$ gravity Starobinsky model in comparison with the $\Lambda$CDM model 
obtained by using different sets of parameters as $r_\text{s}=25-100$ Mpc, 
$l_\text{c}=0.025-0.5$ Mpc, $B=10-80$ nG and $E_\text{c}=1.8-4.5$ EeV.}
\label{fig12}
\end{figure} 
\begin{figure}[h!]
\centerline{
\includegraphics[scale=0.4]{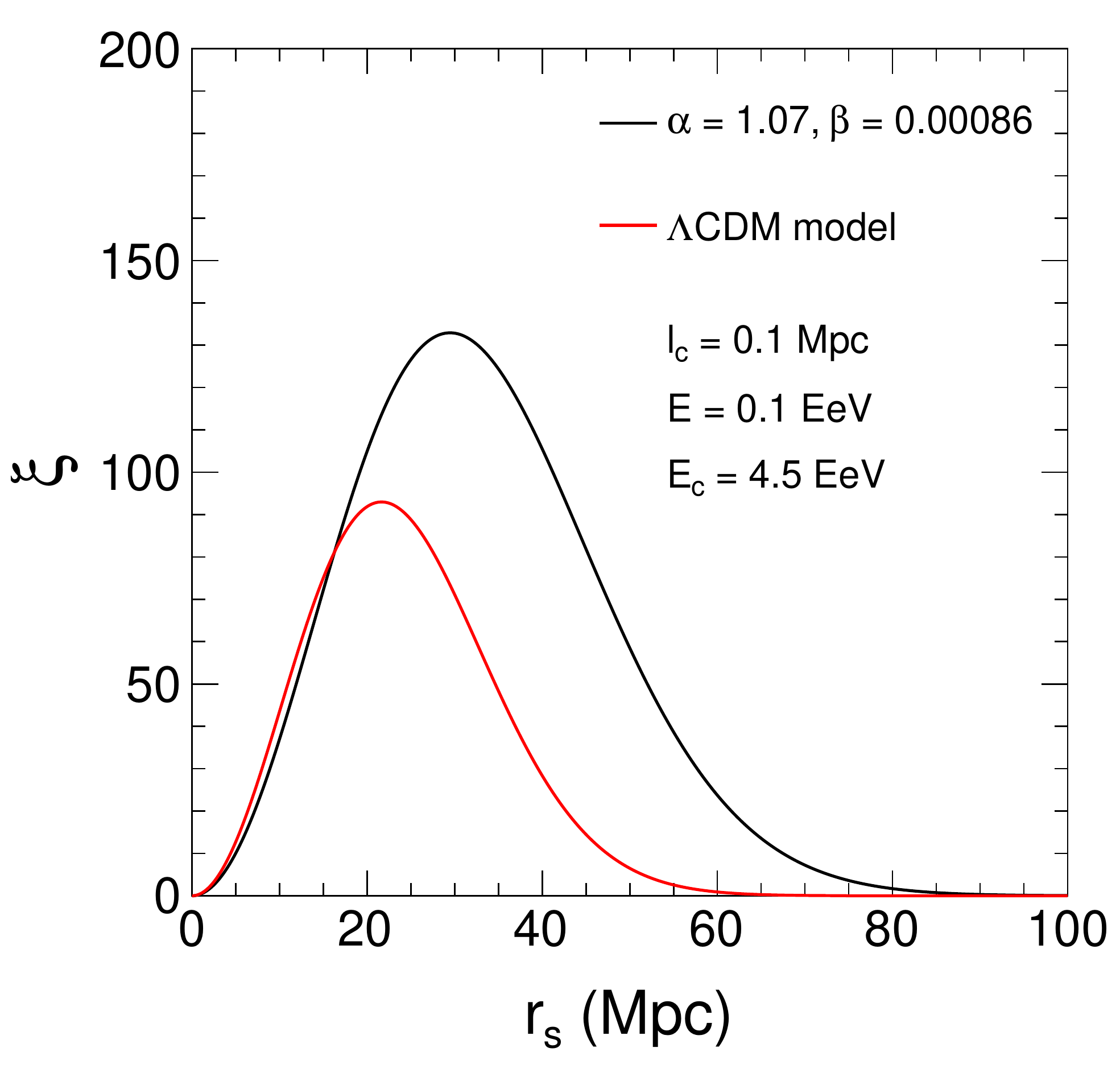}\hspace{0.5cm}
\includegraphics[scale=0.4]{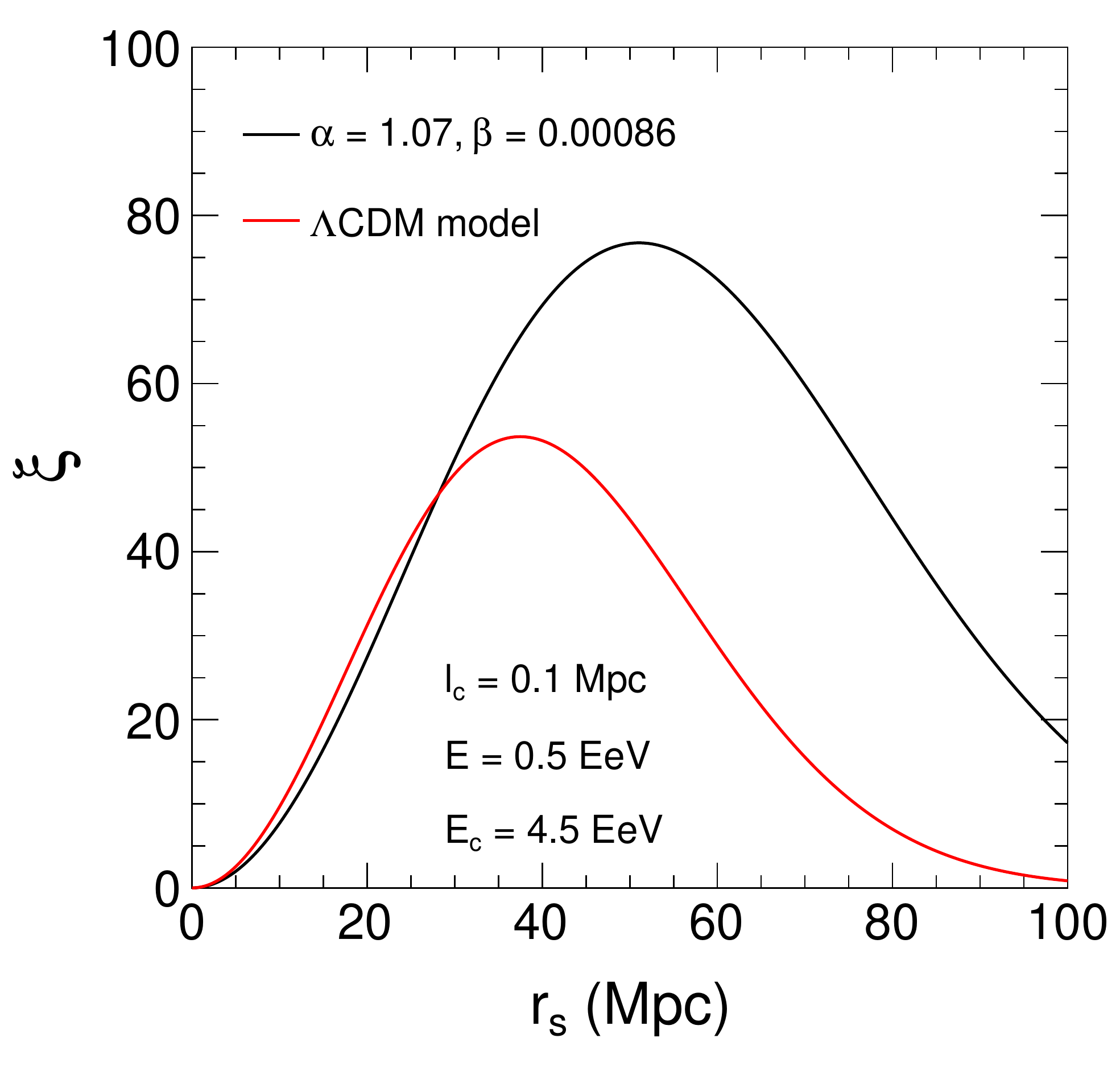}}
\centerline{
\includegraphics[scale=0.4]{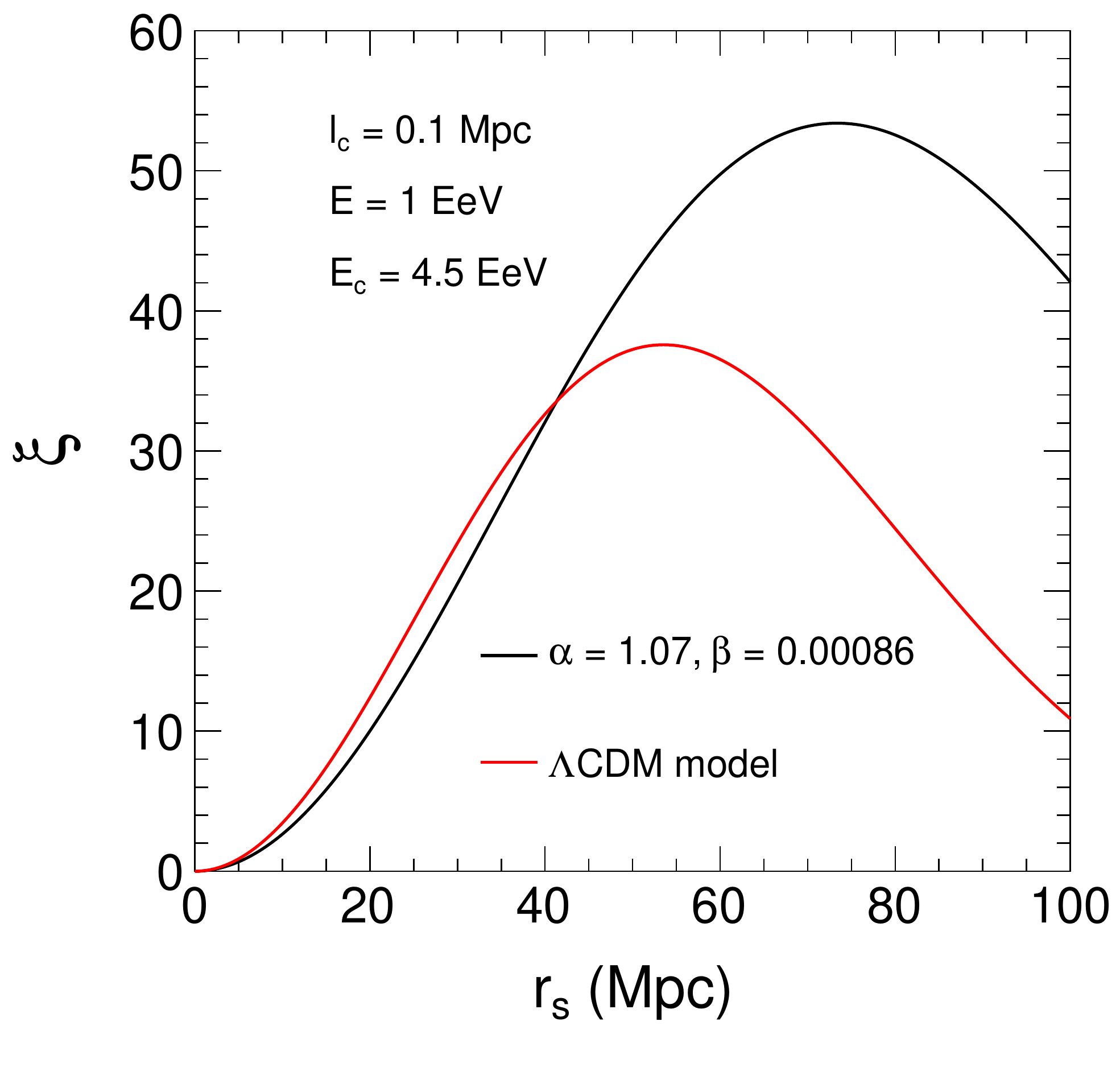}\hspace{0.5cm}
\includegraphics[scale=0.4]{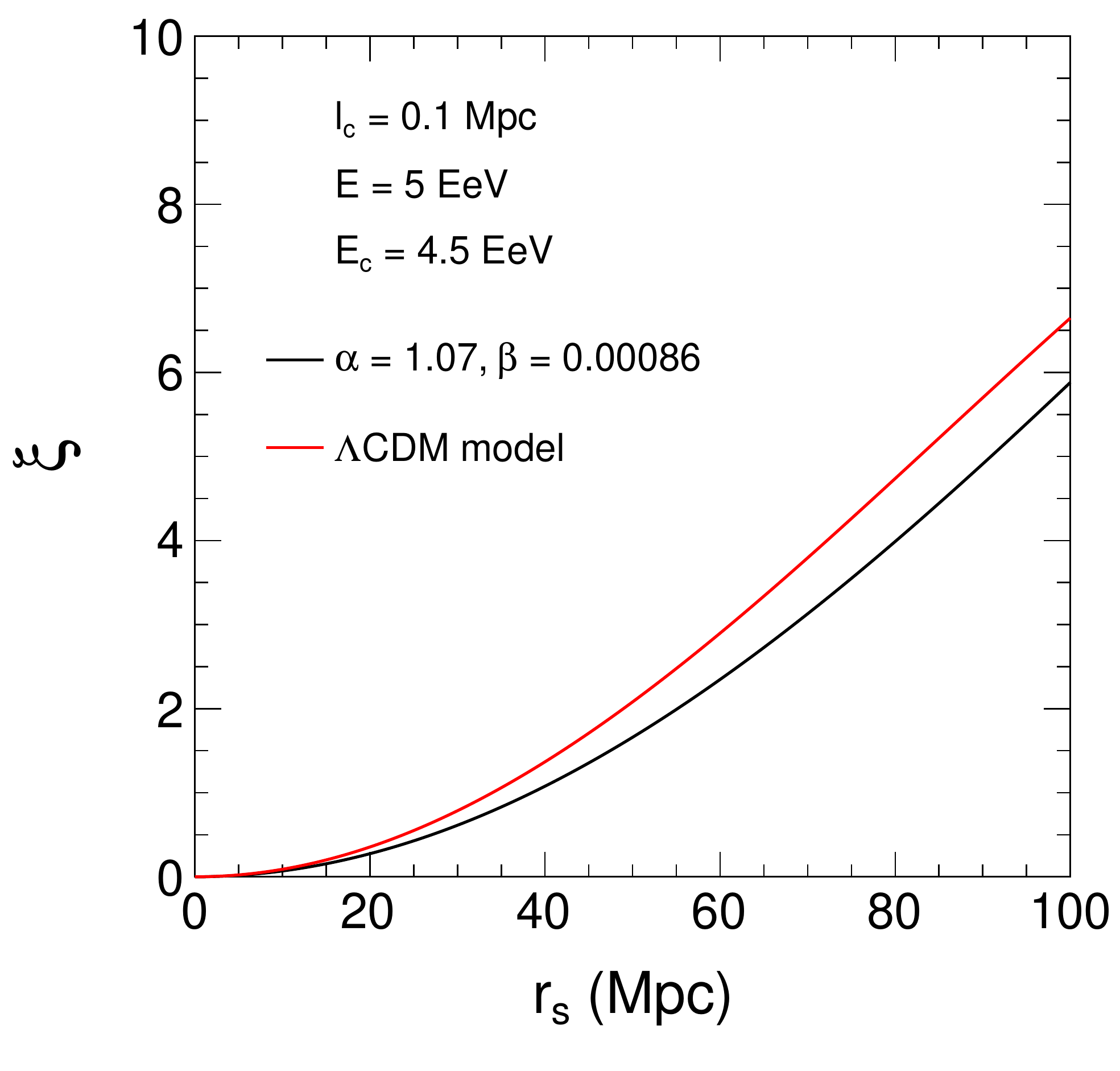}}
\vspace{-0.3cm}
\caption{Variation of $\xi$ with source distance $r_\text{s}$ obtained by 
considering the parameters $l_\text{c}=0.1$ Mpc and 
$E_\text{c} = 4.5 EeV$ with $E = 0.1$ EeV (upper left panel), $0.5$ EeV 
(upper right panel), $1$ EeV (lower left panel), and $5$ EeV (lower right 
panel) for both Starobinsky and $\Lambda$CDM models.}
\label{fig13}
\end{figure}
\begin{figure}[h!]
\centerline{
\includegraphics[scale=0.8]{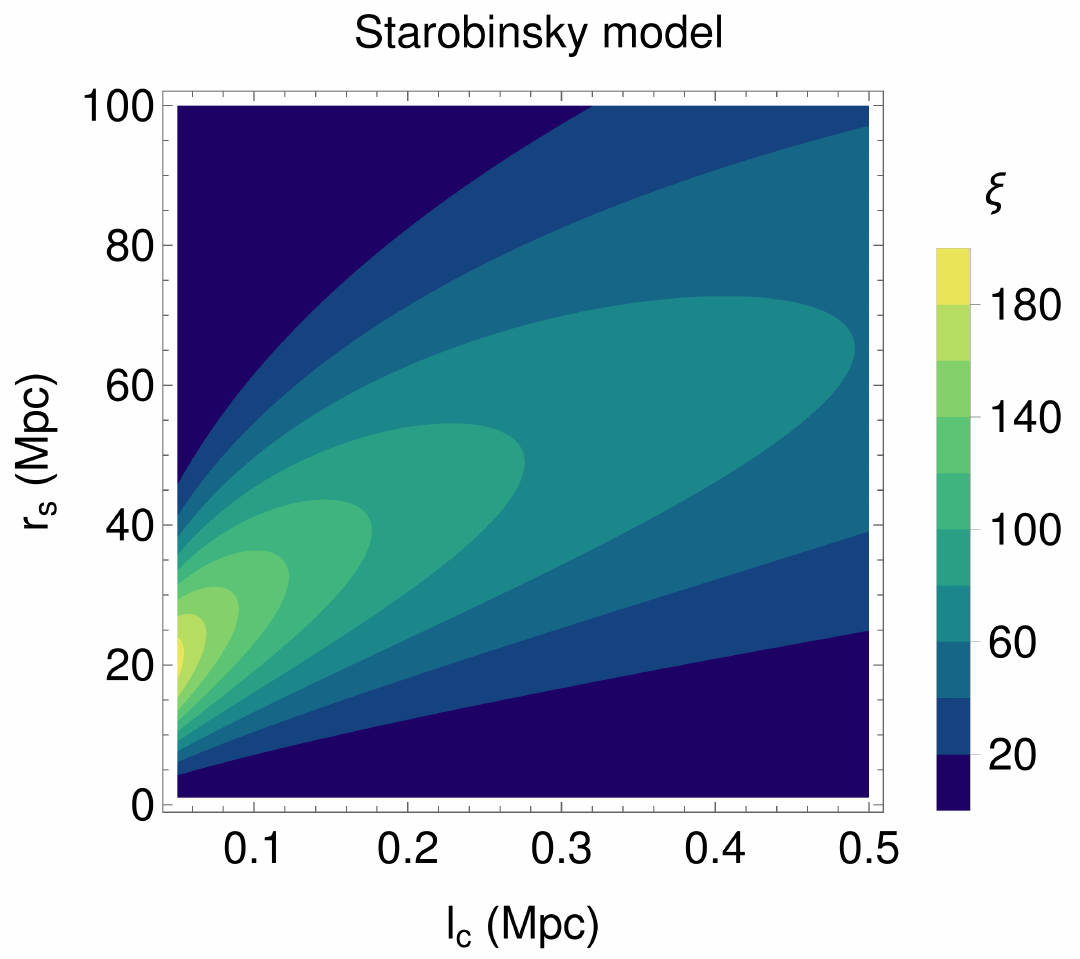}
\includegraphics[scale=0.8]{lambda1.pdf}}
\caption{Contour plots of variation of density enhancement factor $\xi$ 
with respect to source distance $r_\text{s}$ and coherence length 
$l_\text{c}$ obtained by considering $E = 0.1$ EeV and 
$E_\text{c} = 4.5$ EeV for the Starobinsky model and the 
$\Lambda$CDM model.}
\label{fig13a}
\end{figure}
Similarly, using Eqs.\ \eqref{density}, \eqref{degde} and \eqref{lambda_staro} 
in Eq.\ \eqref{enhancement} we calculate the density enhancement 
factor $\xi(E, r_\text{s})$ of UHE particles for the Starobinsky 
model, which can be written as
\begin{equation}\label{enhance_staro}
\xi(E, r_\text{s})= 4\pi r_\text{s}^2H_0^{-1}\! \int_{0}^{z_{i}}\!\! dz\, (1+z)^{-1}\! \left[ \frac{3\, \Omega_{\text{m}0} (1+z)^3 + 6\, \Omega_{\text{r}0} (1+z)^4 + \frac{\alpha R + \beta R^2}{H_0^2}}{6(\alpha + 2 \beta R)\Bigl\{ 1-\frac{9\beta H_0^2 \Omega_{\text{m}0} (1+z)^3}{\alpha(\alpha+2\beta R)} \Bigl\}^2 }\right]^{-\frac{1}{2}}\!\!\!\!\! \frac{\exp [-r_\text{s}^2/4 \lambda^2]}{(4\pi \lambda^2)^{3/2}}\, \frac{dE_\text{g}}{dE}.
\end{equation}
Considering the source distances $r_\text{s}=25$ Mpc and $50$ Mpc, 
coherence lengths $l_\text{c}=0.5$ Mpc and $0.1$ Mpc, and field strengths 
$B=10$ nG and $50$ nG, we plot the density enhancement factors as a function 
of energy $E$ 
for both Starobinsky model and $\Lambda$CDM model in the left panel of 
Fig.\ \ref{fig12} and that for $r_\text{s}=75$ Mpc and $100$ Mpc, 
$l_\text{c}=0.025$ Mpc and $0.05$ Mpc, and $B=40$ nG and $80$ nG in the 
right panel of Fig.\ \ref{fig12}. Note that in the figure, we constrain the 
critical energy i.e., $E_\text{c}=4.5$ EeV and $E_\text{c}=1.8$ EeV 
for the left and the right panel respectively. One can see that the 
enhancement of density precisely relies on the parameters we consider and for 
the different parameters we find a very distinct result in each of the 
cases. The distinction between enhancement factors for the Starobinsky model 
and $\Lambda$CDM model is clearly visible. The Starobinsky model gives a higher 
peak and wider range of the enhancement factor than that given by the 
$\Lambda$CDM model. Moreover, for smaller to medium values of $r_\text{s}$ 
the difference between the two models on the higher energy side is very small, 
while for higher values of $r_\text{s}$, it is very small on the lower 
energy side of the enhancement factor plots.    

\begin{figure}[h!]
\centerline{
\includegraphics[scale=0.4]{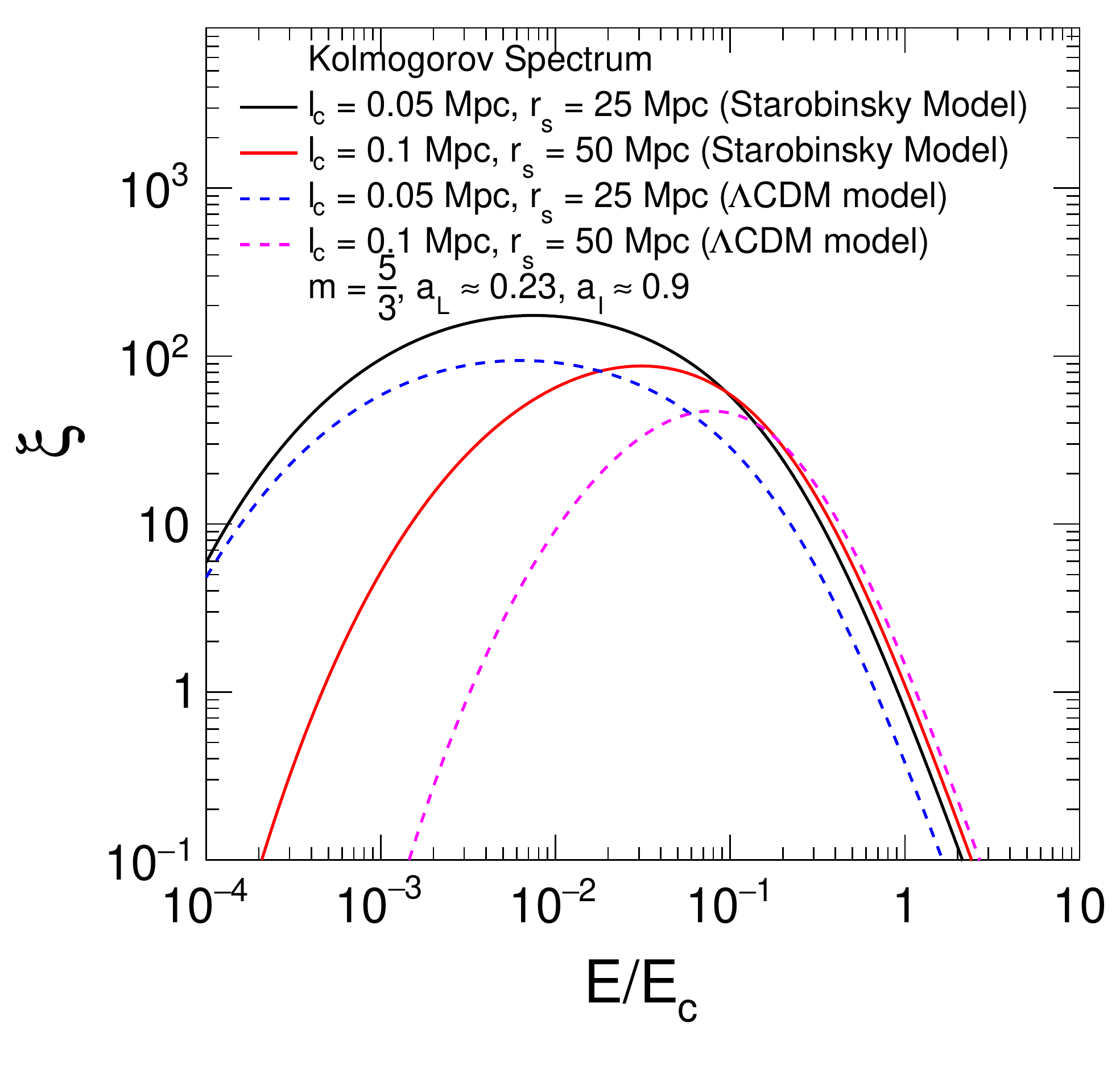}\hspace{0.5cm}
\includegraphics[scale=0.4]{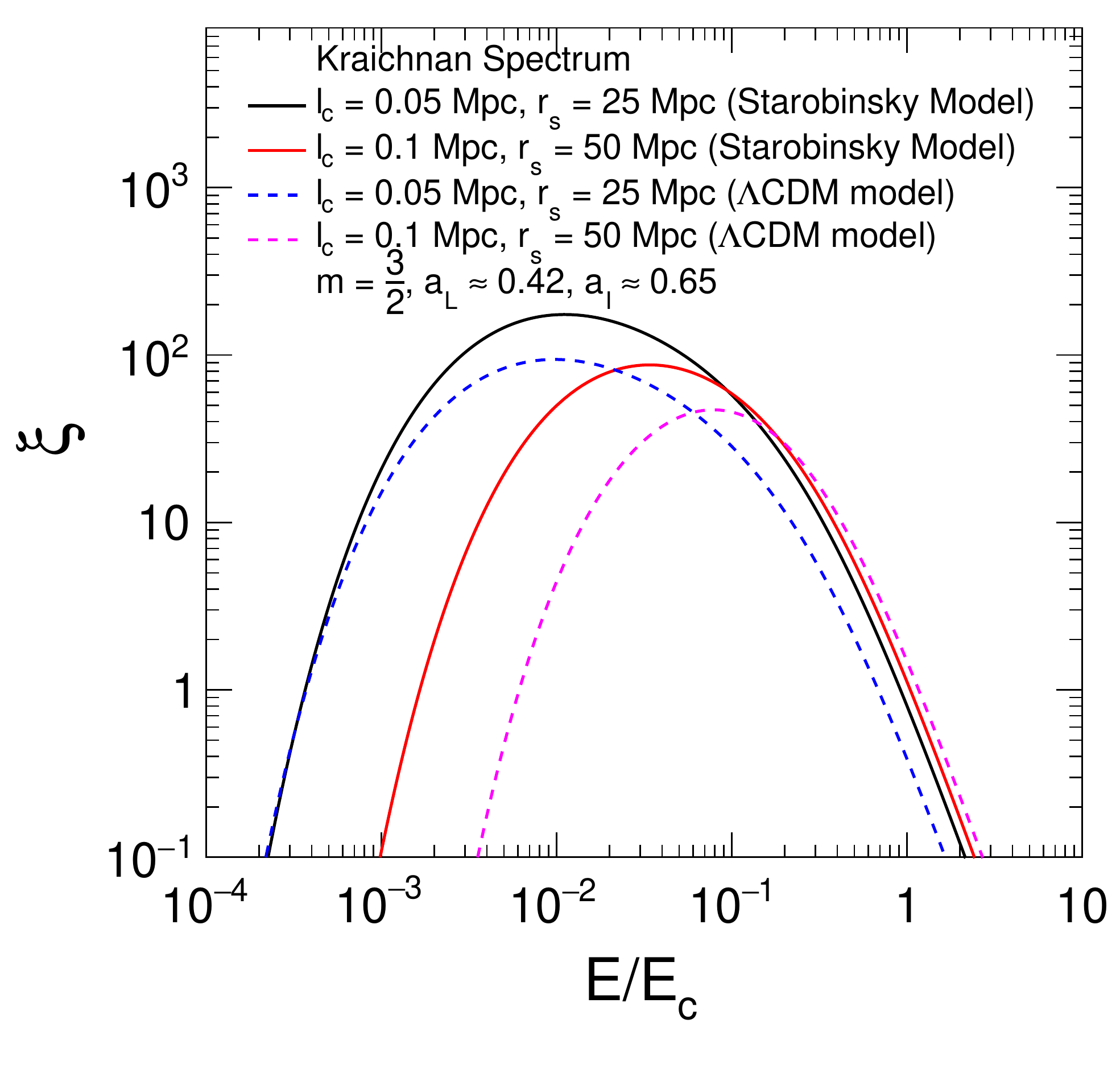}}
\vspace{-0.2cm}
\caption{Variation of density enhancement $\xi$ with $E/E_\text{c}$ for 
the Starobinsky model in comparison with the $\Lambda$CDM model. The left 
panel is for the Kolmogorov spectrum while the right panel is for the 
Kraichnan spectrum obtained by considering different sets of coherence 
length $l_\text{c}$ and source distance $r_\text{s}$.}
\label{fig14}
\end{figure}

Similar to the case of the $f(R)$ gravity power-law model, here also we plot 
the density enhancement factor $\xi$ with respect to the source distance 
$r_\text{s}$ by keeping fixed the coherence length $l_\text{c} =0.1$ 
Mpc for $E = 0.1$ EeV 
(upper left panel), $E =0.5$ EeV (upper right panel), $E = 1$ EeV (lower left 
panel) and $E = 5$ EeV (lower right panel) in Fig.\ \ref{fig13} to understand 
the propagation of UHECR protons in the light of the Starobinsky model in 
comparison with the $\Lambda$CDM model. From this figure, one can see that 
similar to the power-law model the peak of the enhancement is higher for 
smaller values of $E/E_\text{c}$, whereas the peak of the distribution is 
higher for the Starobinsky model than that of the $\Lambda$CDM model. Also 
similar to the power-law model the $\xi$ distribution becomes wider and the 
peak of it is shifted away from the source for higher $E/E_\text{c}$ 
values. As in the previous case for a clear understanding of the 
diffusive propagation, here also we draw contour plots in 
Fig.\ \ref{fig13a} for enhancement by keeping a range for the coherence 
length $l_\text{c}$ from $0.05 - 0.5$ Mpc and that of the source 
distance $r_\text{s}$ from $0 - 100$ Mpc for $E = 0.1$ EeV and 
$E_\text{c} = 4.5$ EeV. We see that for increasing the value of 
$l_\text{c}$ the enhancement decreases and also shifts its 
maximum value from the sources similar to the Power-law model case.

For a more distinct observation of the density enhancement features, 
we plot the density enhancement as a function of $E/E_\text{c}$ in 
Fig.\ \ref{fig14} for the Starobinsky model as well as for the $\Lambda$CDM 
model. Using $l_\text{c}= 0.05$ Mpc, $r_\text{s}=25$ Mpc (solid line) 
and $l_\text{c}= 0.1$ Mpc, $r_\text{s}=50$ Mpc (dotted line), 
the Kolmogorov spectra are shown in the left panel for both models. 
A remarkable variation is observed for $E/E_\text{c} < 0.1$ in both the 
sets of values, although for $E/E_\text{c} > 0.1$ quite similar results we 
obtained. Using the same sets of parameters, Kraichnan spectra are also 
plotted in the right panel of Fig.\ref{fig14}. The peaks of both spectra 
are almost in the same energy range but the variation in lower 
$E/E_\text{c}$ is quite different.

\begin{figure}[h!]
\centerline{
\includegraphics[scale=0.42]{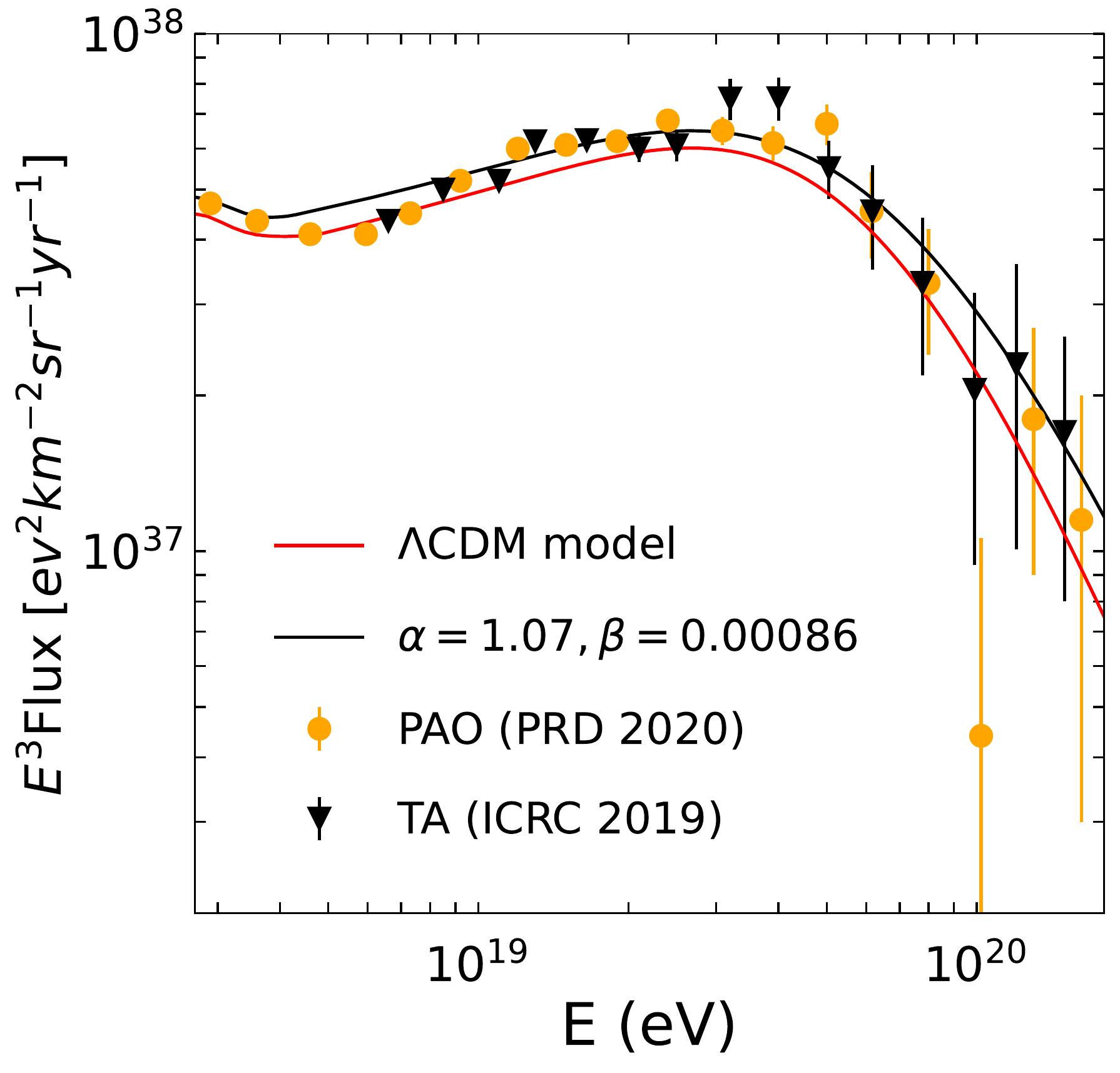}}
\vspace{-0.3cm}
\caption{UHECR protons flux is shown for the $f(R)$ gravity Starobinsky model 
and compared with the experimental data of TA 
experiment \cite{ta2019} and PAO \cite{augerprd2020} along with the 
flux for the $\Lambda$CDM model.}
\label{fig15}
\end{figure}
The diffuse UHECR protons flux for the $f(R)$ gravity Starobinsky 
model can be expressed as
\begin{equation}\label{flux2}
J_\text{p} (E)= \frac{c\,H_0}{4\pi}\, \mathcal{L}_0 K\! \int_{0}^{z_{max}}\!\!\!\!     dz\,(1+z)^{\delta-1} \left[ \frac{3\, \Omega_{\text{m}0} (1+z)^3 + 6\, \Omega_{\text{r}0} (1+z)^4 + \frac{\alpha R + \beta R^2}{H_0^2}}{6(\alpha + 2 \beta R)\Bigl\{ 1-\frac{9\beta H_0^2 \Omega_{\text{m}0} (1+z)^3}{\alpha(\alpha+2\beta R)} \Bigl\}^2 }\right]^{-\frac{1}{2}}\!\!\!\!\!\! q_\text{gen}(E_\text{g})\, \frac{dE_\text{g}}{dE}.
\end{equation}
In Fig.\ \ref{fig15}, we plot this flux \eqref{flux2} as a function of energy 
by considering the Starobinsky model parameters as we discuss in 
Section \ref{secIV}. From the figure, we can see that the Starobinsky 
model predicts a spectrum that is also in good agreement with 
the TA experiment and PAO data. However, within 4 EeV to 8 
EeV, the Starobinsky model's predicted spectrum remains
slightly above the 
observational data range. It also gives noticeably higher flux in 
comparison to the $\Lambda$CDM model over the almost entire energy range 
considered and the 
difference increases with increasing energy. Moreover, it also predicts the 
ankle of the spectrum at a lower energy of 
around $3.5$ EeV than data at the energy of around $4.5$ EeV, but 
the position of the predicted instep 
remains as that of the data. A detailed comparison of the diffuse fluxes for 
all the models considered in this work will be discussed in the next section.
\begin{figure}[h!]
\centerline{
\includegraphics[scale=0.43]{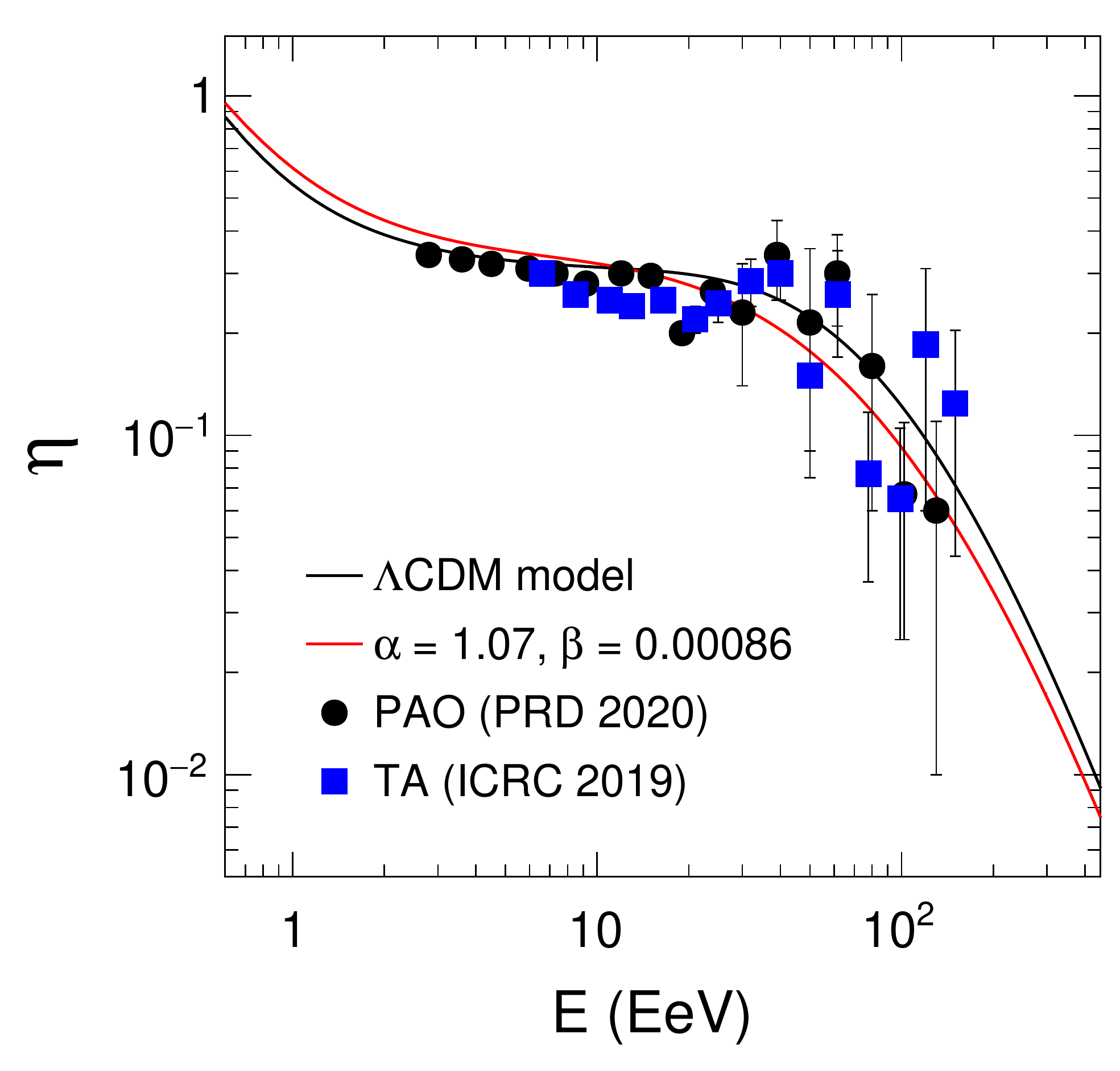}}
\vspace{-0.3cm}
\caption{Spectrum of the modification factor for the $f(R)$ 
gravity Starobinsky model along with that for the $\Lambda$CDM model with 
$\gamma_g=2.7$, which is in comparison with the experimental 
data of TA experiment \cite{ta2019} and PAO \cite{augerprd2020}.}
\label{fig16}
\end{figure}

Finally, for the calculation of the modification factor $\eta$ of the energy 
spectrum, the unmodified flux of UHECR protons for the Starobinsky model is 
given by
\begin{equation}
J_\text{p}^\text{unm}(E)=\frac{c\,H_0}{4\pi}\,\mathcal{L}_0 (\gamma_g -2) E^{-\gamma_g} \int_{0}^{z_\text{max}} \!\!\! dz \,(1+z)^{-\gamma_g} \left[ \frac{3\, \Omega_{\text{m}0} (1+z)^3 + 6\, \Omega_{\text{r}0} (1+z)^4 + \frac{\alpha R + \beta R^2}{H_0^2}}{6(\alpha + 2 \beta R)\Bigl\{ 1-\frac{9\beta H_0^2 \Omega_{\text{m}0} (1+z)^3}{\alpha(\alpha+2\beta R)} \Bigl\}^2 }\right]^{-\frac{1}{2}}\!\!\!\!\!\!.
\end{equation}
Fig.\ \ref{fig16} shows the behaviour of the modification factor for the 
Starobinsky model along with that of the $\Lambda$CDM model, and is 
compared with experimental data as in the previous case. 
%One can observe that for 
%$E < 0.9$ EeV the modification factor $\eta > 1$, which signifies the 
%presence of other components for the galactic origin of CRs like the 
%$f(R)$ gravity power-law model. 
The observational data have given a good agreement with the calculated 
modification factor spectrum with the ankle as well as the 
instep for the Starobinsky model similar to the $\Lambda$CDM 
model. It is also clear that the modification factor is very weakly model 
dependent as seen in the case of the power-law model also.

\section{Discussions and conclusions} \label{secVI}
The believable sources of UHECRs are extragalactic in origin 
\cite{harari, biermann_2012}. 
Accordingly, the propagation mechanisms of UHECRs through the extragalactic 
space have been one of the prime issues of study for the past several decades. 
It can be inferred that in the propagation of UHECRs across the extragalactic 
space, the TMFs that exist in such spaces and the current accelerated 
expansion of the Universe might play a crucial role. Thus this idea led us to 
study the propagation of UHECRs in the TMFs in the extragalactic space in the 
light of $f(R)$ theory of gravity and to compare the outcomes with the 
experimental data of two world-class experiments on UHECRs. The $f(R)$ 
theory of gravity is the simplest and one of the most successful MTGs that 
could explain the current accelerated expansion of the Universe. To this end, 
we consider two $f(R)$ gravity models, viz., the power-law model and the 
Starobinsky model. The Starobinsky model of $f(R)$ gravity is the most
widely used and one of the most viable models of the theory 
\cite{staro, d_gogoi, gogoi_model}. 
Similarly, the power-law model is also found to be suitable in various 
cosmological and astrophysical perspectives \cite{d_gogoi}. The basic 
cosmological equations for these two $f(R)$ gravity models, 
which are required for this study are taken from the Ref.\ \cite{d_gogoi}. 
Independent parameters of the models are first constrained by using the recent 
observational Hubble data. The relation between the redshift $z$ and 
the evolution time $t$ is calculated for both models. The UHECRs density 
$\rho(E,r_\text{s})$ and hence the enhancement factor of the 
density $\xi(E,r_\text{s})$ are obtained and 
they are calculated numerically for both the models of $f(R)$ gravity.

\begin{figure}[h!]
\centerline{
\includegraphics[scale=0.42]{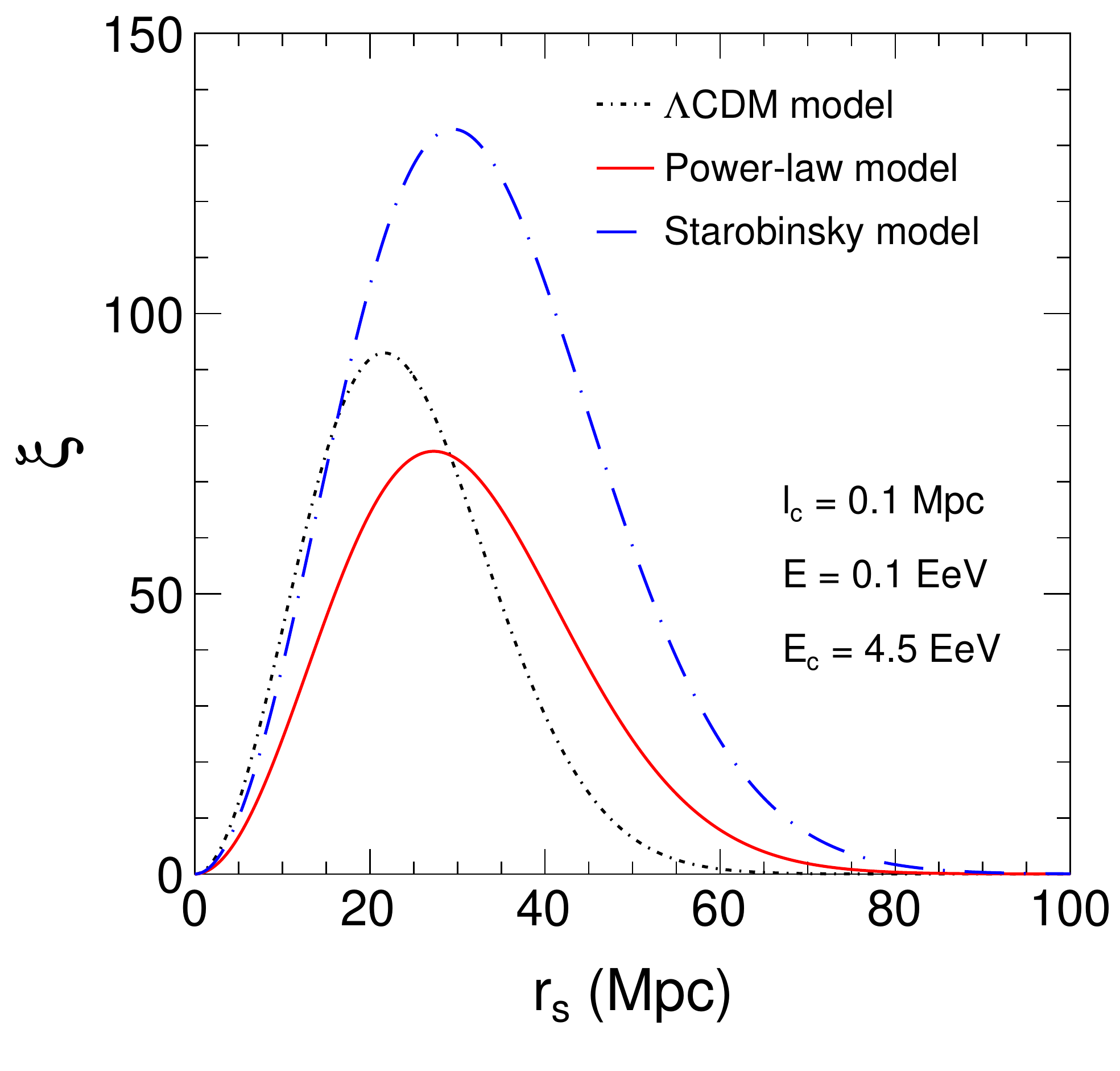}}
\vspace{-0.4cm}
\caption{Density enhancement factor $\xi$ as a function of source distance 
$r_\text{s}$ is shown for the power-law ($n = 1.4$) and the Starobinsky 
models of $f(R)$ gravity in comparison with that for the $\Lambda$CDM 
model by considering $l_\text{c}=0.1$ Mpc, $E = 0.1 EeV$ and 
$E_\text{c}= 4.5$ EeV. }
\label{fig17}
\end{figure}
\begin{figure}[h!]
\centerline{
\includegraphics[scale=0.4]{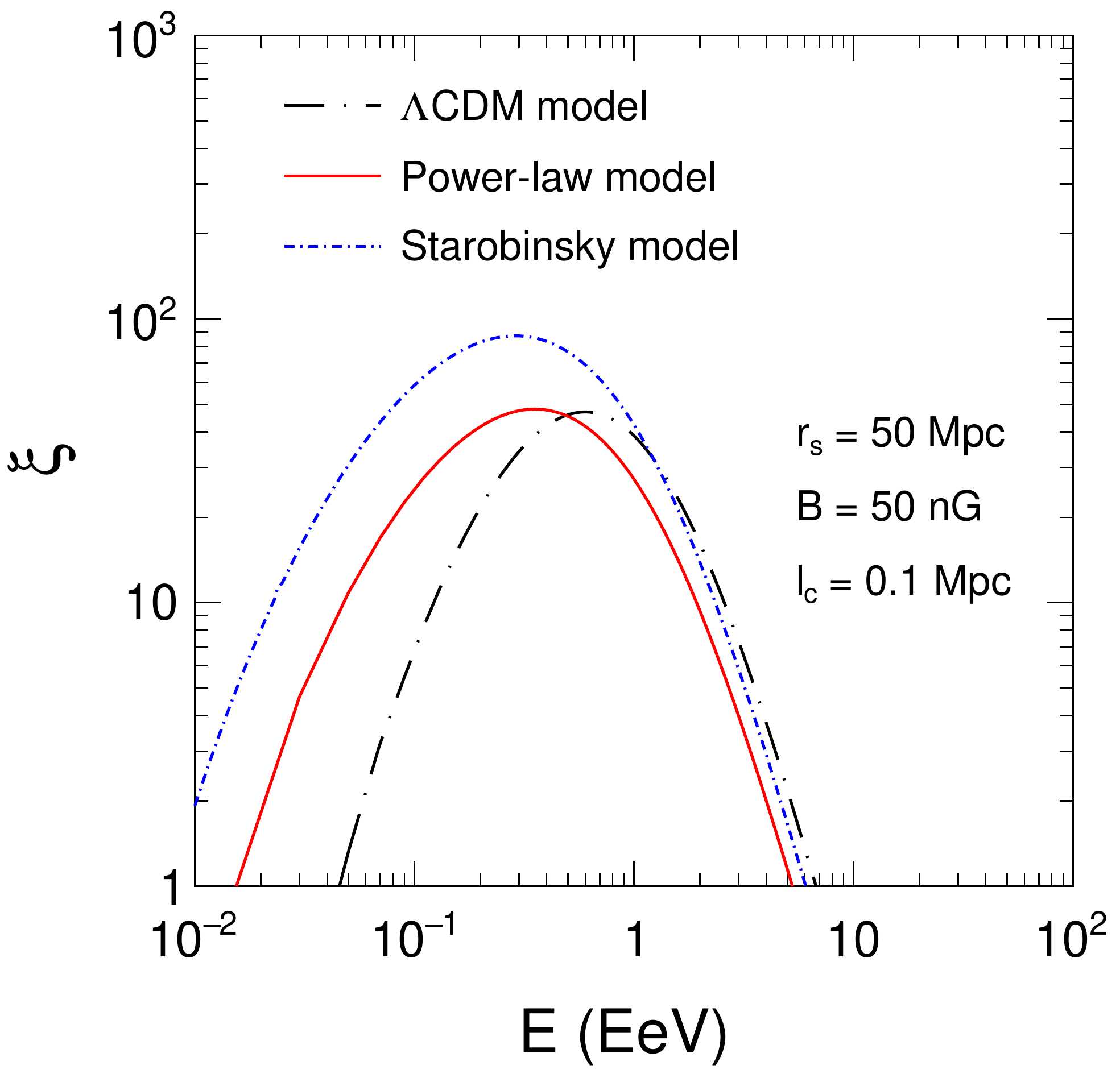}\hspace{0.5cm}
\includegraphics[scale=0.4]{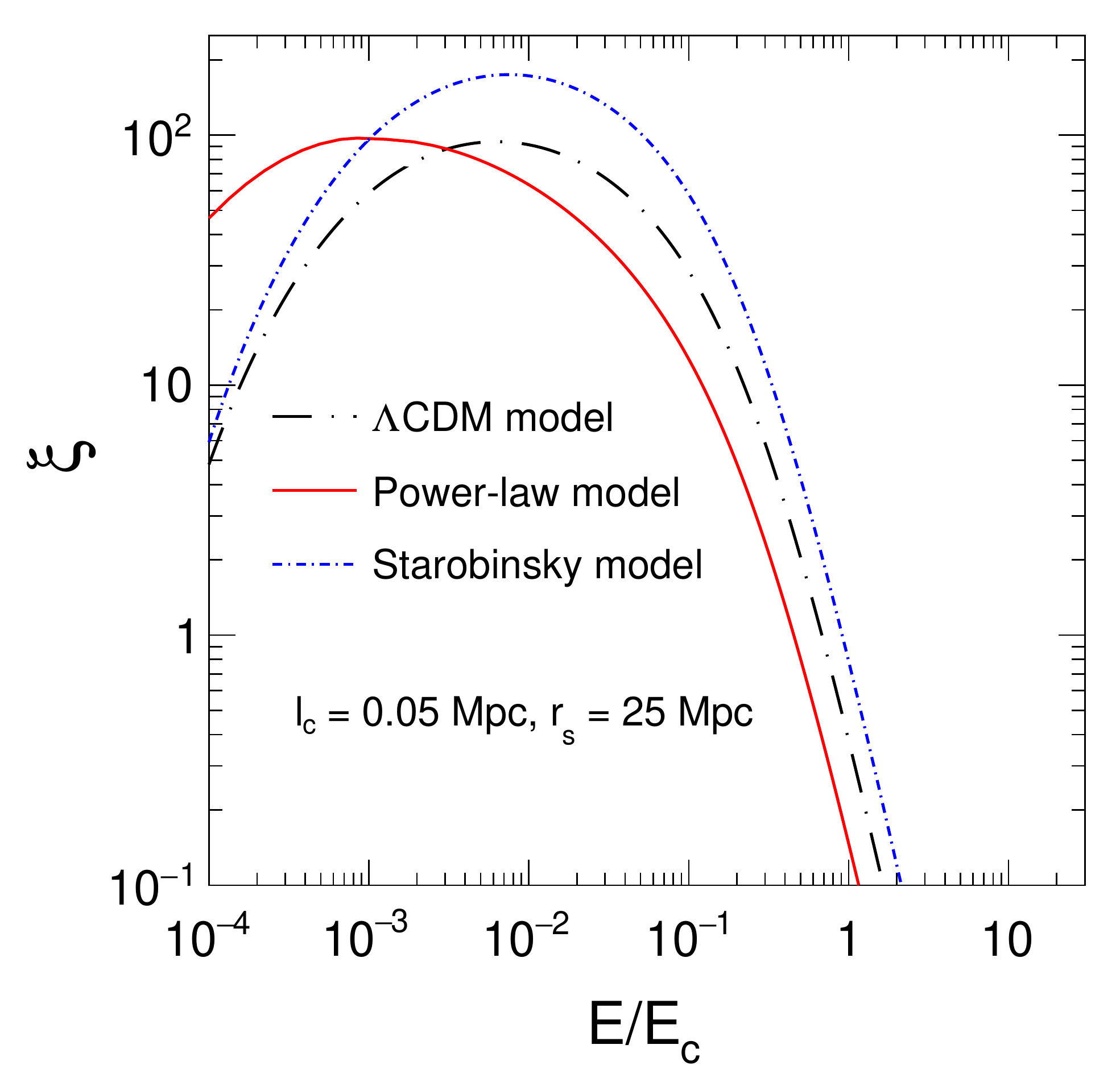}}
\vspace{-0.2cm}
\caption{Density enhancement factor $\xi$ as a function of energy $E$ of 
UHECR protons obtained by considering $r_\text{s}=50$ Mpc, 
$l_\text{c}=0.1$ Mpc, $B=50$ nG and $E_\text{c}=4.5$ EeV (left panel), 
and also as a function of $E/E_\text{c}$ of the same particles at
$r_\text{s}=25$ Mpc with $l_\text{c}=0.05$ Mpc (right panel). 
Both panels are shown for the power-law model ($n = 1.4$), the Starobinsky model, 
and the $\Lambda$CDM model.}
\label{fig18}
\end{figure}

A comparative analysis has been performed between the predictions of the 
power-law model and Starobinsky model of $f(R)$ gravity along with the same of 
the $\Lambda$CDM model for the density enhancement factor $\xi$ as a function 
of source distance $r_\text{s}$ in Fig.\ \ref{fig17}. In this analysis, 
we consider the coherence length $l_\text{c}=0.1$ Mpc and the fraction of 
energy and critical energy $E/E_\text{c}=0.02$. One can observe that at 
$r_\text{s} < 20$ Mpc, the variation of 
$\xi$ for the Starobinsky model and the $\Lambda$CDM is not very different 
but at the far distance from the source, the behaviour of these two models is 
quite different in terms of the peak position of the enhancement and the 
range of the source distance where the enhancement takes place. In the case of 
the $f(R)$ power-law model, the enhancement is less than the Starobinsky model 
and the $\Lambda$CDM model, but it gives the density enhancement in a much 
wider range than the $\Lambda$CDM model. In fact, it gives the same range of 
source distance distribution in the enhancement and gives the peak of 
enhancement at the same distance as that of the Starobinsky model although the 
enhancement is comparatively low. Another comparative analysis has been done in 
Fig.\ \ref{fig18} (left panel) for the CRs density enhancement with energy. 
For this purpose, we take the parameters as $r_\text{s}=50$ Mpc, 
$l_\text{c}=0.1$ Mpc, $B=50$ nG and $E_\text{c}=4.5$ EeV.  
The Starobinsky model has given the best results as compared with the other two 
models. From the left panel of Fig.\ \ref{fig18}, we see that at lower 
energies i.e., below $1$ Eev, the enhancement is different for different 
energy values including the peaks for all three models. But if we take a 
look at the higher values of energy, all three models depict almost 
similar results in the enhancement. One can say that the maximum value of 
enhancement for the power-law model and the $\Lambda$CDM model is 
approximately the same but the power-law model has covered a wider range of 
energy values than the $\Lambda$CDM model. While the Starobinsky model gives 
the highest enhancement value as well as the enhancement in a much wider range 
of energy values. The right panel of Fig.\ \ref{fig18} is plotted to show the 
variation of density enhancement as a function of $E/E_\text{c}$. In this 
panel, we consider the coherence length $l_\text{c}=0.05$ Mpc and source 
distance $r_\text{s}=25$ Mpc to demonstrate the behaviour of enhancement 
with the per unit increase of energy with respect to the critical energy. It 
is seen that at $E/E_\text{c} = 10^{-4}$, the values of 
enhancement for the Starobinsky model and $\Lambda$CDM are 
approximately the same, while the $f(R)$ power-law model has shown a higher 
value of enhancement at this point. But as the fraction of energy is 
increased, the Starobinsky model has given a better result of enhancement as 
compared to the other two models. 

\begin{figure}[h!]
\centerline{
\includegraphics[scale=0.428]{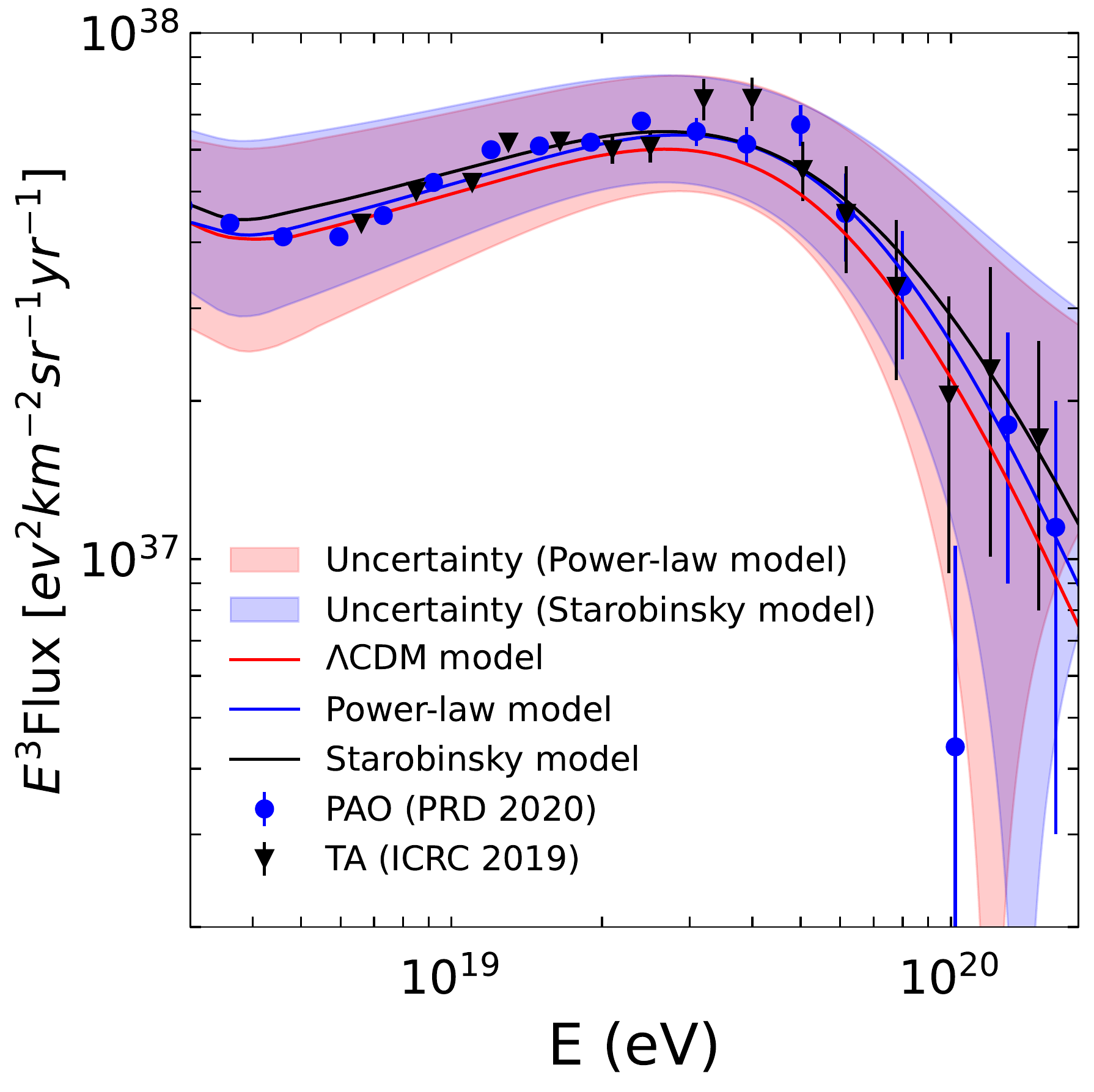}\hspace{0.5cm}
\includegraphics[scale=0.417]{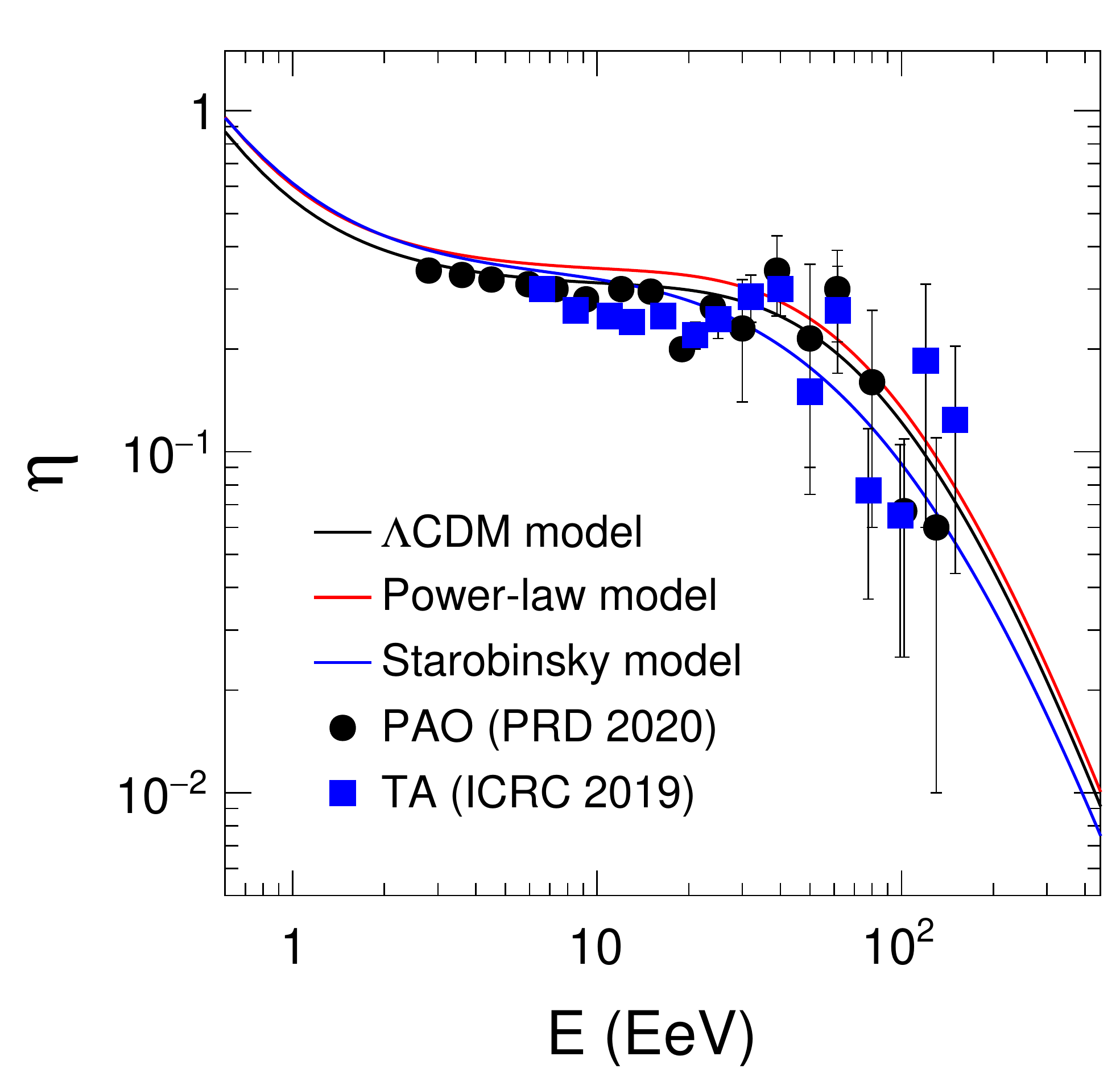}}
\vspace{-0.3cm}
\caption{Calculated $E^3$ magnified spectra of UHECR protons for the 
$\Lambda$CDM model, the $f(R)$ gravity power-law model ($n = 1.4$), and the Starobinsky 
model in comparison with the data of TA experiment 
\cite{ta2019} and PAO \cite{augerprd2020} with the uncertainty regions for the 
considered cosmological models (left panel). The modification factors 
of these spectra are shown in comparison with the TA 
experiment and PAO data in the right panel.}
\label{fig19}
\end{figure}
We calculate the $E^3$ magnified flux numerically for the both $f(R)$ 
gravity power-law and Starobinsky models and plot them along with that for 
the $\Lambda$CDM model in Fig.\ \ref{fig19} (left panel). We compare our 
calculations with the available observational datasets of the TA 
experiment \cite{ta2019} and PAO \cite{augerprd2020} consisting of 15 and 18 
numbers of data points respectively. All of these models 
have shown a very good agreement with the observational data of both the UHECRs 
experiments in predicting the signatures of UHECRs energy spectra. The 
Starobinsky model spectrum has shown a higher flux throughout the energy ranges 
considered. However, around $30$ EeV it gives the flux very near to that of the
power-law model. While the power-law model gives almost the same flux as that
of the $\Lambda$CDM model below $4$ EeV, above this energy the 
power-law model 
gives gradually higher flux than the $\Lambda$CDM model. The shaded regions 
have depicted the uncertainties in predicting the fluxes by the power-law and 
Starobinsky models. It is seen that the uncertainty regions in the plot are 
confined within the error bars of the observational data range. 

To test the goodness of fit of our model's predictions to the
experimental data, we implement the $\chi^2$ test defined as
\begin{equation}
    \chi^2 = \sum_{i} \frac{(\text{F}^i_\text{th}-\text{F}^i_\text{obs})^2}{\sigma^2},
\end{equation}
where $\text{F}^i_\text{th}$ is the $i$th theoretical value of flux 
that we obtained from a cosmological model and 
$\text{F}^i_\text{obs}$ is the $i$th observational value of flux 
obtained from the TA experiment or PAO. $\sigma$ is 
the standard deviation of the correspodng observed data. The values 
of $\chi^2$ along with their 
critical values are shown in Table \ref{table2}. It is seen that the 
$\chi^2$ value of the fit of predicted fluxes for each model 
to the data set is small in comparison to the corresponding critical 
value. Hence, this justifies the trustability of the model's predictions. 
It is to be noted that the critical value is calculated for the 95\% 
confidence level of each dataset using the Python scipy library 
\cite{scipy}.

\begin{table}
\caption{$\chi^2$ values of the fits of the predicted UHECR protons 
energy spectra by the considered $f(R)$ gravity models and the 
$\Lambda$CDM model to the data of TA experiment and 
PAO, and their associated critical values}.
\vspace{0.35cm}
    \centering
    \begin{tabular}{c c c c c c}
     \hline
     & Model \hspace{5pt} &\hspace{5pt} Data \hspace{5pt} & \hspace{5pt} $\chi^2$ value\hspace{5pt} &\hspace{5pt} Critical value &\\[3pt]
     \hline    
&        Power-law   & PAO   &    2.52     &      24.99 & \\[3pt]     
&        Power-law   & TA    &    2.74     &      23.68 & \\[3pt]      
&        Starobinsky & PAO   &    2.12     &      24.99 & \\[3pt]      
&        Starobinsky & TA    &    1.47     &      23.68 & \\[3pt] 
&        $\Lambda$CDM       & PAO   &    2.79     &      24.99 & \\[3pt]    
&        $\Lambda$CDM       & TA    &    2.84     &      23.68 & \\[2pt]        
    \hline
    \end{tabular}
    \label{table2}
\end{table}

The analysis of the ankle and instep is more convenient with 
respect to the modification factor and we have compared it with the available 
data. Both the considered $f(R)$ gravity models have shown a good 
agreement with the observational data. Thus it can be concluded that the 
$f(R)$ gravity models considered here are found to be noteworthy with some 
limitations depending upon the range of energies in explaining the 
propagations of UHECRs and hence the observed data of their fluxes. 
However, we would like to clarify here that our aim was to study the 
possible effects of $f(R)$ cosmology on UHECRs propagation by considering 
the pure proton composition of UHECRs as a conservative case study and at 
present our results may not be used to favour or disfavour whether it is the 
non-standard or standard cosmology, as we need to do more work to confirm our 
results and to rule out other possible explanations for our findings.
Consequently, it is worth mentioning that by extending the work with these 
models, it would be interesting to study the localised low-scale anisotropies 
of CRs that arise at their highest energies. So, we keep this as one 
of the future prospects of study.

\section*{Acknowledgements} UDG is thankful to the Inter-University Centre for 
Astronomy and Astrophysics (IUCAA), Pune, India for awarding the Visiting 
Associateship of the institute.

\appendix
\section {Parametric function for the generation energy $\boldmath{\text{E}_\text{g}}$} \label{append}
Owing to the complex nature of the dependence of the generation energy 
$E_\text{g}$ on the energy $E$ of UHECR particles, we consider a 
parametric function for the generation energy in this work as given by 
$$F(c_1,c_2,c_3,c_4)\equiv c_1 E + c_2 E^2 \exp{\left(-\frac{c_3}{E}\right)}+(1+c_4 E)\exp{\left(-\frac{1}{E}\right)},$$ 
\begin{figure}[h!]
\centerline{
\includegraphics[scale=0.35]{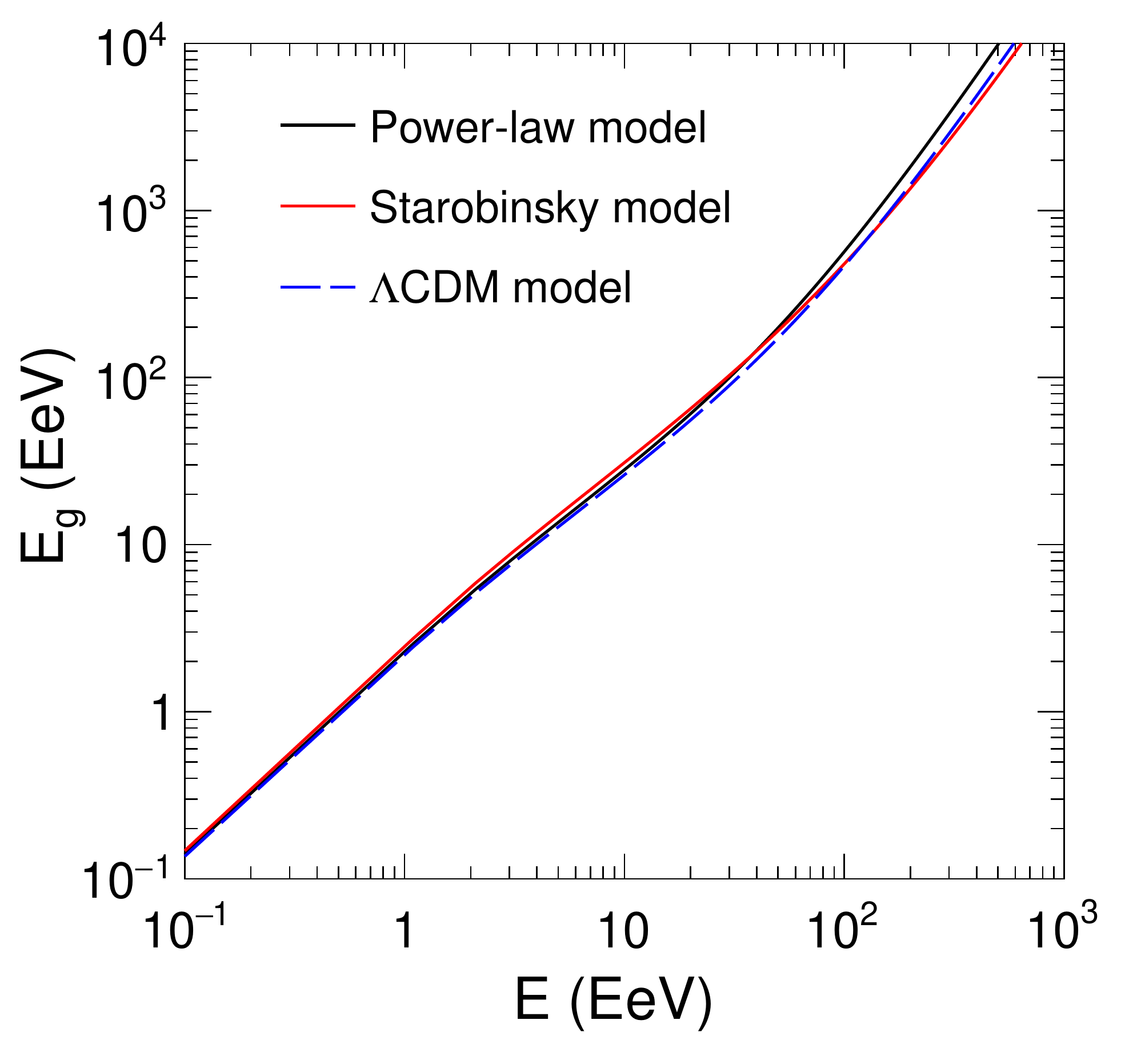}}
\vspace{-0.3cm}
\caption{Variation of generation energy $E_\text{g}$ with respect to 
energy $E$ for the $f(R)$ gravity power-law model ($n = 1.4$) and Starobinsky model 
in comparison with that for the $\Lambda$CDM model.}
\label{fig20}
\end{figure}
where $c_1$, $c_2$, $c_3$ and $c_4$ are constant parameters to be determined. 
In the function, the first term represents the energy loss due to red-shift 
(expansion of the Universe), the second term the energy loss due to the pair 
production process with the CMB and the third term the photopion reaction with 
the CMB that dominates at higher energies \cite{berezinsky_book}. We estimate the $E_\text{g}$ for the power-law model as
\begin{equation}
E_\text{g} \simeq F(1.8,0.04,-20,1.3),
\end{equation}
for the Starobinsky model as
\begin{equation}
E_\text{g} \simeq F(2.2,0.035,-18.5,1.3),
\end{equation}
and for the $\Lambda$CDM model as
\begin{equation}
E_\text{g} \simeq F(1.6,0.025,-18.5,1.25),
\end{equation}
In Fig.\ \ref{fig20}, a variation of $E_\text{g}$ with respect to $E$ is 
plotted for the power-law model and the Starobinsky model along with for the
$\Lambda$CDM model. For $E < 1$ EeV, the variation is linear, and above this 
energy $E_\text{g}$ is increasing non-linearly with the energy $E$. The 
difference between the estimated $E_\text{g}$ by the power-law 
model and the Starobinsky model is noticeable at higher energies above $1$ EeV.
Again at higher energies, the prediction of the Starobinsky model is 
nearly similar to that of the $\Lambda$CDM model.

\end{document}